\documentclass[11pt]{article}
\usepackage[left=1in,top=1in,right=1in,bottom=1in,head=0in,nofoot]{geometry}

\usepackage{setspace,url,bm,amsmath} 
\usepackage{scalerel}

\usepackage{titlesec} 
\titlelabel{\thetitle.\quad} 
\titleformat*{\section}{\bf\Large\center}

\usepackage{float}
\usepackage{bbm}
\usepackage{latexsym}
\usepackage{caption}

\usepackage[utf8]{inputenc}
\usepackage{amsmath, amssymb, amsthm, bbm, mathrsfs, mathtools}
\usepackage[colorlinks,citecolor=blue,urlcolor=blue, hypertexnames=false]{hyperref}
\usepackage[linesnumbered,ruled,vlined]{algorithm2e}
\usepackage{verbatim}
\usepackage{multirow}
\usepackage{booktabs, array, threeparttable}

\usepackage{rotating}
\usepackage{graphicx}
\usepackage{subfigure}
\usepackage{float}
\usepackage{caption}
\usepackage{xcolor}
\usepackage{amsfonts}
\usepackage{xr}
\usepackage{tabularx, multicol}
\usepackage{enumitem}

\usepackage{etoolbox}
\usepackage{tabularx, booktabs, graphicx}

\makeatletter
\newcommand{\ostar}{\mathbin{\mathpalette\make@circled\star}}
\newcommand{\make@circled}[2]{%
  \ooalign{$\m@th#1\smallbigcirc{#1}$\cr\hidewidth$\m@th#1#2$\hidewidth\cr}%
}
\newcommand{\smallbigcirc}[1]{%
  \vcenter{\hbox{\scalebox{0.77778}{$\m@th#1\bigcirc$}}}%
}
\makeatother

\newcommand{\lt}{\left}
\newcommand{\rt}{\right}

\newcommand{\commenting}[1]{}

\renewcommand{\hat}{\widehat}
\renewcommand{\tilde}{\widetilde}

\newcommand{\Var}[1]{{\operatorname{Var}\left\{#1\right\}}}

\newcommand{\E}[1]{{\bbE\left\{#1\right\}}}
\newcommand{\Prob}[1]{{\bbP\left\{#1\right\}}}

\newcommand{\ind}[1]{\boldsymbol{1}\left\{#1\right\}}

\ifdef{\see}{\renewcommand{\see}[1]{\text{ (#1)}}}{\newcommand{\see}[1]{\text{ (#1)}}}

\newcommand{\diag}[1]{\mathrm{Diag}\left\{#1\right\}}

\def\boxit#1{\vbox{\hrule\hbox{\vrule\kern6pt\vbox{\kern6pt#1\kern6pt}\kern6pt\vrule}\hrule}}

\newcolumntype{P}[1]{>{\centering\arraybackslash}p{#1}}
\newcolumntype{M}[1]{>{\centering\arraybackslash}m{#1}}
\newcolumntype{L}[1]{>{\raggedright\arraybackslash}m{#1}}

\newcommand{\cE}{{\mathcal{E}}}

\newcommand{\cL}{{\mathcal{L}}}
\newcommand{\cT}{{\mathcal{T}}}

\newcommand{\cN}{{\mathcal{N}}}

\newcommand{\cK}{{\mathcal{K}}}

\newcommand{\bbP}{{\mathbb{P}}}

\newcommand{\bbR}{{\mathbb{R}}}

\newcommand{\bbE}{{\mathbb{E}}}

\newcommand{\bbM}{{\mathbb{M}}}
\newcommand{\bbK}{{\mathbb{K}}}

\newcommand{\bone}{{\boldsymbol{1}}}

\newcommand{\bse}{{\boldsymbol{e}}}

\newcommand{\bD}{{\boldsymbol{D}}}

\newcommand{\bP}{{\boldsymbol{P}}}
\newcommand{\bQ}{{\boldsymbol{Q}}}

\newcommand{\bSigma}{{\boldsymbol{\Sigma}}}

\newcommand{\hgamma}{{\hat{\gamma}}}

\newcommand{\hepsilon}{{\hat{\epsilon}}}
\newcommand{\htau}{{\hat{\tau}}}

\newcommand{\hmu}{{\hat{\mu}}}

\newcommand{\hv}{{\hat{v}}}

\newcommand{\hS}{{\hat{S}}}
\newcommand{\hV}{{\hat{V}}}

\newcommand{\hY}{{\hat{Y}}}

\newcommand{\hGamma}{{\hat{\Gamma}}}

\newcommand{\GG}[1]{}

\usepackage{amsthm}
\usepackage{amssymb}
\usepackage{amsmath}
\usepackage{color}

\usepackage{comment}
\theoremstyle{definition}
\newtheorem{assumption}{Assumption}
\newtheorem*{theorem*}{Theorem}
\newtheorem{theorem}{Theorem}
\newtheorem*{rmk*}{remark}
\newtheorem{proposition}{Proposition}
\newtheorem{lemma}{Lemma}
\newtheorem{example}{Example}

\newtheorem{condition}{Condition}

\newtheorem{definition}{Definition}
\newtheorem{remark}{Remark}
\newtheorem{corollary}{Corollary}
\newtheorem*{corollary*}{Corollary}
\newtheorem{strategy}{Strategy}

\usepackage{natbib} 
\bibpunct{(}{)}{;}{a}{}{,} 

\usepackage{etoolbox} 
\apptocmd{\sloppy}{\hbadness 10000\relax}{}{} 

\usepackage{multibib}
\newcites{sec}{References}

\usepackage{listings}

\usepackage{lscape}

\usepackage{tikz}
\usetikzlibrary{shapes.geometric, arrows}
\tikzstyle{startstop} = [rectangle, rounded corners, minimum width=3cm, minimum height=2cm,text centered, text width = 3cm, draw=black, fill=red!30]
\tikzstyle{io} = [trapezium, trapezium left angle=70, trapezium right angle=110, minimum width=3cm, minimum height=1cm, text centered, draw=black, fill=blue!30]
\tikzstyle{process} = [rectangle, minimum width=3.7cm, minimum height=2cm, text centered, draw=black, text width=3.5cm, fill=orange!30]
\tikzstyle{decision} = [diamond, minimum width=3cm, minimum height=1cm, text centered, draw=black, text width = 3cm, fill=green!30]
\tikzstyle{arrow} = [thick,->,>=stealth]

\mathtoolsset{showonlyrefs=true}

\RequirePackage[normalem]{ulem}

\allowdisplaybreaks

\begin{document}
\doublespacing

\title{\bf  
 Forward selection and post-selection inference in factorial designs
} 
\author{Lei Shi, Jingshen Wang and Peng Ding
\footnote{Lei Shi, Division of Biostatistics, University of California, Berkeley, CA 94720  (E-mail: leishi@berkeley.edu). Jingshen Wang, Division of Biostatistics, University of California, Berkeley, CA 94720  (E-mail: jingshenwang@berkeley.edu). Peng Ding, Department of Statistics, University of California, Berkeley, CA 94720 (E-mail: pengdingpku@berkeley.edu).
}
}
\date{}
 
\maketitle

\begin{abstract}
 Ever since the seminal work of R. A. Fisher and F. Yates, factorial designs have been an important experimental tool to simultaneously estimate the effects of multiple treatment factors. In factorial designs, the number of treatment combinations grows exponentially with the number of treatment factors, which motivates the forward selection strategy based on the sparsity, hierarchy, and heredity principles for factorial effects. Although this strategy is intuitive and has been widely used in practice, its rigorous statistical theory has not been formally established. To fill this gap, we establish design-based theory for forward factor selection in factorial designs based on the potential outcome framework. We not only prove a consistency property for the factor selection procedure but also discuss statistical inference after factor selection. In particular, with selection consistency, we quantify the advantages of forward selection based on asymptotic efficiency gain in estimating factorial effects. With inconsistent selection in higher-order interactions, we propose two strategies and investigate their impact on subsequent inference. Our formulation differs from the existing literature on variable selection and post-selection inference because our theory is based solely on the physical randomization of the factorial design and does not rely on a correctly specified outcome model.
\end{abstract}


\newpage

\section{Introduction}\label{sec:introduction}
\subsection{Factorial experiments: opportunities and challenges}

Ever since the seminal work of \citet{Fisher35} and \citet{yates1937design}, factorial designs have been widely used in many fields, including agricultural, industrial, biomedical, and social sciences \citep[e.g.,][]{box2005statistics, wu2011experiments,  gerber2012field}. 
Factorial experiments are popular because they can simultaneously accommodate multiple factors and offer opportunities to estimate not only the main effects of factors but also their interactions.

We focus on the replicated $2^K$ factorial design in which $K$ binary factors are randomly assigned to $N$ experimental units, and each treatment combination contains at least two replicates. Classical factorial experiments are usually conducted with a small $K$ so that we can simultaneously estimate the $2^K-1$ main effects and interactions. For example, Chapter 4 of \cite{wu2011experiments} discussed many full factorial experiments, all of which involve less than four factors ($K\le 4$). 
However, many modern factorial experiments are conducted on a much larger scale for exploring complex research questions. For example, in political science and market research, powered by the development of computers and web-based technology, conjoint survey experiments \citep{luce1964simultaneous, caughey2019item, hainmueller2014causal, zhirkov2022estimating}, which can be viewed as a special type of factorial experiments, are popular for analyzing the effects of many factors together. In Table \ref{tab:factorial-setup-rp}, we list several concrete conjoint experiments in the literature and their corresponding setups. These modern factorial experiments involve a large number of treatment combinations, which motivate us to move beyond the classical small $K$ regimes and develop methodology and theory for large $K$ regimes. 
\begin{table}[ht!]
\caption{Some conjoint survey experiments and the corresponding setup}
\label{tab:factorial-setup-rp}
\centering
\resizebox{0.85 \textwidth}{!}{
\begin{threeparttable}
  \begin{tabular}{P{7cm}ccccc}
        \toprule
        \textbf{Experiment}  & \textbf{Reference}  & $K$ & $Q$ & $N$ & $N_0$ \\
        \midrule
        Immigrant admission experiment & \cite{zhirkov2022estimating} & $6$ & $2^6 = 64$ & $\sim 28,000$ & $\sim 430$ \\
        U.S. presidential election  &  \cite{caughey2019item} & $12$ & $ 2^{12} = 4096 $ & $\sim 30,000 $ & $ \sim 8 $ \\
        Aluminum packaging characteristics & \cite{li2013conjoint} & $7$ & $2^6 * 4 = 256$ & $\sim 15,000$ & $\sim 60$ \\
        \bottomrule
        \end{tabular}
  \begin{tablenotes}
    \item Note: $K$ is the number of factors, $Q$ is the number of treatment combinations, $N$ it the number of units (hypothetical profiles), $N_0$ is the average replications per arm. See Section \ref{Sec:setup} for rigorous definition. 
  \end{tablenotes}
\end{threeparttable}
}
\end{table}

A large number of factors pose new challenges to the analysis of factorial experiments. First, estimation and inference of the causal effects fall into new regimes, especially when $K$ is large and each treatment combination only contains limited replications. Second, a large $K$ results in a large set of factorial effects, which complicate the interpretation of the results. 
This motivates us to conduct factor selection based on \textit{sparsity, hierarchy, and heredity} principles for factorial effects to reduce the dimensionality of the problem and facilitate the interpretation of the results. \citet{wu2011experiments} summarized these three principles as below:
\begin{enumerate}
\item
[(a)]
(sparsity) The number of important factorial effects is small.
\item
[(b)]
(hierarchy) Lower-order effects are more important than higher-order effects, and effects of the same order are equally important.
\item
[(c)]
(heredity) Higher-order effects are important only if their corresponding lower-order effects are important. 
\end{enumerate}

The sparsity principle motivates conducting factor selection in factorial designs. The hierarchy principle motivates the forward selection strategy that starts from lower-order effects and then moves on to higher-order effects. The heredity principle motivates using structural restrictions to select higher-order effects based on the selected lower-order effects. Due to its simplicity and computational efficiency, the forward selection strategy has been widely used in data analysis \citep{wu2011experiments, espinosa2016bayesian}. However, its design-based theory under the potential outcome framework has not been formally established. Moreover, it is often challenging to understand the impact of factor selection on the subsequent statistical inference. The overarching goal of this manuscript is to fill these gaps.

\subsection{Our contributions and literature review}\label{sec:contribution-lit-review}

We summarize our contributions from the following three perspectives.  

First, our study adds to the growing literature of factorial designs with a growing number of factors under the potential outcome framework \citep{dasgupta2015causal, branson2016improving, lu2016randomization, espinosa2016bayesian, egami2019causal, blackwell2023noncompliance, zhao2021regression, pashley2019causal, wu2022non}. To deal with a large number of factors, \citet{espinosa2016bayesian} and \citet{egami2019causal} informally used factor selection without studying its statistical properties, whereas \citet{zhao2021regression} discussed parsimonious model specifications that are chosen a priori and independent of  data. The rigorous theory for factor selection is missing in this literature, let alone the theory for statistical inference after factor selection. At a high level, our paper fills the gaps.

Second, we formalize forward factor selection and establish its consistency under the design-based framework without imposing outcome modeling assumptions; see Section \ref{sec:forward-framework}. Factor selection in factorial design sounds like a familiar statistical task if we formulate it as a variable selection problem in a linear model. Thus, forward selection is reminiscent of the vast literature on forward selection. \citet{wang2009forward} and \citet{jerzy2022model} proved the consistency of forward selection for the main effects in a linear model, whereas \citet{hao2014interaction} and \citet{hao2018model} moved further to allow for second-order interactions. Other researchers proposed various penalized regressions to encode the sparsity, hierarchy, and heredity principles \citep[e.g.,][]{yuan2007efficient, zhao2009composite, bickel2010hierarchical, bien2013lasso, lim2015learning, haris2016convex}, without formally studying the statistical properties of the selected model. Our design-based framework departs from the literature without assuming a correctly-specified linear outcome model. This framework is classic in experimental design and causal inference with randomness coming solely from the design of experiments rather than the error terms in a linear model \citep{neyman1923application, kempthorne1952design, freedman2008regression, lin2013agnostic, dasgupta2015causal}. This framework invokes fewer outcome modeling assumptions but consequently imposes technical challenges for developing the theory. \citet{bloniarz2016lasso} discussed the design-based theory for covariate selection in treatment-control experiments, but the corresponding theory for factorial designs is largely unexplored.

Third, we discuss statistical inference after forward factor selection with consistent (see Sections {\ref{sec:inference-perfect}}) and inconsistent selection (see Section \ref{sec:imperfect-ms}). On the one hand, we prove the selection consistency of the forward selection procedure, which ensures that the selected factorial effects are the true, non-zero ones as the sample size grows. With this selection consistency property, we can then proceed as if the selected working model is the true model. This allows us to ignore the impact of forward selection on the subsequent inference, which is similar to the proposal of \citet{zhao2021defense} for statistical inference after Lasso \citep{tibshirani1996regression}. Moreover, we quantify the advantages of conducting forward selection based on the asymptotic efficiency gain for estimating factorial effects. As an application under selection consistency, we discuss statistical inference for the mean outcome under the best factorial combination in Section \ref{sec:select-best} in the appendix \citep{andrews2019inference, guo2021sharp, wei2023inference}. On the other hand, we acknowledge that selection consistency can be difficult to achieve in practice as it requires strong regularity conditions on factorial effects. As a remedy, we propose two strategies to deal with inconsistent selection in higher-order interactions, and study their impacts on post-selection inference. A key motivation for our strategies is to ensure that the parameters of interest after forward factorial selection are not data-dependent, avoiding philosophical debates in the current literature of post-selection inference \citep{fithian2014optimal, kuchibhotla2022post}.


\subsection{Notation}  

We will use the following notation throughout. For asymptotic analyses, $a_N = {O}(b_N)$ denotes that there exists a positive constant $C>0$ such that $a_N \le Cb_N$; $a_N = o(b_N)$ denotes that $a_N/b_N \to 0$ as $N$ goes to infinity; $a_N =  \Theta(b_N)$ denotes that there exists positive constants $c$ and $C$ such that $cb_N\le a_N \le Cb_N$.

For matrix $V$, define $\varrho_{\max}(V)$ and $ \varrho_{\min}(V)$ as the largest and smallest eigenvalues, respectively, and define $\kappa(V) = \varrho_{\max}(V)/\varrho_{\min}(V)$ as its condition number. For two positive semi-definite matrices $V_1$ and $V_2$, we write $V_1 \preccurlyeq V_2$ or $V_2 \succcurlyeq V_1 $ if $V_2 - V_1$ is positive semi-definite, which is called the Loewner order for positive semidefinite matrices.

We will use different levels of sets. For an integer $K$, let $[K] = \{1,\ldots,K\}$. We use $\cK$ in calligraphy to denote a subset of $[K]$. Let $\bbK = \{\cK\mid \cK\subset[K]\}$ denote the power set of $[K]$. We also use blackboard bold font to denote subsets of $\bbK $. For example, $\bbM \subset \bbK$ denotes that $\bbM$ is a subset of  $\bbK$.

We will use $A_i\sim B_i$ to denote the least-squares fit of $A_i$'s on $B_i$'s, which is purely a numerical procedure without assuming a linear model. Let $\xrightarrow{{\mathrm{P}}} $ denote convergence in probability, and $ \rightsquigarrow$ denote convergence in distribution.

\section{Setup of factorial designs}\label{Sec:setup}

This section introduces the key mathematical components of factorial experiments. 
Section \ref{sec::po-factorialeffects} introduces the notation of potential outcomes and the definitions of the factorial effects.
Section \ref{sec::treatment-data-regression} introduces the treatment assignment mechanism, the observed data, and the regression analysis of factorial experiment data. 
Section \ref{sec::example-2-3factorial}  uses  a concrete example of a $2^3$ factorial experiment to illustrate the key concepts.

\subsection{Potential outcomes and factorial effects}\label{sec::po-factorialeffects}

We first introduce the framework of a $2^K$ factorial design, with $K\ge 2$ being an integer. The design has $K$ binary factors, and factor $k$ can take value $z_k\in\{-1,1\}$ for $k=1,\ldots, K$; we use the $\pm 1$ coding of the factors because of its convenience for later parts and can modify the results under the 0-1 coding of the factors. 
Let $\boldsymbol{z}  = (z_1,\dots ,z_K)$ denote the treatment combining all $K$ factors.
 The $K$ factors in total define $Q = 2^K$ treatment combinations, collected in the set below:
\begin{align*}
    \mathcal{T}=\{z=(z_1,\dots ,z_K)\mid z_k\in\{-1,1\} \text{ for } k=1,\cdots,K\} \quad  \text{ with } \quad |\mathcal{T}| = Q. 
\end{align*}

We follow the potential outcome notation of \citet{dasgupta2015causal}  for $2^K$ factorial designs. 
Unit $i$ has potential outcome $Y_i(z)$ under treatment combination $z$. Corresponding to the $Q=2^K$ treatment combinations, unit $i$ has $Q$ potential outcomes, vectorized as $\bm{Y}_i = \{Y_i(z)\}_{z\in\cT}$ using the lexicographic order based on the treatments. Over units $i=1,\ldots, N$, the 
potential outcomes have finite-population mean vector $\overline{Y} = (\overline{Y}(z))_{z\in\cT}$ and covariance matrix $S = (S(z,z'))_{z,z'\in\cT}$, with elements defined as follows:
\begin{gather}\label{eqn:Yz-S}
    \overline{Y}(z) = \frac{1}{N}\sum_{i=1}^N Y_i(z), \quad
    S(z,z') = \frac{1}{N-1}\sum_{i=1}^N (Y_i(z) - \overline{Y}(z))(Y_i(z') - \overline{Y}(z')).
\end{gather}

We then use the potential outcomes to define factorial effects. For a subset $\mathcal{K}\subset [K] $ of the $K$ factors, we introduce the following ``contrast vector" notation to facilitate the presentation. To start with, we define the main causal effect for factor $k$. For a treatment combination $z = (z_1, \ldots, z_K)\in \cT$, we use $g_{\{k\}}(z) = z_k$ to denoted the ``centered" treatment indicator $z_k$. 
We then define a $Q$-dimensional contrast vector $ g_{\{k\}}$ by concatenating these centered treatment variables using the lexicographic order, that is
\begin{align}\label{eq:contrast-main}
    g_{\{k\}} = \{g_{\{k\}}(z)\}_{z\in\cT}, \text{ where } g_{\{k\}}(z) = z_k. 
\end{align}
Next, for the interactions of multiple factors in $\cK$ with $|\cK|\ge 2$, we define the contrast vector $ g_\cK \in \mathbb{R}^{Q}$  as
\begin{align}\label{eq:contrast-vector}
    g_\cK = \{g_\cK(z)\}_{z\in\cT},  \text{ where }  g_{\cK}(z) = \prod_{k\in\cK} g_{\{k\}}(z) = \prod_{k\in\cK} z_k.
\end{align}
Finally, for the average of potential outcomes, we define $g_\varnothing = \bone_{Q}$, which is orthogonal to all the contrast vectors. Stack the $g_\cK$'s into a $Q\times Q$ matrix 
\begin{align}\label{eqn:G}
G = (g_\varnothing, g_{\{1\}} , \ldots, g_{\{K\}} , g_{\{1,2\}},\ldots, g_{\{K-1,K\}}, \ldots,  g_{[K]}  ),
\end{align} 
which has orthogonal columns with  $G^{\top} G = Q \cdot I_Q$. We refer to $G$ as the contrast matrix.

Equipped with the contrast vector notation, we are ready to introduce the main effects and interactions. We define the main causal effect of a single factor and the $k$-way interaction causal effect of multiple factors ($k \ge 2$)  as the inner product of the contrast vector $g_{\mathcal{K}}$, and the potential outcome mean vector $\overline{Y}$:
\begin{align}\label{eq:def-k-way-interaction-effect}
    \tau_{\mathcal{K}} = Q^{-1} \cdot g_{\mathcal{K}}^\top \overline{Y}  \quad \text{ for } \quad {\mathcal{K}}\subset [K]. 
\end{align}
For convenience in description, we use $\tau_\varnothing = Q^{-1}g_\varnothing^\top \overline{Y}$ to denote the average of potential outcomes. 
We call the effect $\tau_\cK$ a \textit{parent} of $\tau_{\cK'}$ if $\cK\subset\cK'$ and $|\cK| = |\cK'| - 1$. We summarize 
the entire collection of causal parameters as 
 \begin{align}\label{eqn:def-factorial-effects}
     \tau = (\tau_{\mathcal{K}})_{\mathcal{K}\subset [K]} = Q^{-1}\cdot {G}^\top \overline{Y} . 
 \end{align}
The above definition for factorial effects differs from \cite{dasgupta2015causal} by a constant of 2, which does not change the problem fundamentally but has a better mathematical structure under our framework.

 \subsection{Treatment assignment, observed data, and regression analysis}
  \label{sec::treatment-data-regression}

Under the design-based framework, the treatment assignment mechanism characterizes the completely randomized factorial design. The experimenter randomly assigns $N(z)$ units to treatment combination $z\in \mathcal{T}$, with $\sum_{z \in \mathcal{T}} N(z) = N$. Assume $N(z) \ge 2$ to allow for variance estimation within each treatment combination. 
Let $Z_i \in \mathcal{T} $ denote the treatment combination for unit $i$. The treatment vector $(Z_1, \ldots, Z_N)$ is a random permutation of a vector with prespecified number $N(z)$ of the corresponding treatment combination $z$, for $z\in \mathcal{T}$.


For each unit $i$, the treatment combination $Z_i$ only reveals one potential outcome.  We use $Y_i = Y_i(Z_i) = \sum_{z\in\cT} Y_i(z)\ind{Z_i = z}$ to denote the observed outcome. We use  $N_i = N(Z_i)$ to denote the number of units for the treatment group to which unit $i$ is assigned to.
The central task of causal inference in factorial designs is to use the observed data $(Z_i, Y_i)_{i=1}^N$ to estimate the factorial effects. 
Define
$$
\hY(z) = N(z)^{-1}  \sum_{i=1}^N \ind{Z_i = z} Y_i  ,\quad
 \hat{S}(z,z) = \{N(z)-1\}^{-1} \sum_{i=1}^N \ind{Z_i = z} (Y_i - \hY(z))^2
$$
as the sample mean and variance of the observed outcomes under treatment $z$. Recalling that $S$ is the finite population covariance matrix of the potential outcomes defined in \eqref{eqn:Yz-S}. Let $D_\hY = \diag{ N(z)^{-1}S(z,z)}_{z\in\cT}$. Vectorize the sample means as $\hY = (\hY(z) )_{z \in \mathcal{T}}$, which has mean $\overline{Y}$ and covariance matrix $ V_\hY = D_\hY - N^{-1}S $ \citep{li2017general}. 
An unbiased estimator for $D_\hY$ is 
$$  
\hV_\hY = \diag{N(z)^{-1}\hS(z,z)}_{z\in\cT},
$$ 
whereas the covariance matrix $S$ does not have an unbiased sample analog because the potential outcomes across treatment combinations are never jointly observed for the same units. Therefore, $\hV_\hY $ is a conservative estimator of the covariance matrix of $\hY$ in the sense that $ \bbE\{\hV_\hY\} = D_\hY \succcurlyeq V_\hY  $.

A dominant approach to estimating factorial effects from factorial designs is through estimating least-squares coefficients based on appropriate model specifications. Let $g_i$ denote the row vector in the contrast matrix $G$ corresponding to unit $i$'s treatment combination $Z_i$, that is, $g_i = \{ g_\cK(Z_i) \}_{\cK\subset [K]}\in\mathbb{R}^{Q}$ with $g_\cK(z)$ defined in \eqref{eq:contrast-vector}. We can run ordinary least squares (OLS) to obtain unbiased estimates for the factorial effects:
\begin{align}\label{eqn:wls-form-saturated}
    \htau = \arg\min_{\tau} \sum_{i=1}^N (Y_i - g_{i}^\top \tau)^2.
\end{align}
With a small $K$, we can simply fit the \textit{saturated regression} by regressing the observed outcome $Y_i$ on the regressor $g_i$. The saturated regression involves $Q = 2^K$ coefficients without any restrictions on the targeted factorial effects. 

In contrast, an \textit{unsaturated regression} with weighted least squares (WLS) involves fewer coefficients by regressing the observed outcome $Y_i$ on $g_{i, \mathbb{M}}$, a subvector of $g_i$, where $\mathbb{M} \subset \mathbb{K}$ is a subset of the power set of all factors:
\begin{align}\label{eqn:wls-form}
    \htau = \arg\min_{\tau} \sum_{i=1}^N w_i(Y_i - g_{i,\bbM}^\top \tau)^2 \text{ with } w_i = 1/N_i.
\end{align}
The above least squares fits in \eqref{eqn:wls-form-saturated} and \eqref{eqn:wls-form} are based on a fact in factor-based regressions: to get unbiased estimates for a set of factorial effects, one can either run OLS/WLS with a saturated model or run WLS including that particular set of effects with an unsaturated model. Such a result is established in, for example, Section 5.4 and Section A.5 of \cite{zhao2021regression}.

For the convenience of description, we will call $\bbM$ a \emph{working model}. We use a working model to generate estimates based on least squares without assuming the corresponding linear model is correct. When $\bbM = \bbK$, \eqref{eqn:wls-form} incorporates the saturated regression \eqref{eqn:wls-form-saturated}. 
Based on the unsaturated regression with working model $\bbM$, let  
$$
\hat{\tau}(\bbM)  = \{\widehat \tau_\cK\}_{\cK\in\bbM} \quad \text{ and } \quad 
\tau(\bbM) = \{\tau_\cK\}_{\cK\in\bbM}
$$
denote the vectors of estimated and true coefficients, respectively. Because $\hat{\tau}(\bbM) $ is a linear transformation of $\widehat{Y}$, we can use the following estimator for its covariance matrix:
\begin{align}\label{eqn:hSigma}
          \hat{\bSigma}({\bbM})  = \frac{1}{Q^2}G(\cdot,{\bbM})^\top  \hV_\hY G(\cdot,{\bbM}).
\end{align} 
See Lemma \ref{lem:wls-property} in  Section \ref{sec:wls} of the supplementary material for more discussions on the above algebraic results for unsaturated regressions. 
 
\begin{remark}\label{rmk:saturated}
    The terminology ``saturated regression'' here means that the factor-based linear regression takes the full set of elements in $g_i$ as regressors, while the ``unsaturated regression'' only takes a subset. In the experimental design literature, the word ``saturated'' has a different meaning in the terminology ``saturated design'', which is used to indicate that the number of observations in the design equals the number of effects
    in the model. 
\end{remark}

\subsection{An example of a $2^3$ factorial design}
\label{sec::example-2-3factorial}

The above notation can be abstract. In this section, we provide an illustrating Example \ref{exp:2-3-fac-design} below with $K=3$ factors. 

\begin{example}[$2^3$ factorial design]\label{exp:2-3-fac-design}
Suppose we have three binary factors $z_1$, $z_2$, and $z_3$. These three factors generate $8$ treatment combinations, indexed by a triplet $(z_1z_2z_3)$ with $z_1,z_2,z_3\in\{-1,1\}$, in the set 
\begin{align*}
\mathcal{T} = \{(---), (--+), (-+-),(-++),(+--), (+-+), (++-),(+++)\}.
\end{align*}
Each unit $i$ has a potential outcome vector $\bm{Y}_i = \{Y_i(z_1z_2z_3)\}_{z_1,z_2,z_3=-1,1}^\top$. The vector of factorial effects is 
\begin{align*}
    {\tau} = \frac{1}{2^3} {G}^\top\overline{Y} \triangleq \left(\tau_\varnothing, \tau_{\{1\}},\tau_{\{2\}},\tau_{\{3\}}, \tau_{\{1,2\}}, \tau_{\{1,3\}}, \tau_{\{2,3\}}, \tau_{\{1,2,3\}}\right)^\top,
\end{align*}
where $G$ is the contrast matrix 
\[
{G} = 
\bordermatrix{
~ & \tau_\varnothing & \tau_{\{1\}} & \tau_{\{2\}} & \tau_{\{3\}} & \tau_{\{1,2\}} & \tau_{\{1,3\}} & \tau_{\{2,3\}} & \tau_{\{1,2,3\}} \cr
(---) & 1 & -1 & -1 & -1 &  1 & 1 & 1 & -1\cr
(--+) & 1 & -1 & -1 & 1  &  1 & -1 & -1 & 1\cr
(-+-) & 1 & -1 & 1  & -1 & -1 & 1 & -1 &  1\cr
(-++) & 1 & -1 &  1  &  1 & -1 & -1 & 1 & -1\cr
(+--) & 1 & 1 & -1  & -1 & -1 & -1 & 1 & 1\cr
(+-+) & 1 & 1 & -1  &  1 & -1 & 1 & -1 & -1\cr
(++-) & 1 & 1 & 1   & -1 &  1 & -1 & -1 & -1\cr
(+++) & 1 & 1 & 1   &  1 &  1 & 1 & 1 & 1
}.
\]
We observe the pair $(Y_i, {Z}_i)$ for unit $i$, where ${Z}_i = (z_{i,1},z_{i,2},z_{i,3})$ is the observed treatment combination for unit $i$. Recall $g_{\{k\}}(Z_i) = z_{i,k}$.  For the factor-based regression, the regressor $g_i$ corresponding to the treatment combination $Z_i$ equals 
\begin{align*}
{g}_i = \Big[&1, ~g_{\{1\}}(Z_i), ~g_{\{2\}}(Z_i), ~g_{\{3\}}(Z_i),
~g_{\{2,3\}}(Z_i), ~g_{\{1,3\}}(Z_i),  ~g_{\{1,2\}}(Z_i), ~ g_{\{1,2,3\}}(Z_i)\Big].
\end{align*}
For instance, when $Z_i = (+-+)$, the regressor $g_i$ corresponds to the row $(+-+)$ of the contrast matrix $G$. Then, a saturated regression is to regress $Y_i$ on $g_i$. 
For the unsaturated regression, if we only include indices $\varnothing$, $\{1\},\{1,2\},\{1,3\},\{1,2,3\}$, we can form the working model $\bbM=\{\varnothing, \{1\},\{1,2\},\{1,3\},\{1,2,3\}\}$ and perform WLS $Y_i\sim g_{i,\bbM},$ where 
\begin{align*} 
g_{i,\bbM} = &\Big[1, ~g_{\{1\}}(Z_i), 
 ~g_{\{1,3\}}(Z_i),  ~g_{\{1,2\}}(Z_i), ~ g_{\{1,2,3\}}(Z_i)\Big] 
\end{align*} 
and the weight for unit $i$ equals $1/N_i = 1/N(Z_i)$. 
\end{example}

\section{Forward selection in factorial experiments} \label{sec:forward-framework}

In factorial designs with small $K$, we can run the saturated regression to estimate all factorial effects simultaneously \citep{lu2016randomization, zhao2021regression}. However, when $K$ is large, estimation and inference of the causal effects fall into new regimes, especially when each treatment combination only contains limited replications. Second, it is difficult to interpret a large number of estimates for factorial effects. As a remedy, forward selection is a popular strategy frequently adopted to analyze data collected from factorial experiments, due to its benefits in ruling out zero nuisance factorial effects. In this section, we formalize forward selection as a principled procedure to select an unsaturated working model $\hat{\bbM}$. We first present a formal version of forward selection and then demonstrate its consistency property.

\subsection{A formal forward selection procedure}\label{sec::forward-ms-alg}

In this subsection, we introduce a principled forward selection procedure that not only respects the effect hierarchy, sparsity, and heredity principles but also results in an interpretable parsimonious model with statistical guarantees. More concretely, the algorithm starts by performing factor selection over lower-order effects, then moves forward to select higher-order effects following the heredity principle. Algorithm \ref{alg:forward-ms} summarizes the forward selection procedure. In what follows, we illustrate why the proposed procedure in Algorithm \ref{alg:forward-ms} respects the three fundamental principles in factorial experiments.

\begin{algorithm}[!ht]

\DontPrintSemicolon
\SetKwInput{KwInput}{Input}                
\SetKwInput{KwOutput}{Output}  
\SetKwFunction{MADES}{MADE-S}
\SetKwProg{Fn}{Function}{:}{\KwRet}
  \KwInput{Factorial data $\{(Y_i, {Z}_i)\}_{i=1}^N$; prespecified integer $D \leq K$; initial working model $\hat{\bbM} = \{\varnothing\}$; prespecified significance levels $\{\alpha_d\}_{d=1}^D$.}
  \KwOutput{Selected working model $\hat{\bbM}$. }
    
    Define an intermediate working model $\hat{\bbM}' = \hat{\bbM}$ for convenience.
    
    \For{$d = 1,\ldots, D$ \label{alg:loop}}
    {
      Update the intermediate working model to include all the $d$-order (interaction) terms: $\hat{\bbM}' =  \hat{\bbM} \cup \{\cK\mid |\cK| = d\} \triangleq\hat{\bbM} \cup \bbK_d$.\label{alg:step-add-d-way} 
         
      Drop indices in $\hat{\bbM}'$ according to either the weak or strong heredity principles
      , and  renew the selected working model as $\hat{\bbM}'$.  \label{alg:step-prune}      
      
      Run the unsaturated regression with the working model $\hat{\bbM}'$:
       \begin{align*}
          Y_i \sim g_{i, \hat{\bbM}'}, \text{ with weights } w_i = N/N_i.
       \end{align*}

      Obtain coefficients $\hat{\tau}(\hat{\bbM}')$ and robust covariance estimation $\hat{\bSigma}(\hat{\bbM}')$ defined in \eqref{eqn:hSigma}. \label{alg:BC-1}

       \label{alg:BC-2} Obtain $\hat{\tau}_{\cK}(\hat{\bbM}')$ and $\hat{\sigma}_\cK(\hat{\bbM}')$ for all $\cK\in\hat{\bbM}'$ with $|\cK| = d$, where 
       $\hat{\sigma}_\cK(\hat{\bbM}')$ is the variance estimator in the diagonal values of $\hat{\bSigma}(\hat{\bbM}')$ corresponding to the factor combination $\cK$. 

      Run marginal $t$-tests using the above $\hat{\tau}_{\cK}(\hat{\bbM}')$ and $\hat{\sigma}_\cK(\hat{\bbM}')$  under the significance level $\min\{\alpha_d/(|\hat{\bbM}'|-|\hat{\bbM}|),1\}$ and remove the non-significant terms from $\hat{\bbM}'\backslash\hat{\bbM}$.  \label{alg:BC-3}

      Set $\hat{\bbM} = \hat{\bbM}'$.
         
    }

    \KwRet{$\hat{\bbM}$.}
    
\caption{Forward factorial selection}
\label{alg:forward-ms}
\end{algorithm}

First, Algorithm \ref{alg:forward-ms}  obeys the hierarchy principle as it performs factor selection in a forward style (coded in the global loop from $d = 1$ to $d = D \le K$ with prespecified $D$, Step \ref{alg:loop} in particular). More concretely, we begin with an empty working model. We then select main effects (Steps 4 and 8) and add them into the working model. Once the working model is updated, we continue to select higher-order interaction effects in a forward style. Such a forward selection procedure is again motivated by the hierarchy principle that lower-order effects are more important than higher-order ones. 

Second, Algorithm \ref{alg:forward-ms} operates under the sparsity principle as it removes potentially unimportant effects using marginal $t$-tests with the Bonferroni correction (see Step \ref{alg:BC-3}). 
This step induces a sparse working model and helps us to identify important factorial effects. The sparsity-inducing step can incorporate many popular selection frameworks, such as marginal $t$-tests, Lasso \citep{tibshirani1996regression}, sure independence selection \citep{fan2008sure}, etc. For simplicity, we present Algorithm \ref{alg:forward-ms}  with marginal $t$-tests and relegate general discussions to Section \ref{sec:general-consistency} of the supplementary material.

Third, Algorithm \ref{alg:forward-ms} incorporates the heredity principle as it rules out the interaction effects \citep{wu2011experiments, hao2014interaction, lim2015learning} when either none of their parent effects is included (weak heredity) or some of their parent effects are excluded (strong heredity) in the previous working model (see Step \ref{alg:step-prune}). 
 
Lastly, Algorithm \ref{alg:forward-ms} enhances the interpretability of the selected working model by iterating between the ``Sparsity-selection" step (Step \ref{alg:BC-3}, called the S-step in the rest of the manuscript), captured by a data-dependent operator $\hat{\texttt{S}} = \hat{\texttt{S}}(\cdot;\{Y_i,Z_i\}_{i=1}^N)$, and the ``Heredity-selection" step (Step \ref{alg:step-prune}, called the H-step in the rest of the manuscript), captured by a deterministic operator $\texttt{H} = \texttt{H}(\cdot)$. Because the working model is updated in an iterative fashion, 
\begin{align}\label{eqn:track-model}
    \hat{\bbM}_{1}
    \stackrel{{\texttt{H}}}{\longrightarrow}     \hat{\bbM}_{2,+} \stackrel{{ \hat{\texttt{S}}}}{\longrightarrow}\hat{\bbM}_{2}
    \cdots\stackrel{{ \hat{\texttt{S}}}}{\longrightarrow}\hat{\bbM}_{d-1}
    \stackrel{{\texttt{H}}}{\longrightarrow}\hat{\bbM}_{d,+}\stackrel{{ \hat{\texttt{S}}}}{\longrightarrow}\hat{\bbM}_{d}\to \cdots \stackrel{{ \hat{\texttt{S}}}}{\longrightarrow} \hat{\bbM}_{D},
\end{align}
the final working model includes effects that fully respect the heredity principle. 

\begin{remark} \label{rmk:fs}
    While forward selection has been set up as a standard tool in the literature (e.g. \cite{wang2009forward, hao2014interaction}), we provided Algorithm \ref{alg:forward-ms} as it is customized for the factorial designs.  More specifically, Algorithm \ref{alg:forward-ms} differs from existing proposals \citep{wang2009forward, hao2014interaction} from several perspectives. (i) The tools for effect selection are different. In Algorithm \ref{alg:forward-ms}, we use factor-based regression and robust variance estimation, which do not assume the true outcome model is linear.  In contrast, the forward regression procedure in \cite{wang2009forward, hao2014interaction} selects the variables by iteratively minimizing the residual sum of squares. The validity of their procedure relies on the linear model assumption and the homoskedasticity of the noise. (ii) The theoretical justification is different. We study the property of forward selection from a design-based perspective, where the randomness originates from the treatment assignment. On the contrary, \cite{wang2009forward, hao2014interaction} focus on linear models as the underlying data-generating process, where the randomness comes from the outcomes. The forward selection procedure requires novel theoretical justification under the design-based framework.
\end{remark}

\subsection{Consistency of forward selection}

We are now ready to analyze the selection consistency property of Algorithm \ref{alg:forward-ms}. 
We shall show that Algorithm \ref{alg:forward-ms} selects the targeted working model up to level $D$ with probability tending to one as the sample size goes to infinity. Here, the targeted working model at level $k\in[K]$, denoted as $\bbM^\star_k$, is the collection of $\cK$'s where $|\cK| = k$ and $\tau_\cK \neq 0$. Define the full targeted working model up to level $D$ as
\begin{align*}
    \bbM^\star_{1:D} = \bigcup_{d=1}^D{\bbM_d^\star}.
\end{align*} 
In particular, when $D = K$, we omit the subscript and simply denote $\bbM^\star =  \bbM^\star_{1:K}$.

We start by introducing the following condition on \textit{nearly uniform designs}:
\begin{condition}[Nearly uniform design]\label{cond:uniform-design} There exists a positive integer $N_0$ and constants $\underline{c} \le \overline{c}$, such that 
\begin{align*}
    N(z) = c(z)  {N}_0 \ge 2,  \text{ where } \underline{c}\le c(z)\le \overline{c},
\end{align*}
where $\underline{c}$ and $\overline{c}$ are universal constants that do not depend on other quantities. 
\end{condition}
Condition \ref{cond:uniform-design} is a finite sample characterization of the design. It allows either $Q$ or $N_0$ (thus $N(z)$ across all treatment combinations) to diverge \citep{shi2022berry}. It
generalizes the classical assumption where $Q$ is fixed, and each treatment arm contains a sufficiently large number of replications \citep{li2017general}. The quantities $Q$,  $N_0$, $c(z)$ can vary for different design settings, which leads to different asymptotic regimes. In general, $N$ has the order of $O(Q N_0)$ under Condition \ref{cond:uniform-design}.

Next, we quantify the order of the true factorial effect sizes $\tau_\cK$'s and the tuning parameters $\alpha_d$'s adopted in the Bonferroni correction. We allow these parameters to change with the sample size $N$: 
\begin{condition}[Order of parameters]\label{cond:order}
The true factorial effects $\tau_\cK$'s and tuning parameters $\alpha_d$'s have the following orders:
\begin{enumerate}[label = {\em (\roman*)}]
    \item True nonzero factorial effects: $|\tau_\cK| =  \Theta(N^\delta)$ for some $-1/2 < \delta \le 0$ and all $\cK\in\bbM^\star_{1:D}$.
    \item Tuning parameters in Bonferroni correction: ${\alpha_d} =  \Theta(N^{-\delta'})$ for all $d\in[D]$ for some $\delta' > 0$.
    \item Size of the targeted working model: $\sum^D_{d=1}|\bbM_d^\star| = \Theta(N^{{\delta''}}) $ for some $0 \le \delta'' < 1/3$.
\end{enumerate} 
\end{condition}

Condition \ref{cond:order}(i) specifies the allowable order of the true factorial effects. If Condition \ref{cond:order}(i) fails with a $\delta \le -1/2$, the effect size is of the same or smaller order as the statistical error and thus is too small to be detected by marginal $t$-test. As a special case, when the number of nonzero factorial effects has a finite upper limit as $N \to \infty$ then Condition \ref{cond:order}(i) is satisfied with $\delta = 0$. Similar conditions are also adopted in the variable selection literature under the linear model, including \cite{zhao2006model} and \cite{jerzy2022model}. Condition \ref{cond:order}(ii) requires the tuning parameter $\alpha_d$ to converge to zero, which ensures that there is no Type I error asymptotically in our procedure as $N$ goes to infinity, which is crucial for the selection consistency. \citet[][Theorems 4.1 and 4.2]{wasserman2009high} assumed similar conditions in the variable selection literature under the linear model. Condition \ref{cond:order}(iii) restricts the size of the targeted working model. The rate is due to our technical analysis. As a special case, when the number of nonzero effects is a constant (i.e., constant sparsity), it suffices to set $\delta'' = 0$. Similar conditions also appeared in \cite{zhao2006model}, \cite{jerzy2022model} and \cite{wasserman2009high}.

The next condition specifies a set of regularity assumptions on the potential outcomes. 
\begin{condition}[Regularity conditions on the potential outcomes]\label{cond:regularity}
The potential outcomes satisfy the following conditions:
\begin{enumerate}[label = {\em (\roman*)}]
    \item Let $V^\star$ be the correlation matrix of $\hat{Y}$. There exists $\sigma > 0$ such that the condition number of  $V^\star$ is smaller than or equal to $\sigma^{2}.$ 
    \item There exists a universal constant $\nu>0$ and $\underline{S}>0$ such that 
    \begin{align*}
        {\max_{i\in[N],q\in[Q]} |Y_i(q) - \overline{Y}(q)|} < \nu, \quad 
        \min_{q\in[Q]}S(q,q) >\underline{S}.
    \end{align*}
\end{enumerate}
    
\end{condition}
Condition \ref{cond:regularity}(i) requires the correlation matrix of $\hY$ to be well-behaved.  Condition \ref{cond:regularity}(ii) imposes a universal bound on potential outcomes and their variances, which is a sufficient condition by \cite{shi2022berry} to prove the Berry--Esseen bound based on Stein's method.  

Lastly, we impose the following structural conditions on the factorial effects:
\begin{condition}[Hierarchical structure in factorial effects]\label{cond:heredity} The nonzero true factorial effects obey the effect heredity principle:
\begin{itemize}
\item Weak heredity: $\tau_\cK\neq 0$ only if there exists $\cK'\subset\cK$ with $|\cK'| = |\cK| - 1$ such that $\tau_{\cK'}\neq 0$.

\item Strong heredity: $\tau_\cK\neq 0$ only if $\tau_{\cK'}\neq 0$ for all $\cK'\subset\cK$ with $|\cK'| = |\cK| - 1$. 
\end{itemize}
\end{condition}

Finally, we present the selection consistency property of Algorithm \ref{alg:forward-ms}: 
\begin{theorem}[Consistent selection property]\label{thm:marginal-t}
Under Conditions \ref{cond:uniform-design}-\ref{cond:heredity}, 
\begin{align*}
    \underset{N\to\infty}{\lim}~\Prob{\hat{\bbM} = \bbM^\star_{1:D}  }
     = 1.
\end{align*}
\end{theorem}
Theorem \ref{thm:marginal-t} guarantees that the working model selected by Algorithm \ref{alg:forward-ms} converges to the targeted working model with probability one as the sample size goes to infinity. Here we used the terminology ``consistent selection'' that is widely adopted (e.g. \citep{shao1997asymptotic}). A closely related property is ``screening consistency'', which is a terminology by \cite{fan2008sure}. It refers to the fact that the selected model should cover the true model with a high probability and allow for over-selection.

\section{Inference under selection consistency}\label{sec:inference-perfect}

Statistical inference is relatively straightforward under the selection consistency of the factorial effects as ensured by Theorem \ref{thm:marginal-t}. If forward selection correctly identifies the true, nonzero factorial effects with probability approaching one, we can proceed as if the selected working model is not data-dependent. In Section \ref{Sec:inference-perfect-screeenig}, we present the point estimators and confidence intervals for general causal parameters. In Section \ref{sec::advantages-forward-selection}, we study the advantages of forward selection in terms of asymptotic efficiency in estimating general causal parameters, compared with the corresponding estimators without forward selection. We relegate the extensions to vector parameters to Section \ref{sec:extend-to-vector} of the supplementary material since it is conceptually straightforward.

\subsection{Post-selection inference for general causal parameters}\label{Sec:inference-perfect-screeenig}

Define a general causal parameter of interest as a weighted combination of average potential outcomes:
\begin{align}
\gamma = \sum_{z\in\cT}  f(z)\overline{Y}(z) \triangleq  f^{\top} \overline{Y}, \label{eqn:target-gamma}
\end{align}
where $ f = \{  f(z) \}_{z\in\cT}$ is a pre-specified weighting vector.  For example, if one is interested in estimating the main factorial effects, $ f$ can be taken as the contrast vectors $g_{\{k\}}$ given in \eqref{eq:contrast-main}. If one wants to estimate interaction effects, then $ f$ can be constructed from \eqref{eq:contrast-vector}. However, we allow $ f$ to be different from the contrast vectors $g_\cK$. For instance, if we focus on the first two arms in factorial experiments and estimate the average treatment effect, we shall choose
\begin{align*}
     f = (1,-1,0,\dots,0)^\top.
\end{align*}
In general, researchers may tailor the choice of $ f$ to the specific research questions of interest.

Without factor selection, the plug-in estimator of $\gamma$ is to replace $\overline{Y}$ with its sample analogue \citep{li2017general, zhao2021regression, shi2022berry}: 
\begin{align}\label{eqn:hgamma}
    \hgamma  =  f^\top \hY = \sum_{z\in\cT}  f(z)\hat{Y}(z). 
\end{align}
Under regularity conditions in \cite{shi2022berry}, the plug-in estimator $\hat{\gamma}$ satisfies a central limit theorem $(\hgamma - \gamma) / v  \rightsquigarrow \cN(0,1)$ with the variance $v^2 =  f^\top V_{\hY}  f$.
When $N(z) \ge 2$, its variance can be estimated by: 
\begin{align}\label{eqn:hv2R}
    \hat{v}^2 =  f^\top \hV_{\hY}  f = \sum_{z\in\cT}  f(z)^2 N(z)^{-1} \hat{S}(z,z)  .
\end{align}

With factor selection, based on the selected working model  $ \hat{\bbM}$, we consider a potentially more efficient estimator of $\overline{Y}$ via the restricted least squares (RLS)
\begin{align}\label{eqn:RLS-1}
    \hY_{\textsc{r}} = {\arg\min_{\mu\in\bbR^Q}} \Big\{ \|\hY - \mu\|_2^2:\ G(\cdot,  {\hat{\bbM}^c})^\top \mu = 0 \Big\} ,
\end{align}
which leverages the information that the nuisance effects $G(\cdot,  {\hat{\bbM}^c})^\top\overline{Y}$ are all zero. The
$\hY_\textsc{r}$ in \eqref{eqn:RLS-1} has a closed form solution (see Lemma \ref{lem:RLS} in the supplementary material):
\begin{align*}
    \hY_{\textsc{r}} = Q^{-1}G(\cdot,  {\hat{\bbM}})G(\cdot,  {\hat{\bbM}})^\top \hY.
\end{align*}
Under selection consistency, $\hY_{\textsc{r}}$ is also a consistent 
estimator for $\overline{Y}$, so $  \hgamma_{\textsc{r}} =  f^\top \hY_{\textsc{r}} $ is also consistent for $\gamma$. Introduce the following notation  
\begin{align}\label{eqn:fM}
     f[\bbM] = Q^{-1}G(\cdot,  {{\bbM}})G(\cdot,  {{\bbM}})^\top  f
\end{align}
to simplify $\hgamma_{\textsc{r}} $ and its variance estimator as 
\begin{align}\label{eqn:RLS-3}
    \hgamma_{\textsc{r}} =    f[\hat{\bbM}]^\top\hY  \quad\text{ and }\quad  \hv_{\textsc{r}}^2 =  f[\hat{\bbM}]^\top \hV_\hY  f[\hat{\bbM}]. 
\end{align} 
Construct a Wald-type level-$(1-\alpha)$ confidence interval for $\gamma$: 
\begin{align}\label{eqn:confidence-interval}
    \Big[ \hgamma_{\textsc{r}} \pm z_{1-\alpha/2}\times \hv_{\textsc{r}}  \Big], 
\end{align}
where $z_{1-\alpha/2}$ is $(1-\alpha/2)$th quantile of a standard normal distribution. We can also obtain point estimates and confidence intervals handily from  WLS of $Y_i$ on $g_{i, \hat{\bbM}}$ with weights $1/N_i$. See Section \ref{sec:wls} in the supplementary material for more details.

In the following subsection, we provide the theoretical properties of $\hgamma_{\textsc{r}} $ and $\hv_{\textsc{r}}^2$, and compare their asymptotic behaviors with the plug-in estimators $\hgamma $ and $\hv^2$ in various settings.

\subsection{Theoretical properties under selection consistency}\label{sec::advantages-forward-selection}

In this subsection, we first present the asymptotic normality result for  $\hat{\gamma}_{\textsc{r}}$. To simplify the discussion, we denote $ f^\star  =  f[\bbM^\star]$. Given $\bbM^\star$ is the true  working model,
we have $  ( f^{\star})^{\top}\overline{Y} =  f^\top\overline{Y}, ~\text{for all }  f\in\bbR^Q$ (see Lemma \ref{lem:fstar-f} in the supplementary material).

We are now ready to present the asymptotic properties of $\hgamma_{\textsc{r}}$ and $\hv_{\textsc{r}}^2$:  
\begin{theorem}[Statistical properties of $\hgamma_{\textsc{r}}$ and $\hv_{\textsc{r}}^2$]\label{thm:be-perfect-ms}
Let $N\to\infty$. Assume Conditions \ref{cond:uniform-design}-\ref{cond:heredity}. 
We have
\begin{align*}
    \frac{\hgamma_\textsc{r} - \gamma}{v_{\textsc{r}}} \rightsquigarrow \cN(0,1)
\end{align*}
where $v^{2}_{\textsc{r}} =  f^{\star\top} V_\hY  f^\star$.  Further assume $\| f^\star\|_\infty = O(Q^{-1})$. The variance estimator $\hv_{\textsc{r}}^2$ is conservative in the sense that:
    \begin{align*}
        N(\hv_{\textsc{r}}^2 - v_{\textsc{r},\mathrm{lim}}^2) \xrightarrow{{\mathrm{P}}} 0, \quad v_{\textsc{r},\mathrm{lim}}^2 \ge v_{\textsc{r}}^2,
    \end{align*}
    where $v_{\textsc{r},\mathrm{lim}}^2 =   f^{\star\top} D_\hY  f^\star$ is the limiting value of $\hv_{\textsc{r}}^2$. 
\end{theorem}

Theorem \ref{thm:be-perfect-ms} above guarantees that the proposed confidence interval in \eqref{eqn:confidence-interval} for $\gamma$ attains the nominal coverage probability asymptotically. Below we add some detailed discussion for Theorem \ref{thm:be-perfect-ms}.

First, the asymptotic regime of Theorem \ref{thm:be-perfect-ms} is that $N\to\infty$, which is equivalent to $QN_0\to\infty$ based on Condition \ref{cond:uniform-design}. This covers the classical regime where $Q$ is fixed and $N_0$ converges to $\infty$. In this regime, enough replications within each arm guarantee that the point estimates for all arms converge jointly to a multivariate normal distribution and that the variance estimators converge in probability. However, when $N_0$ is small but $Q$ diverges, the joint normality fails. In this case, the asymptotic properties of the point estimates and variance estimators are guaranteed by pooling the outcomes from a large number of treatment combinations due to a large $Q$.  Interestingly, both small and large $Q$ regimes are unified by the finite sample probability results provided in Section \ref{sec:preliminary-pf}.

Second, we add some explanation for the condition $\| f^\star\|_\infty = O(Q^{-1})$. The condition requires each element of $f^\star$ has order $Q^{-1}$, which averages the outcome information over $Q$ treatment combinations. Averaging outcomes across different treatment levels is especially important for guaranteeing the convergence of the point estimates and variance estimates when $Q$ diverges. This condition holds under many settings and is motivated by some specific examples of $ f $. One special case is that $ f = Q^{-1}g_\cK $ for some $\cK\in\bbM^\star$, which gives $\|f^\star\|_\infty = Q^{-1}$. As another special case, when $ f = (1,0,\dots,0)^\top $, we have $ \|f^\star\|_\infty = Q^{-1}|\bbM^\star| $ and the condition holds when $|\bbM^\star|$ has constant order. This example is important in the application of best arm identification, which we shall discuss in Appendix \ref{sec:select-best}. 

Third, Theorem \ref{thm:be-perfect-ms} allows us to compare the conditions for reaching asymptotic normality of $\hat{\gamma}$, which we formalize in the following remark: 
\begin{remark}[Comparison of conditions for asymptotic normality]\label{rmk:clt-conditions}
Without factor selection, the simple plug-in estimator $\hat{\gamma}$ in \eqref{eqn:hgamma} satisfies a central limit theorem if
\begin{align}\label{eqn:asp-condition-3}
    N_0^{-1/2}\cdot \frac{\|f\|_\infty}{\|f\|_2} \to 0
\end{align}
recalling the definition of $N_0$ in Condition \ref{cond:uniform-design} (See Theorem 1 of \citet{shi2022berry}).
Condition \eqref{eqn:asp-condition-3} fails when $N_0$ is small and $f$ is sparse. Besides, it does not incorporate the sparsity information in the structure of factorial effects. 
With factor selection, however, we can borrow the benefit of a sparse working model and overcome the above drawbacks. Therefore, factor selection broadens the applicability of our proposed estimator $\hgamma_\textsc{r}$ by weakening the assumptions for the Wald-type inference.
\end{remark}

To elaborate on the benefits of conducting forward factorial selection in terms of asymptotic efficiency, we compare the asymptotic variances of $\hgamma$ and $\hgamma_\textsc{r}$ in Proposition \ref{prop:ARE-comparison} below. 
In the most general setup, there is no ordering relationship between $v_\textsc{r}^2$ and $v^2$. That is, the RLS-based estimator may have a higher variance than the unrestricted OLS estimator. This is a known fact due to heteroskedasticity and the use of sandwich variance estimators \citep{meng2014got, zhao2021regression}. Nevertheless, in many interesting scenarios, we can prove an improvement in efficiency by factor selection. Two conditions are summarized in Proposition \ref{prop:ARE-comparison}:

\begin{proposition}[Asymptotic relative efficiency comparison between $\hat{\gamma}$ and $\hat{\gamma}_{\textsc{r}}$]\label{prop:ARE-comparison} Assume that both $\hat{\gamma}$ and $\hat{\gamma}_{\textsc{r}}$ converge to normal distributions with variances $v^2$ and $v_\textsc{r}^2$ as $N\to\infty$.
\begin{enumerate}[label =  {\em (\roman*)}]
    \item If the covariance matrix $V_\hY$ is $\sigma^2 I_Q$, then
 \begin{align*}
    \frac{v_\textsc{r}^2}{v^2} \le 1. 
\end{align*}
    \item Let $s^\star$ denote the number of nonzero elements in $f$. Then the asymptotic relative efficiency between $\hat{\gamma}$ and $\hat{\gamma}_{\textsc{r}} $  is upper bounded by 
\begin{align*}
    \frac{v^2_{\textsc{r}}}{v^2} \le  \kappa(V_\hY) \cdot\frac{s^\star |\bbM^\star|}{Q}.
\end{align*}
\end{enumerate}
\end{proposition}

Proposition \ref{prop:ARE-comparison}(i) gives a sufficient condition assuming that the potential outcomes are homoskedastic and uncorrelated.  Proposition \ref{prop:ARE-comparison}(ii) studies a general heteroskedastic setting with sparse weighting vector $f$ and small working model size $|\bbM^\star|$. The condition number $\kappa(V_\hY)$ captures the variability of the variances of $\hat{Y}(z)$ across multiple treatment combination groups in $\mathcal{T}$. When the variability is upper bounded by $\kappa(V_\hY)< Q/(s^\star |\bbM^\star|)$, the RLS-based estimator is  more efficient than  $\hat{\gamma}$. As an application, in Section \ref{sec:select-best} we use Proposition \ref{prop:ARE-comparison}(ii) to solve the problem of inferring the best arm in factorial experiments.  Moreover, we emphasize that Proposition \ref{prop:ARE-comparison} is only a set of sufficient conditions. 

There are also interesting examples that demonstrate the advantage of factor selection but are not covered by Proposition \ref{prop:ARE-comparison}. One concrete problem of interest is testing the \textit{sharp null hypothesis} of zero effects in uniform factorial designs (with $N_0$ replications in each arm), i.e.,
\begin{align*}
    \mathrm{H}_{0\textsc{F}} : Y_i(z) = Y_i \text{ for all } i\in[N] \text{ and } z\in\cT.
\end{align*}
Under $\mathrm{H}_{0\textsc{F}}$, we have 
\begin{gather*}
    V_\hY = N_0^{-1} \sigma^2 \cdot I_Q - N^{-1} \sigma^2 \bone_{Q}\bone_{Q}^\top,
\end{gather*}
where $\sigma^2 = {(N-1)}^{-1}\sum_{i=1}^N (Y_i - \overline{Y})^2$  and $\overline{Y} = {N}^{-1}\sum_{i=1}^N{Y_i}$. 
We can verify that $V_\hY$ has eigenvalue decomposition 
$$ V_\hY = N_0^{-1}\sigma^2 G \diag{0,1,\dots,1} G^\top$$
where $G$ is the contrast matrix \eqref{eqn:G}.
In this case, the proposed RLS-based estimator $\hgamma_\textsc{r}$ is more efficient than the plug-in estimator $\hat{\gamma}$ for testing the sharp null. 

Last but not least, Proposition \ref{prop:ARE-comparison} can be extended to compare the length of the confidence intervals as well. The conclusion is similar. See Proposition \ref{prop:ASY-CI-LEN-comparison} in the supplementary material for the details.

\section{Post-selection inference under inconsistent selection}\label{sec:imperfect-ms}

Similar to many other consistency results for variable selection, the selection consistency property in Theorem \ref{thm:marginal-t} can be difficult to achieve due to the strong regularity conditions. This is because the selection consistency property of forward selection requires the non-zero effects to be well separated from zero. Such a theoretical requirement can be stringent for higher-order factorial effects. In other words, as implied by the hierarchy principle, while main factorial effects and lower-order factorial effects are more likely to have non-negligible effect sizes, higher-order factorial effects tend to have smaller effect sizes. The selection consistency property is less likely to hold when applied to select those higher-order effects. 
More rigorously, when Condition \ref{cond:order}(i) is violated, Algorithm \ref{alg:forward-ms} may no longer enjoy the selection consistency property.

Statistical inference without selection consistency is not a trivial problem in factorial designs. If we do not impose any restrictions on the factorial selection procedure, the selected model can be arbitrary, even without a stable limit. Classical strategies for post-selection inference \citep{kuchibhotla2022post} have various drawbacks in our current setup. For example, data splitting \citep{wasserman2009high} is a widely used strategy to validate inference after variable selection due to its simplicity. However, it relies on the independent sampling assumption, which is violated in our setting. Also, data splitting faces the conceptual challenge of inference of a random parameter. Alternatively, selective inference \citep{fithian2014optimal} is another widely studied strategy, which can deliver valid inference for data-dependent parameters. However, it cannot be directly applied to analyze data collected in factorial designs. Because the selective inference strategy often relies on specific selection methods and parametric modeling assumptions on the outcome. 

\begin{figure}[!htbp]
\centering
\resizebox{0.9\textwidth}{!}{%
\begin{tikzpicture}[node distance = 2cm]
\node (pro1) [process] {Select the first $d^\star$ levels};
\node (dec2) [startstop, right of=pro1, xshift=2cm] {Are higher-order effects important? };
\node (pro4) [process, right of=dec2, xshift = 2.5cm, yshift = -1.5cm] {Select higher-order effects by heredity (Strategy \ref{str:select-by-heredity})};
\node (pro5) [process, right of=dec2, xshift = 2.5cm,  yshift = 1.5cm] {Exclude higher-order effects (Strategy \ref{str:exclude-all})};
\node (pro6) [process, right of=pro5, xshift = 2.5cm,  yshift = 0cm] {Under-selection with the targeted working model $ \underline{\bbM}^\star$};
\node (pro7) [process, right of=pro4, xshift = 2.5cm,  yshift = 0cm] {Over-selection with the targeted working model $\overline{\bbM}^\star $};

\draw [arrow] (pro1) -- (dec2);
\draw [arrow] (dec2) -- node[anchor=north] {yes} (pro4);
\draw [arrow] (dec2) -- node[anchor=south] {no} (pro5);
\draw [arrow] (pro4) -- node[anchor=north] {} (pro7);
\draw [arrow] (pro5) -- node[anchor=south] {} (pro6);

\end{tikzpicture}
}
\caption{Two strategies for factorial selection: Strategy 1 under-selects whereas Strategy 2 over-selects, depending on whether higher-order effects are important or not. }
\label{fig:general-strategy}
\end{figure}
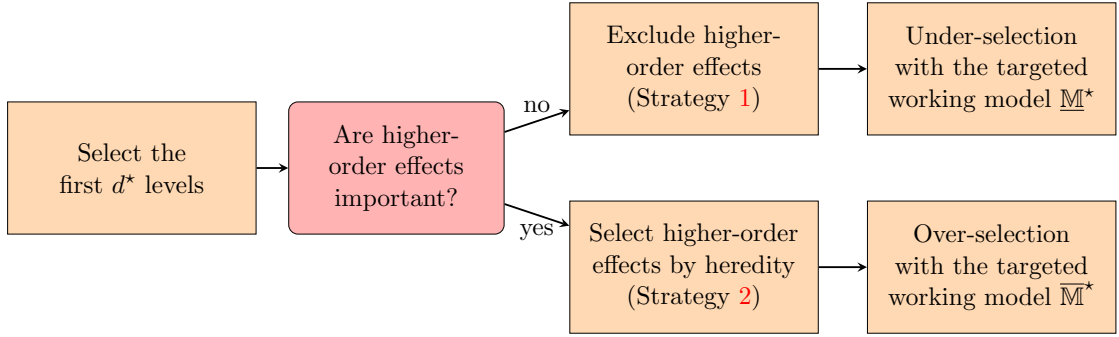

Rather than directly generalizing classical post-selection inference methods to factorial experiments, in this section, we shall discuss two alternative strategies  (summarized in Figure \ref{fig:general-strategy}) leveraging the special structures in factorial experiments and discuss the corresponding statistical inference results.


\subsection{Two strategies for inconsistent selection and statistical inference}\label{sec:imperfect-selection}

We propose two strategies based on the assumption that selection consistency is more plausible for selecting the main factorial effects and lower-order factorial effects up to level $d^\star$ than the high-order effects. We will add more discussion on $d^\star$ after presenting these two strategies. 

In certain research questions, high-order interactions are nuisance parameters or not of interest, or there is domain knowledge indicating that higher-order interactions are negligible. Then we can stop our forward selection procedure in Algorithm \ref{alg:forward-ms} at $d = d^* < D$ (instead of $d=D$). 
Such a strategy focuses on recovering a targeted working model $ \underline{\bbM}^\star$ up to level $d^\star$, that is,
\begin{align*}
    \underline{\bbM}^\star =  \cup_{d=1}^{d^\star} {\bbM_{d}^\star} \subseteq{\bbM}^\star,
\end{align*}
which leads to an under-selected working model. We summarize this strategy below. 

\begin{strategy}[Under-selection by excluding high-order interactions]\label{str:exclude-all}
In Algorithm \ref{alg:forward-ms}, we stop the selection procedure at $d = d^*$. Equivalently, we set $\alpha_d = \infty$ for $d\ge d^\star + 1$ so that no effects beyond level $d^\star$ will be selected and $\underline{\hat{\bbM}} = \cup_{d=1}^{d^\star}\hat{\bbM}_{d}$. 
\end{strategy}

In Strategy \ref{str:exclude-all}, $\alpha_d = \infty$ leads to never rejecting the null hypothesis of zero effects, which excludes the high-order interactions. Given the selected working model $\underline{\hat{\bbM}}$, we can again construct an estimator of $\gamma = f^{\top} \overline{Y}$ (defined in Section \ref{Sec:inference-perfect-screeenig}) based on RLS: 
\begin{align}\label{eqn:RU}
 \hgamma_{\textsc{ru}} =  f[\hat{\underline{\bbM}}]^\top\hY, \quad\text{ and }\quad  \hv_{\textsc{ru}}^2 =  f[\underline{\hat{\bbM}}]^\top \hV_\hY  f[\underline{\hat{\bbM}}].
\end{align}

For Strategy 2, rather than excluding all higher-order interactions with negligible effects, we may further leverage the heredity principle and continue our selection procedure beyond level $d^\star$. This means that instead of selecting the higher-order interactions via marginal $t$-test and Bonferroni correction, we select the higher-order interaction terms whenever either all of their parent effects are selected (strong heredity), or one of their parent effects is selected (weak heredity). Such a strategy takes higher-order factorial effects into account and  targets a working model $\overline{\bbM}^{\star}$ that includes the true model ${\bbM}^{\star}$, that is,
\begin{align*}
   {\bbM}^{\star}\subseteq \overline{\bbM}^{\star} = \bigcup_{d=1}^D \overline{\bbM}^{\star}_d,
    \text{ where }
    \overline{\bbM}^{\star}_d = \left\{
    \begin{array}{cc}
        \bbM^{\star}_d, & d\le d^\star; \\
        \mathrm{H}^{(d-d^\star)}(\bbM^{\star}_{d^\star}), & d^\star+1\le d\le D.
    \end{array}
    \right.
\end{align*}
Here $\mathrm{H}^{(d-d^\star)}(\cdot)$ means applying the $\mathrm{H}(\cdot)$ operator $(d-d^\star)$ times. This strategy is expected to yield an over-selected model that includes ${\bbM}^{\star}$. We summarize this strategy as follows: 
\begin{strategy}[Over-selection by including higher-order interactions through the heredity principle]\label{str:select-by-heredity}
In Algorithm \ref{alg:forward-ms}, set $\alpha_d = 0,d\ge d^\star+1$ and apply a heredity principle (either weak or strong, depending on the knowledge of the structure of the effects). Then the high-order effects beyond level $ d^\star$ are selected merely by the heredity principle and 
$$
\hat{\overline{\bbM}} = \cup_{d=1}^{D}\hat{\bbM}_{d} \text{ where }\hat{\bbM}_{d} = \left\{
\begin{array}{cc}
    \text{\em Algorithm \ref{alg:forward-ms} Output}, & d\le d^\star; \\
    \mathrm{H}^{(d-d^\star)}(\hat{\bbM}_{d^\star}), & d^\star+1\le d\le D.
\end{array}
\right.
$$
Here $\mathrm{ H}^{(d-d^\star)}$ is the $(d-d^\star)$-order composition of $\mathrm{ H}$, meaning applying $\mathrm{ H}$ for $(d-d^\star)$ times.
\end{strategy}

In Strategy \ref{str:select-by-heredity}, $\alpha_d = 0$ means always rejecting the null hypothesis of zero effect size, which corresponds to always including high-order interactions, as this gives a threshold of $\infty$ for marginal $t$-tests. Given the selected working model $\hat{\overline{\bbM}}$, similarly, we can construct an estimator of $\gamma =  f^{\top} \overline{Y}$ based on RLS: 
\begin{align}\label{eqn:RO}
 \hgamma_{\textsc{ro}} =  f[\hat{\overline{\bbM}}]^\top\hY, \quad\text{ and }\quad  \hv_{\textsc{ro}}^2 =  f[\hat{\overline{\bbM}}]^\top \hV_\hY  f[\hat{\overline{\bbM}}].
\end{align}


In real-world factorial experiments, how should practitioners choose between Strategy \ref{str:exclude-all} and Strategy \ref{str:select-by-heredity}? This relies on the domain knowledge and the research question of interest. Strategy 1 is more suitable when there is domain knowledge that higher-order interactions are negligible, or when the research question only involves lower-order factorial effects. Moreover, Strategy 1 is more suitable when the number of active lower-order interactions is large and Strategy 2 cannot be applied. Meanwhile, Strategy 2 works better when domain knowledge suggests non-negligible higher-order interactions or the research question targets a more general parameter beyond factorial effects themselves. Strategy \ref{str:select-by-heredity} also works well when the number of active lower-order interactions is small, and we can include all the corresponding high-order terms according to the heredity principle. 

Another component that appears in Strategy \ref{str:exclude-all} and \ref{str:select-by-heredity} is the parameter $d^\star$. In practice, $d^\star$ should be determined by the research question of interest as well as the domain knowledge.  For example, if the research question involves only main effects and two-way interactions, we can take $d^\star = 2$. As another example, one common practice in analyzing factorial experiments is to assume away all the high-order interactions beyond a certain level \citep[say for $d \ge 3$; see][]{egami2019causal, zhao2021regression, hao2014interaction}. This is usually supported by the domain knowledge that high-order interaction signals beyond some level $d^\star$ are weak, which supports the choice of $d^\star = 2$. In general, it is an interesting question to propose some data-adaptive procedure for selecting $d^\star$. We save this as a future effort.

In the following subsection, we study the statistical properties of $(\hgamma_\textsc{ro}, \hv_\textsc{ro})$ and $(\hgamma_\textsc{ru}, \hv_\textsc{ru})$ and demonstrate the trade-offs between the two strategies. 

\subsection{Theoretical properties under inconsistent selection}\label{sec::theory-imperfect-selection}

Throughout this subsection, we discuss the scenario where selection consistency is hard to achieve. We relax Condition \ref{cond:order} as follows:
\begin{condition}[Order of parameters up to level $d^\star$]\label{cond:under-selection}
Condition \ref{cond:order} holds with $D = d^\star$. 
\end{condition}
Condition \ref{cond:under-selection} no longer imposes any restriction on the order of the parameters beyond level $d^\star$. By Theorem \ref{thm:marginal-t},  Condition \ref{cond:under-selection} guarantees that Algorithm \ref{alg:forward-ms} perfectly screens the first $d^\star$ levels of factorial effects in the sense that $\mathrm{P}\{\hat{\bbM}_d = \bbM_d^\star \text{ for } d=1,\ldots,  d^\star\} \to 1$.

We start by analyzing the statistical property of $\hgamma_\textsc{ru}$ with $\underline{\hat{\bbM}}$ obtained from the under-selection Strategy \ref{str:exclude-all}. Because the selected working model can deviate from the truth beyond level $d^\star$, $\hgamma_\textsc{ru}$ may not be a consistent estimator of $\gamma$. Therefore, we focus on  weighting vectors $ f$ that satisfy certain orthogonality conditions as introduced in Theorem \ref{thm:strategy-I} below:
\begin{theorem}[Guarantee for Strategy \ref{str:exclude-all}]\label{thm:strategy-I}
Recall \eqref{eqn:fM} and define $\underline{ f}^\star  =  f[\underline{\bbM}^\star] = Q^{-1}G(\cdot,\underline{\bbM}^\star)G(\cdot,\underline{\bbM}^\star)^\top f$. Assume Conditions \ref{cond:uniform-design}, \ref{cond:regularity}, \ref{cond:heredity}, \ref{cond:under-selection}, and $ f$ satisfies the following orthogonality condition: 
\begin{align}\label{eqn:orthogonality}
    G(\cdot, \bbM_d^{\star})^\top f = 0 \text{ for } d^\star +1 \le d \le K.
\end{align}
Let $N\rightarrow \infty$. We have
\begin{align*}
    \frac{\hgamma_\textsc{ru}-\gamma}{v_\textsc{ru}} \rightsquigarrow \cN(0,1), 
\end{align*}
where $v^{2}_{\textsc{ru}} = \underline{ f}^{\star\top} V_\hY \underline{ f}^\star$. Further assume $\|\underline{ f}^\star\|_\infty = O(Q^{-1})$. The variance estimator $\hv_{\textsc{ru}}^2$ is conservative in the sense that:
    \begin{align*}
        N(\hv_{\textsc{ru}}^2 - v_{\textsc{ru},\mathrm{lim}}^2) \xrightarrow{{\mathrm{P}}} 0, \quad v_{\textsc{ru},\mathrm{lim}}^2 \ge v_{\textsc{ru}}^2,
    \end{align*}
    where $v_{\textsc{ru},\mathrm{lim}}^2 =  \underline{ f}^{\star\top} D_\hY \underline{ f}^\star$ is the limiting value of $\hv_{\textsc{ru}}^2$.
\end{theorem}

Now we add some discussion on a key condition \eqref{eqn:orthogonality} in Theorem  \ref{thm:strategy-I}. The orthogonality condition in \eqref{eqn:orthogonality} restricts the weighting vector $ f$ to be orthogonal to the higher-order contrasts. Intuitively, because the higher-order interactions are excluded from the model, making inferences on a weighted combination of those excluded interactions is infeasible. 
One set of weighting vectors satisfying  \eqref{eqn:orthogonality} is the contrast vectors of nonzero canonical lower-order interactions, given by $ f = g_\cK$ for some $\cK\in \cup_{d=1}^{d^\star}\bbM^\star_d$. In large $K$ settings, the lower-order interactions can also grow polynomially fast in $K$ and add difficulty for interpretation. As an example, when $K = 10$, for the first two levels of factorial effects without selection, there are a total of more than $50$ estimates. It can still greatly benefit the analysis and interpretation to filter out the insignificant ones and obtain a parsimonious working model.

Without the condition in \eqref{eqn:orthogonality}, Strategy \ref{str:exclude-all} could lead to biased estimates when nonzero higher-order interactions are excluded. An inconsistent model $\bbM_{\text{im}}$ would miss a set of true effects $\bbM_{\text{miss}} = \bbM^\star\backslash\bbM_{\text{im}}$ and lead to the bias:
\begin{align}\label{eqn:bias}
    \text{Bias} = Q^{-1}f^\top G(\cdot, \bbM_{\text{miss}})\tau(\bbM_{\text{miss}}).
\end{align}
From \eqref{eqn:bias}, the bias is determined by two parts. The first part is the size of the missing nonzero effects, given by $\tau(\bbM_{\text{miss}})$. If the excluded effects are large, then the bias will be large. The second part depends on $f^\top G(\cdot, \bbM_{\text{miss}})$. If the linear coefficient vector $f$ is closer to the span of the excluded contrasts $G(\cdot, \bbM_{\text{miss}})$, the bias will also be larger.


For Strategy \ref{str:select-by-heredity}, similarly, we have the following results: 
\begin{theorem}[Guarantee for Strategy \ref{str:select-by-heredity}]\label{thm:strategy-II}
Recall \eqref{eqn:fM} and define $ \overline{ f}^\star =   f[\overline{\bbM}^\star] = Q^{-1}G(\cdot,\overline{\bbM}^\star)G(\cdot,\overline{\bbM}^\star)^\top f$.
Assume Conditions \ref{cond:uniform-design}, \ref{cond:regularity}, \ref{cond:heredity} and \ref{cond:under-selection}.   Let $N\rightarrow \infty$. If $|\overline{\bbM}^\star| / N \rightarrow 0$, 
then 
\begin{align*}
    \frac{\hgamma_\textsc{ro}-\gamma}{v_\textsc{ro}} \rightsquigarrow \cN(0,1),
\end{align*}
where $v^{2}_{\textsc{ro}} = \overline{ f}^{\star\top} V_\hY \overline{ f}^\star$. Further  assume $\|\overline{ f}^\star\|_\infty = O(Q^{-1})$. The variance estimator $\hv_{\textsc{ro}}^2$ is conservative in the sense that:
    \begin{align*}
        N(\hv_{\textsc{ro}}^2 - v_{\textsc{ro},\mathrm{lim}}^2) \xrightarrow{{\mathrm{P}}} 0, \quad v_{\textsc{ro},\mathrm{lim}}^2 \ge v_{\textsc{ro}}^2,
    \end{align*}
    where $v_{\textsc{ro},\mathrm{lim}}^2 =  \overline{ f}^{\star\top} D_\hY \overline{ f}^\star$ is the limiting value of $\hv_{\textsc{ro}}^2$. 
\end{theorem}

There is an additional technical requirement in Theorem \ref{thm:strategy-II} for over-selection: $|\overline{\bbM}^\star|/N \to 0$, which is a sufficient condition for the central limit theorem. The reason is that we need to control the size of the target model $\overline{\bbM}^\star$ compared with the sample size $N$ to infer a general causal parameter.  


When analyzing Strategies \ref{str:exclude-all} and \ref{str:select-by-heredity},  Algorithm \ref{alg:forward-ms} recovers a targeted model with high probability. Both strategies have advantages and disadvantages:
\begin{itemize}
    \item \textit{Estimation bias.} Under-selection can induce more bias for certain weighting vectors (quantified by Equation \eqref{eqn:bias}) while over-selection helps reduce or remove the bias. 
    \item \textit{Variance.} The constructed estimator typically enjoys a smaller variance with under-selection compared with over-selection. To understand this trade-off quantitatively, if we consider the homoskedasticity condition that $V_\hY$ equals $\sigma^2 I_Q$ for some $\sigma > 0$, we can prove $v^2_{\textsc{ru}} \le v^2_{\textsc{ro}}$. Therefore, in this case, by excluding higher-order terms and pursuing under-selection, we can obtain an equal or smaller asymptotic variance compared with over-selection.  In general, due to heteroskedasticity, the order of $v^2_{\textsc{ru}}$ and $v^2_{\textsc{ro}}$ depends on the choice of target weighing vector $ f$. Here we take a sparse $ f = \bse_1 = (1,0,\dots,0)^\top$ as an example. We can show that 
\begin{align*}
    \frac{v^2_{\textsc{ru}}}{v^2_{\textsc{ro}}} \le \kappa(V_\hY) \cdot \frac{\big|\underline{\bbM}^\star\big|}{\big|\overline{\bbM}^\star\big|}.
\end{align*}
When the variability of $V_\hY$ between treatment arms is small in the sense that $\kappa(V_\hY) <  {\big|\overline{\bbM}^\star\big|}/{\big|\underline{\bbM}^\star\big|}$, under-selection leads to smaller asymptotic variance for inferring $\bse_1^\top\overline{Y}$.
    \item \textit{Interpretability.} Under-selection leads to simple models that are easy to interpretable, while over-selection may not be practical if there are too many nonzero lower-order terms which can result in many redundant terms in the selected model.
\end{itemize}

In practice, if higher-order interactions are not crucial, Strategy \ref{str:exclude-all} should be applied. If high-order interactions are of interest and hard to select, one could pursue Strategy \ref{str:select-by-heredity} as a practically useful and interpretable solution.


\section{Simulation}\label{sec:simulation}

In this section, we use simulation to demonstrate the finite-sample performance of the proposed forward selection framework and the inferential properties of the RLS-based estimator. Our simulation results verify the following properties of the proposed procedure and estimators: 
\begin{enumerate}[label = \textbf{(G\arabic*)}]
    \item\label{goal:RLS} The RLS-based estimator $\hgamma_\textsc{r}$ demonstrates efficiency gain (in terms of improved power and shortened confidence interval) compared with the simple plug-in estimator $ \hgamma $ for general causal parameters defined by sparse weighting vectors. 
    \item\label{goal:forward} The factorial forward selection procedure provided in Algorithm \ref{alg:forward-ms} can improve the performance of effect selection compared to naive procedure (i.e., selection without leveraging the heredity principle).
\end{enumerate}
\ref{goal:RLS} echoes our discussion on the comparison of conditions and asymptotic variance for central limit theorems in Remark \ref{rmk:clt-conditions} and Proposition \ref{prop:ARE-comparison}. \ref{goal:forward} verifies the results in Theorem \ref{thm:marginal-t} and \ref{thm:be-perfect-ms} and checks the finite sample behaviors of the proposed procedures.  
For both \ref{goal:RLS} and \ref{goal:forward}, we will vary the sample size and effect size to provide a comprehensive understanding of their performance. 

\subsection{Simulation setup}\label{sec:simulation-setup}
We set up a $2^K$ factorial experiment, with $N_0$ units in each treatment arm where $K$ and $N_0$ are varied across settings. 
We generate independent potential outcomes from a shifted exponential distribution:
\begin{align*}
    Y_i(z) \sim \mathrm{EXP}(1) - 1 + \mu(z).
\end{align*}
Here $\mu(z)$ are super population means of potential outcomes under treatment $ z $. We choose $\mu(z) $ such that the factorial effects satisfy the following structure:
\begin{itemize}
    \item Main effects: the main effects corresponding to the first five factors, $\tau_{\{1\}},\dots, \tau_{\{5\}}$, are nonzero; the rest three main effects, $\tau_{\{6\}},\dots, \tau_{\{8\}}$, are zero. 
    \item Two-way interactions: the two-way interactions associated with the first five factors are nonzero, i.e., $\tau_{\{kl\}} \neq 0$ for $k\neq l, k,l\in[5] $. The rest of the two-way interactions are zero. 
    \item Higher-order interactions: all the higher-order interactions $\tau_\cK$ are zero if $|\cK|\ge 3$. 
\end{itemize}
The above setup of factorial effects guarantees that they are sparse and follow the strong heredity principle. In the provided simulation results, we will vary the number of units in each treatment arm, the size of the nonzero factorial effects, and the number of factors. More details can be found in the \texttt{R} code attached to the support materials. 

\subsection{Simulation results supporting \ref{goal:RLS}}\label{sec:simulation-G1}
\begin{figure}[ht!]
\centering
\begin{subfigure}
  \centering
  \includegraphics[scale = 0.4]{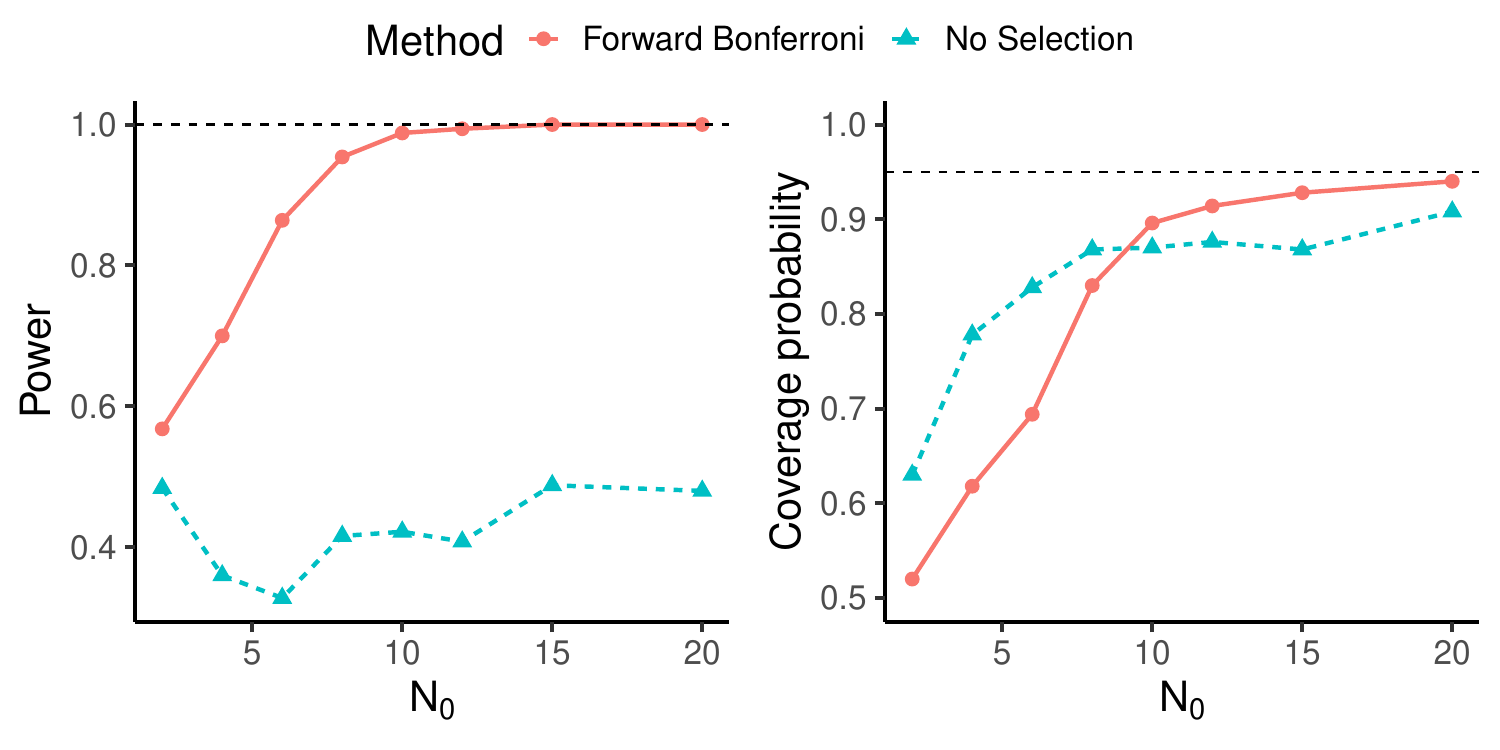}
\end{subfigure}%
\begin{subfigure}
  \centering
  \includegraphics[scale = 0.4]{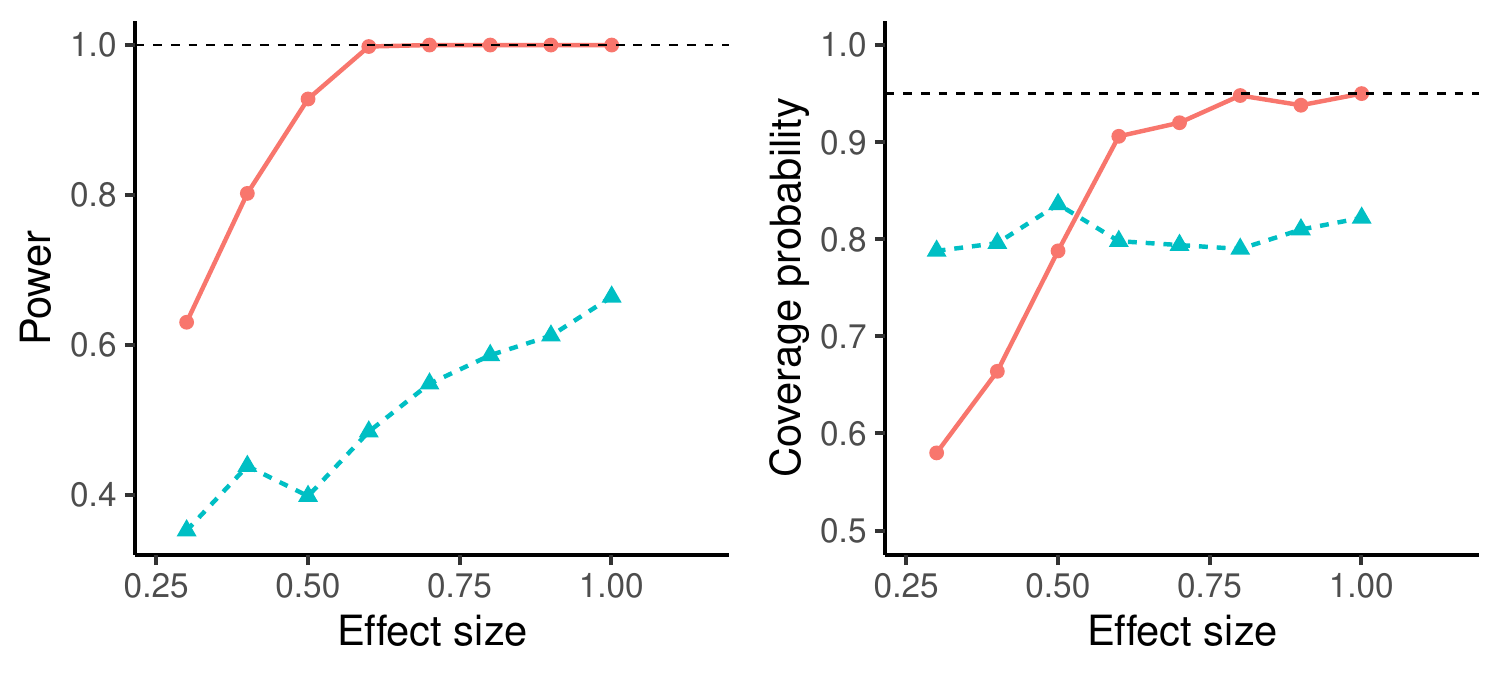}
\end{subfigure}
\begin{subfigure}
  \centering
  \includegraphics[scale = 0.4]{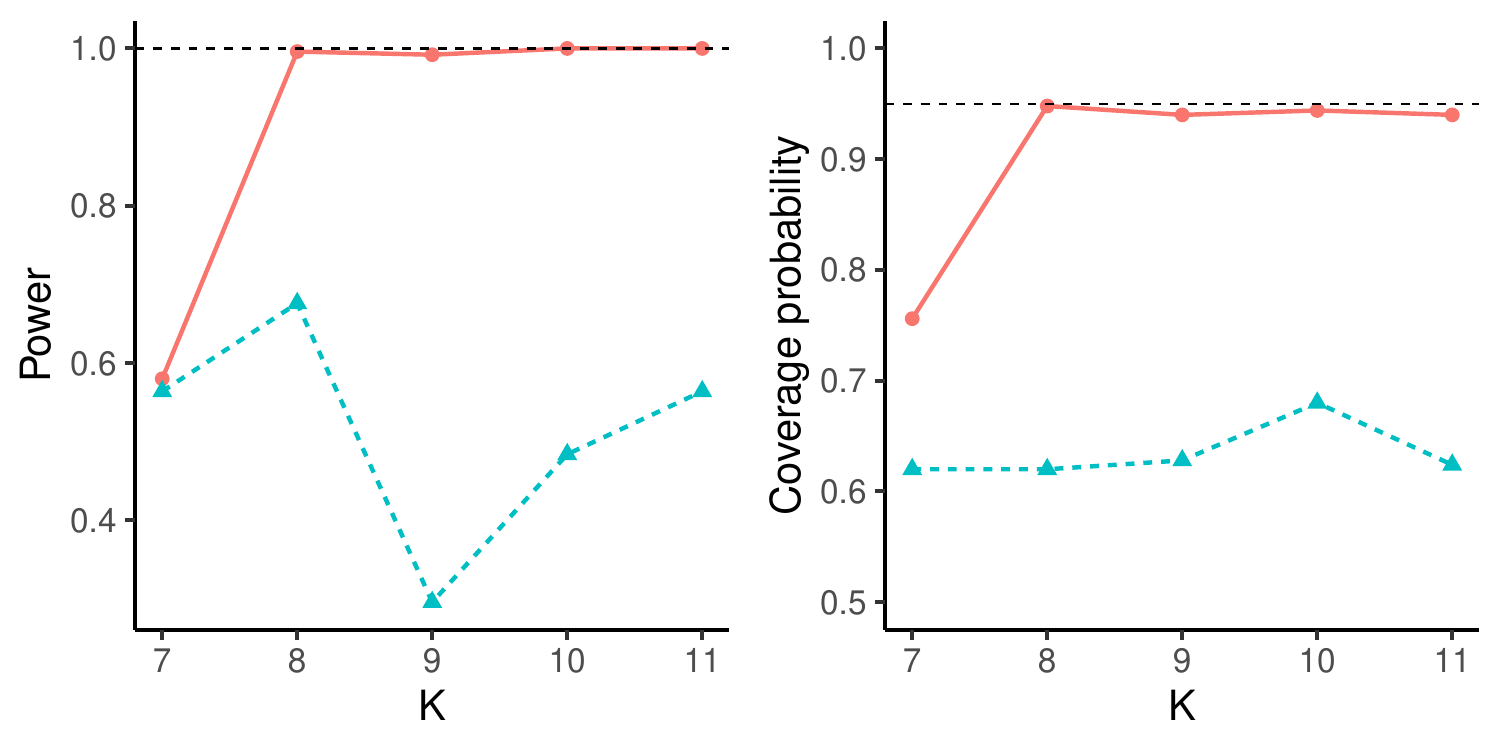}
\end{subfigure}
\caption{Simulation results supporting \ref{goal:RLS}. (i) Top left panel: power curve with varying $N_0$; (ii) Top right panel: coverage probability with varying $N_0$; (iii) Middle left panel: power curve with varying effect size $\gamma_{\text{target}}$; (iv) Middle right panel: coverage probability with varying effect size $\gamma_{\text{target}}$.  (v) Bottom left panel: power curve with varying number of factors $K$; (vi) Bottom right panel: coverage probability with varying number of factors $K$.}
\label{fig:G1}
\end{figure}

In this subsection, we evaluate the performance of the RLS-based estimators $(\hgamma_{\textsc{r}},\hv_\textsc{r})$ compared to $(\hgamma,\hv)$  for testing a causal effect $\gamma_{\text{target}} =  f^\top\overline{Y}$ specified by a sparse vector: $ f = (0,\dots,0,1)^\top\in\bbR^Q$. Intuitively, $ \gamma_{\text{target}} $ measures the average of potential outcomes in the last level. 
For each estimator, we report: (i) power for testing $\mathrm{H}_0:\gamma_{\text{target}} = 0$.
(ii) coverage probability of the confidence intervals for $\gamma_{\text{target}}$ at level $0.95$. Figure \ref{fig:G1} summarizes the results. 

Figure \ref{fig:G1} demonstrates that the RLS-based estimator $\hgamma_{\textsc{r}}$ has much higher power compared with the simple moment estimator $\hgamma $ for inferring $\gamma_{\text{target}}$ for all considered simulation settings. This echoes our conclusion in Proposition \ref{prop:ARE-comparison} that the RLS-based estimator has reduced variance compared with the simple moment estimator. {Moreover, while the RLS-based estimator attains near nominal coverage probability with reasonably large $N_0$ and $\gamma_{\texttt{target}}$, the simple moment estimator tends to provide under-covered confidence intervals in all cases. } 



\subsection{Simulation results supporting \ref{goal:forward}}\label{sec:simulation-G2}
In this subsection, we compare the performance of four candidate effect selection methods:
\begin{itemize}
    \item \textit{Forward Bonferroni}. Forward selection  based on Bonferroni corrected marginal $t$-tests; 
    \item \textit{Forward Lasso}. Forward selection based on Lasso;
    \item \textit{Naive Bonferroni}. selection with the full working model based on Bonferroni corrected margin $t$-tests;
    \item \textit{Naive Lasso}. selection with the full working model based on Lasso.
\end{itemize}

\begin{figure}[ht!]
\centering
\begin{subfigure}
  \centering
  \includegraphics[scale = 0.35]{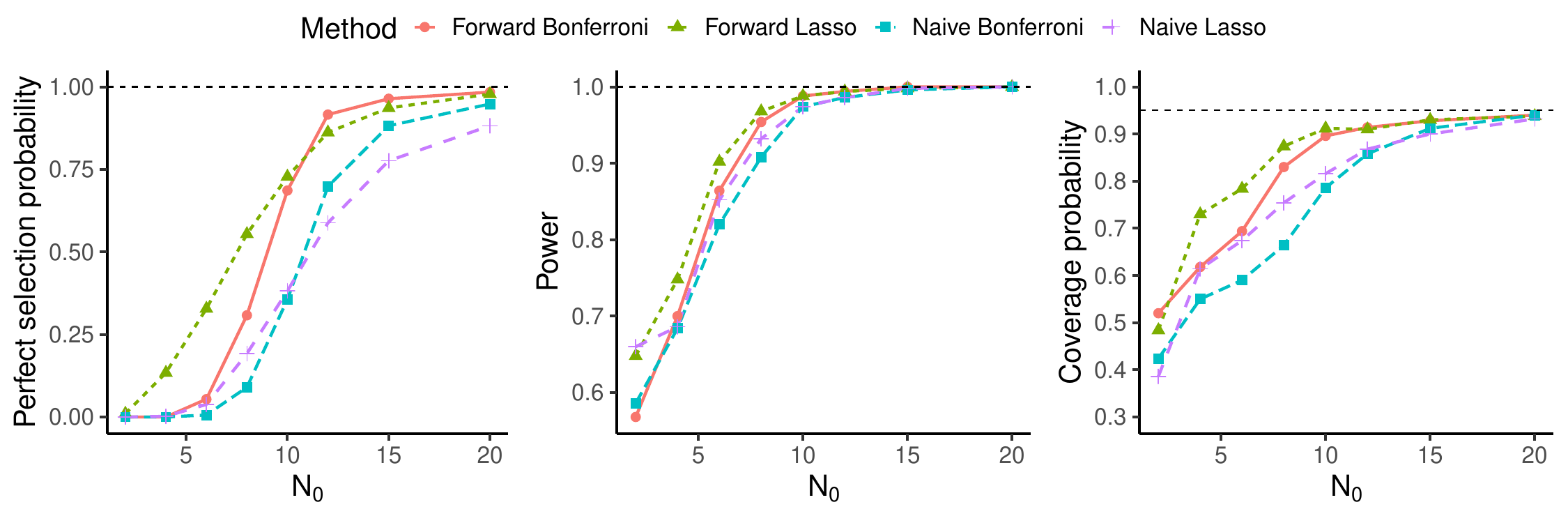}
\end{subfigure}%
\begin{subfigure}
  \centering
  \includegraphics[scale = 0.35]{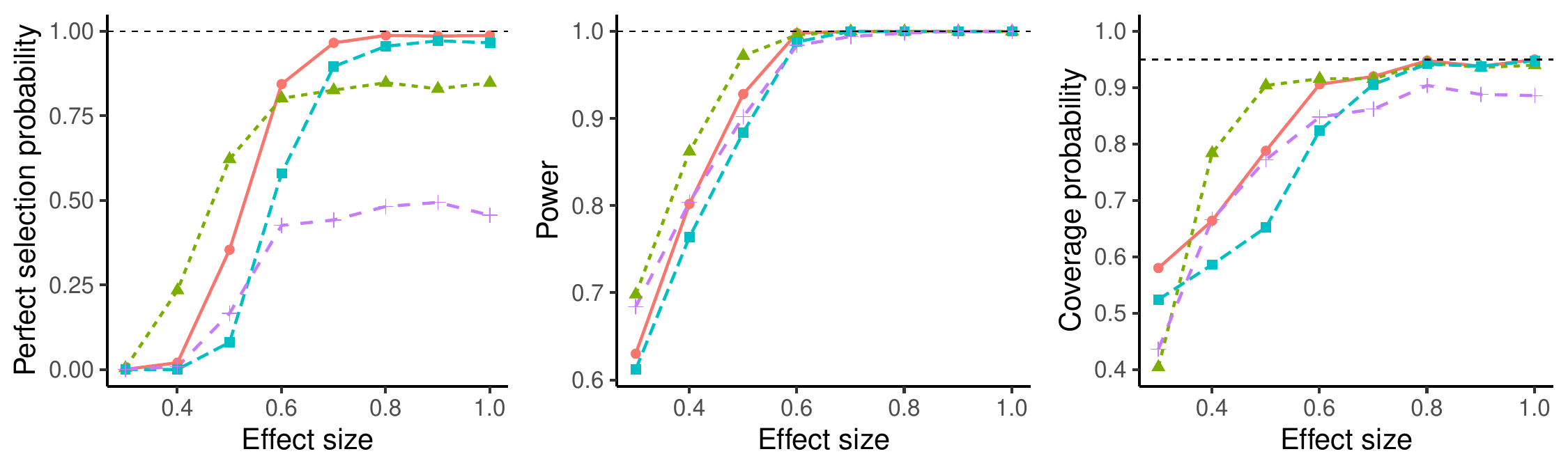}
\end{subfigure}
\begin{subfigure}
  \centering
  \includegraphics[scale = 0.35]{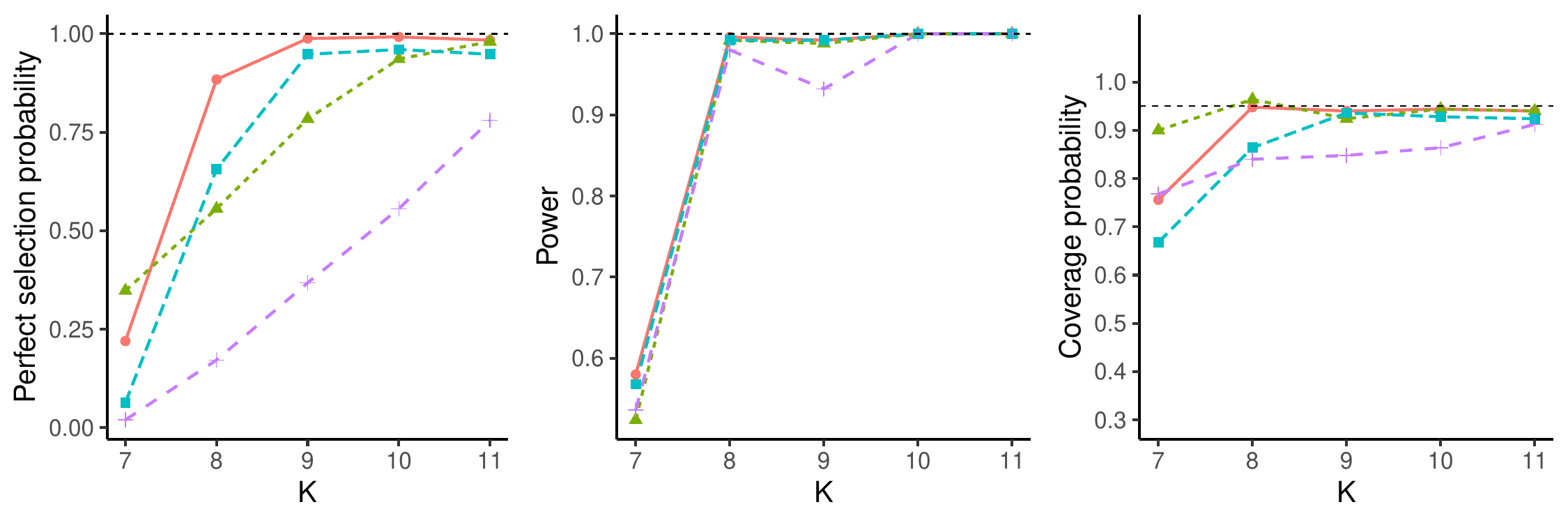}
\end{subfigure}
\caption{Simulation results supporting \ref{goal:forward}.  (i) Top row left panel: selection consistency probability with a small fixed effect size $\gamma_{\text{target}} = 0.20$, a fixed number of factors $K=8$ and varying $N_0$;  (ii) Top row middle panel: power curve with a small fixed effect size $\gamma_{\text{target}} = 0.20$, a fixed number of factors $K=8$ and varying $N_0$; (iii) Top row right panel: coverage probability with a small fixed effect size $\gamma_{\text{target}} = 0.20$, a fixed number of factors $K=8$ and varying $N_0$; (iv) Middle row left panel: selection consistency probability with a small fixed replication $N_0=2$, a fixed number of factors $K=8$ and varying effect size $\gamma_{\text{target}}$;  (v) Middle row middle panel: power curve with a small fixed replication $N_0=2$, a fixed number of factors $K=8$ and varying effect size $\gamma_{\text{target}}$; (vi) Middle row right panel: coverage probability with a small fixed replication $N_0=2$, a fixed number of factors $K=8$ and varying effect size $\gamma_{\text{target}}$. (vii) Bottom row left panel: selection consistency probability with a small fixed replication $N_0=2$, an effect size $\gamma_{\text{target}} = 0.50$ and a varying number of factors;  (viii) Bottom row middle panel: power curve with a small fixed replication $N_0=2$, an effect size $\gamma_{\text{target}} = 0.50$ and a varying number of factors; (ix) Bottom row right panel: coverage probability with a small fixed replication $N_0=2$, an effect size $\gamma_{\text{target}} = 0.50$ and a varying number of factors.}
\label{fig:G2}
\end{figure}

For each selection method, we evaluate their performance with three measures: (i) selection consistency probability ${\mathrm{P}}\{\hat{\bbM} = \bbM^\star\}$, (ii) power of $\hgamma_\textsc{r}$ for testing $\mathrm{H}_0:\gamma_{\text{target}} = 0$ for the same $\gamma_{\text{target}}$ defined in the previous section, and (iii) coverage probability of the RLS-based confidence interval for $\gamma_{\text{target}} $ with the nominal level at $0.95$. The results are summarized in Figure \ref{fig:G2}.

From Figure \ref{fig:G2}, all four effect selection methods lead to selection consistency with high probability as $N_0$ or $\gamma_{\texttt{target}}$ increases. Nevertheless, with the forward selection procedure, the probability of selection consistency is higher than the naive selection procedure. Besides, forward selection complies with the heredity structure and demonstrates higher interpretability than the naive selection methods.  In terms of the power of $\hgamma_\textsc{r}$ and $\hv_\textsc{r}$ for testing  $\mathrm{H}_0:\gamma_{\text{target}} = 0 $, while all four methods have power approaching one as $N_0$ and $\gamma_{\texttt{target}}$ increases,  forward selection based procedures possess higher power with small $N_0$ and $\gamma_{\texttt{target}}$.  Lastly, we can see an improvement in the coverage probability of the RLS-based confidence intervals with the forward selection procedure.

\subsection{Violations of conditions}\label{sec:violation-of-conditions}
In this subsection, we discussed the impact of violation of conditions on the performance of Algorithm \ref{alg:forward-ms}. We highlight some simulation studies where some conditions break down. 

\textbf{Small effect size.} Condition \ref{cond:order} assumes that the effect sizes should be of a certain order with respect to the sample size $N$. Small effect sizes will impact the performance of the algorithm, in terms of selection consistency probability, coverage,  power for post-selection inference, etc. For example, the middle panels in Figure \ref{fig:G2} show how the selection consistency probability and coverage/power for inference vary with effect sizes in a factorial experiment with $K=8$ factors and $N_0 = 2$ replications on each arm. We can conclude that if the effect sizes are too small compared to the sample size, selection consistency is hard to achieve and post-selection inference is also hurt by the bias generated from model misspecification. Nevertheless, the issue is mitigated as the effect size increases to a larger level. 
        
\textbf{Failure of heredity.} Condition \ref{cond:heredity} assumes that the factorial effects follow a weak/strong heredity structure.  For effects selection,  failure of effect heredity typically leads to under-selection in the interaction terms, because the effects that violate heredity will be ruled out by the heredity step in Algorithm \ref{alg:forward-ms}. For post-selection inference, failure of heredity will lead to bias for the RLS-based estimator \eqref{eqn:RLS-1} and impact coverage and power for hypothesis testing. 
        
\begin{figure}[ht!]
\centering
\includegraphics[width = 0.9\textwidth]{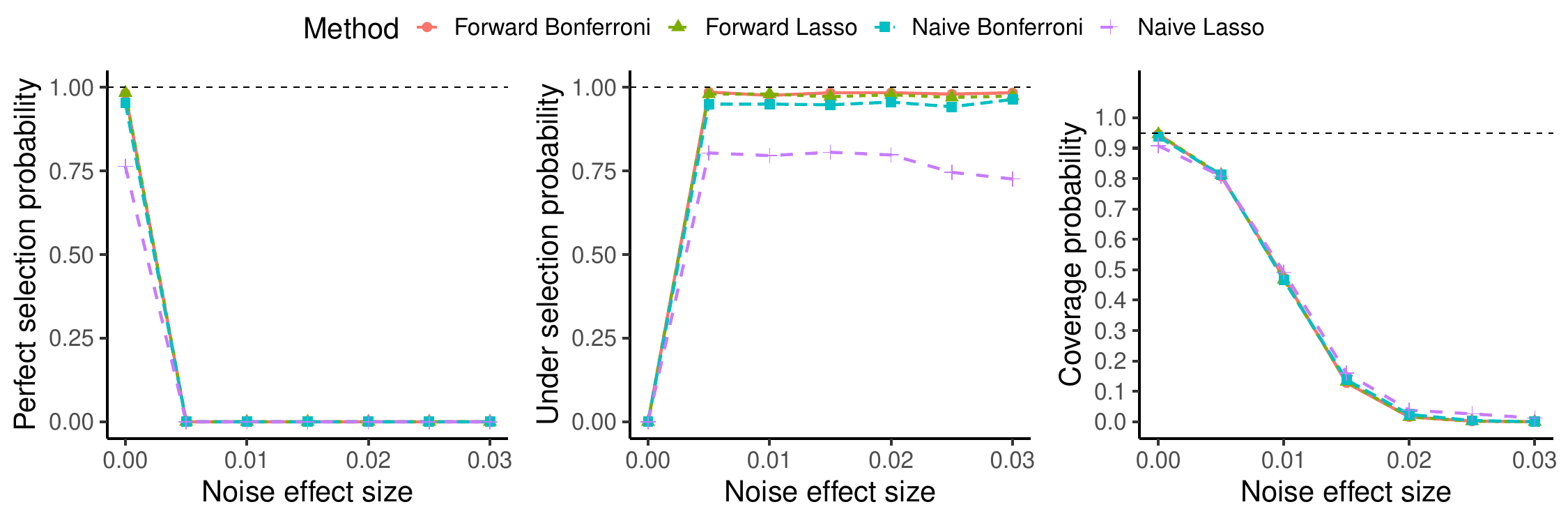}
\caption{Failure of heredity conditions and its impact on effect selection and inference. (i) left panel: how consistent effect selection probability changes with noise effect size. (ii) middle panel: how under-selection probability changes with noise effect size. (iii) right panel: how coverage probability changes with noise effect size.}
\label{fig:failure-of-heredity}
\end{figure}

To demonstrate this, we conduct a simulation study with $K=8$ factors. The setup for potential outcomes follows that in Section \ref{sec:simulation-setup}. We let the first $5$ main effects be nonzero with absolute size $0.50$ and the two-way interactions among the first five factors be $0.25$. At the same time, we set the rest of the two-way interactions to have size $\tau_{\text{noise}}$, which takes values in the set $\{0, 0.005, 0.010, 0.015, 0.020, 0.025, 0.030\}$. In particular, when $\tau_{\text{noise}} = 0$, the effects follow the strong heredity principle; otherwise the heredity structure is violated in different magnitudes. Figure \ref{fig:failure-of-heredity} reports the simulation results. From the left panel, selection consistency property is impacted greatly even with mild violation of heredity. Nevertheless, under-selection is still achieved with a high probability based on the middle panel. In the right panel, the coverage probability is also impacted and gradually decreases with more severe violations. 

\section{Case study: conjoint survey experiment regarding U.S. presidential candidates}\label{sec:real-data}

In this section, we apply the forward selection method to a real data example. In particular, we analyze a conjoint survey experiment regarding U.S. citizens’ preferences across presidential candidates studied by \cite{hainmueller2014causal}. The study focuses on how the candidates' traits impact citizen's preferences. The original experiment involves eight attributes of the imaginary candidate profiles: military service ($z_1$), religion ($z_2$), college education ($z_3$), annual income ($z_4$), racial/ethnic background ($z_5$), age ($z_6$), gender ($z_7$), and profession ($z_8$), where military service and gender are binary factors while the rest six factors have six levels. To fit into our framework, we drop the profession factor and collapse the other six-level factors into binary ones. The outcome is a rating of the candidate profile on a one-to-seven scale, representing the levels of absolute support or opposition to each profile separately. The final dataset contains a total of $K = 7$ factors (with $Q = 2^7 = 128$ treatment combinations) and $N = 3456$ profiles. Each treatment combination contains $27$ respondents. 

We applied the forward selection procedure to analyze the data. For each level, we apply LASSO to penalize the factor-based regression and select a working model. Here we applied LASSO instead of marginal $t$-tests for the convenience of tuning parameter selection based on existing packages. For the LASSO implementation, we apply cross-validation to decide the level of penalization. Between levels, we incorporate the heredity structure. For comparison, we also report the selected working model from forward selection without heredity as well as that from a full LASSO that does not proceed in a forward style. Table \ref{tab:model-selection-results} reports the effect selection results based on these strategies. 
\begin{table}[ht!]
    \centering
    \caption{Working Model Selection Results for the Presidential Candidate Experiment Based on Different Strategies}
    \label{tab:model-selection-results}
    \begin{tabular}{P{3.5cm}P{10cm}}
    \toprule
        Selection Strategy & Selected Working Model \\
    \midrule 
        Forward + Strong Heredity & $\tau_2$, $\tau_3$, $\tau_4$, $\tau_6$, $\tau_7$, $\tau_{23}$, $\tau_{36}$, $\tau_{46}$, $\tau_{47}$, $\tau_{67}$, $\tau_{467}$   
        \\
        Forward + Weak Heredity & $\tau_2$, $\tau_3$, $\tau_4$, $\tau_6$, $\tau_7$, $\tau_{12}$, $\tau_{23}$, $\tau_{13}$, $\tau_{35}$, $\tau_{36}$, $\tau_{14}$, $\tau_{46}$, $\tau_{47}$, $\tau_{56}$, $\tau_{67}$, $\tau_{57}$ \\
        Forward + No Heredity & $\tau_2$, $\tau_3$, $\tau_4$, $\tau_6$, $\tau_7$, $\tau_{12}$, $\tau_{23}$, $\tau_{13}$, $\tau_{35}$, $\tau_{36}$, $\tau_{14}$, $\tau_{46}$,
        $\tau_{47}$, $\tau_{56}$, $\tau_{67}$, $\tau_{57}$
        \\
        No Forward & $\tau_3$, $\tau_6$, $\tau_{14}$ \\
    \bottomrule 
    \end{tabular}
\end{table}

From Table \ref{tab:model-selection-results}, we can draw the following conclusions:
\begin{itemize}
    \item \textit{Forward versus non-forward selection.} A forward selection procedure selects more terms into the working model, while the full LASSO procedure selects an overly sparse one. This is because the scale of the factorial effect sizes for different levels can vary, and the forward selection procedure can pick different penalization levels to adapt to the hierarchy. Besides, forward selection can incorporate heredity structure into the selected working model and full LASSO can not include such consideration. Therefore, the forward selection procedure provides a more comprehensive and interpretable view of the role of the factors as well as their interactions.

    \item \textit{Strong heredity, weak heredity versus no heredity.} In this application, strong heredity produces a more parsimonious working model for the two-way interactions and also discovers one three-way interaction ($\tau_{467}$). Compared with the routine of assuming away three-way or higher-order interactions in practice, this result suggests that forward selection with heredity makes it possible to gain scientific insights beyond lower-order effects.  Moreover, weak heredity also leads to a sparse and interpretable working model for two-way interactions. In this case, the result based on forward selection with weak heredity coincides with that of forward selection with no heredity, which can be viewed as a validation for the plausibility of the heredity principle. 
\end{itemize}

\section{Discussion}\label{sec:discussion}

We have discussed the theory for forward selection and post-selection inference in $2^K$ factorial designs. The method and theory are especially relevant when the number of factors, $K$, is large and diverges with the sample size. With a large $K$, fractional factorial designs \citep{wu2011experiments} are attractive alternatives in the design stage if some higher-order interactions are absent and the designer has correct prior knowledge on them \citep{wu2011experiments, pashley2019causal}. The trade-off between full and fractional factorial designs is well documented: the fractional factorial design is less costly, whereas the full factorial design allows for exploring higher-order interactions. Moreover, the design-based theory for factor selection and post-selection inference for full factorial designs serves as a stepstone for the corresponding theory for fractional factorial designs. We leave it to future research. 

There are several further directions for exploration. It is conceptually straightforward to extend the theory to general factorial designs with multi-valued factors under more complicated notations, and we thus omit the technical details to simplify the theoretical discussion. Another important direction is covariate adjustment in factorial experiments.  \citet{lin2013agnostic}, \citet{lu2016covariate} and \citet{liu2022randomization} demonstrated the efficiency gain of covariate adjustment with small $K$. \citet{zhao2023covariatefactorial} discussed covariate adjustment in factorial experiments with factors and covariates selected independent of data. Moreover, it is interesting to extend the framework to observational studies by incorporating propensity score and outcome model estimation and exploring the properties of the procedure, such as double robustness, etc. Besides, in the current work, we focus more on the analysis part instead of the design part. If we understand the properties of factor screening, then it is possible to have a better experimental design for factorial experiments, say, by introducing a pilot study. We leave it to future research to establish the theory for factor selection and covariate selection in factorial designs. 



\bibliographystyle{apalike}
\bibliography{ref.bib}

\appendix

\newpage 

\setcounter{theorem}{0}
\renewcommand{\thetheorem}{S\arabic{theorem}}
\renewcommand\theHtheorem{S\thetheorem}

\setcounter{lemma}{0}
\renewcommand{\thelemma}{S\arabic{lemma}}
\renewcommand{\theHlemma}{S\thelemma}

\setcounter{proposition}{0}
\renewcommand{\theproposition}{S\arabic{proposition}}
\renewcommand\theHproposition{S\theproposition}

\renewcommand{\thecorollary}{S\arabic{corollary}}
\setcounter{corollary}{0}
\renewcommand\theHcorollary{S\thecorollary}

\renewcommand{\thepage}{S\arabic{page}}
\setcounter{page}{1}

\renewcommand{\theequation}{S\arabic{equation}}
\setcounter{equation}{0}
\renewcommand\theHequation{S\theequation}

\renewcommand{\theassumption}{S\arabic{assumption}}
\setcounter{assumption}{0}
\renewcommand\theHassumption{S\theassumption}

\begin{center}
\Huge 
Supplementary material
\end{center}

Section \ref{sec:more-discussion} provides more discussions and extensions to the results introduced in the main paper. Section \ref{sec:wls} presents a detailed discussion of the use of WLS in factorial experiments.  Section \ref{sec:extend-to-vector} extends the inference results in Section \ref{sec:inference-perfect} to a vector of causal effects. 
Section \ref{sec:general-consistency} presents general results on the consistency of forward factor selection. Theorem \ref{thm:marginal-t} is a corollary of the results in Section \ref{sec:general-consistency}. Section \ref{sec:select-best} presents one concrete application of the methods and theory for performing inference on the best arm in factorial experiments. 

Section \ref{sec:technical-pfs} contains the technical details of the paper. Section \ref{sec:preliminary-pf} presents some preliminary probabilistic results in randomized experiments. Section \ref{sec:pf-ms-consistency} - \ref{sec:infer-order} presents proofs of all the theoretical results in both the main paper and the Appendix.

\appendix

\section{Additional results} \label{sec:more-discussion}


This section provides extensions to the results in the main paper. Section \ref{sec:wls} discusses the use of WLS in analyzing factorial experiments.  Section \ref{sec:extend-to-vector} extends the inference results under selection consistency (see Section \ref{sec:inference-perfect}) to a vector of causal effects.

\subsection{WLS for estimating factorial effects}\label{sec:wls}

In this subsection, we briefly state and prove some useful facts about WLS in estimating factorial effects. More discussions can be found in \cite{zhao2021regression}.  
Denote the design matrix as $ X = (g_{1,\bbM},\dots, g_{N,\bbM})^\top $. Let $ W = \diag{w_i}_{i=1}^N$. The problem \eqref{eqn:wls-form} has closed-form solution:
\begin{align}
    \htau &= (X^\top W X)^{-1} (X^\top W Y) \see{closed form solution of WLS}\\
    &= \{G(\cdot, \bbM)^\top G(\cdot, \bbM)\}^{-1} \{G(\cdot, \bbM)^\top \hY\} \\
    &\see{units under the same treatment arm share the same regressor} \\
    & = Q^{-1} G(\cdot, \bbM)^\top \hY. \label{eqn:htau}
\end{align}
The closed form \eqref{eqn:htau} motivates the variance estimation:
\begin{align}\label{eqn:hV-1}
    \hV_{\htau} = Q^{-2} G(\cdot, \bbM)^\top \hV_\hY G(\cdot, \bbM).
\end{align}
Alternatively, one can use the Eicker--Huber--White (EHW) variance estimation with the HC2 correction \citep{angrist2009mostly}:
\begin{align} \label{eqn:hV-2}
    \hV_{\mathrm{EHW}} =  (X^\top W X)^{-1} X^\top W \diag{\frac{\hepsilon_i^2}{1-N_i^{-1}}} W X(X^\top W X)^{-1}, ~ \hepsilon_i = Y_i - g_{i,\bbM}^\top \htau.
\end{align}
Again, because units under the same treatment arm share the same regressor, $\hV_{\mathrm{EHW}}$ simplifies to 
\begin{gather}
    \hV_{\mathrm{EHW}} = Q^{-2} G(\cdot, \bbM)^\top \hV'_\hY G(\cdot, \bbM),\label{eqn:hV-22}
    \end{gather}
    where 
    $$
    \quad \hV'_\hY = \diag{N(z)^{-1}\hS'(z,z)}_{z\in\cT} \text{ with }
    \hS'(z,z) = \frac{1}{N(z) - 1} \sum_{Z_i=z} (Y_i - g_{i,\bbM}^\top \htau)^2. 
$$
Following some algebra, we can show
\begin{align}
     \hS'(z,z) & = \frac{1}{N(z) - 1} \sum_{Z_i=z} (Y_i - \hY(z))^2 + \frac{N(z)}{N(z) - 1}\{\hY(z) - G(z,\bbM) \htau\}^2\\
     &= \hS(z, z) + \frac{N(z)}{N(z) - 1}\{\hY(z) - G(z,\bbM) \htau\}^2. 
\end{align}
Hence $\hS'(z,z) \ge \hS(z,z) $. In general $ \hY(z) \neq G(z,\bbM) \htau $, 
so the difference is not negligible. The following Lemma \ref{lem:wls-property} formally summarizes the statistical property of $\htau$ and its two variance estimators, $\hV_\htau$ and $ \hV_{\mathrm{EHW}} $. The proof can be done by utilizing the moment results from Sections C.2 and C.3 of \cite{shi2022berry}, which we omit here.
\begin{lemma}\label{lem:wls-property}
    Assume Conditions \ref{cond:uniform-design} and \ref{cond:regularity}. For the WLS in \eqref{eqn:wls-form}, we have
    \begin{enumerate}
        \item $\htau = Q^{-1} G(\cdot, \bbM)^\top \hY$ is unbiased  for the true factorial effects $\tau(\bbM)$; i.e., $\E{\htau} =  \tau(\bbM)$. 
        
        \item Both variance estimators are conservative: $N(\hV_{\htau} - V_{\htau,\lim}) = o_{\mathrm{P}}(1) $, $N(\hV_{\mathrm{EHW}} - V_{\mathrm{EHW},\lim}) = o_{\mathrm{P}}(1) $, with $V_{\htau,\lim} \succcurlyeq V_{\htau}$ and $V_{\mathrm{EHW}} \succcurlyeq V_{\htau}$, where 
$$
            V_{\htau,\lim} = Q^{-2}G(\cdot,\bbM)^\top D_\hY G(\cdot,\bbM), 
            $$
            and
            $$
            V_{\mathrm{EHW},\lim} = 
            Q^{-2}G(\cdot,\bbM)^\top \diag{\frac{1-N^{-1} }{N(z) - 1}S(z,z) + \frac{1}{N(z) - 1}\{\overline{Y}(z) - G(z,\bbM)\tau(\bbM)\}^2 } G(\cdot,\bbM).
$$ 

        \item The EHW variance estimator is more conservative than the direct variance estimator: $\hV_{\mathrm{EHW}} \succcurlyeq \hV_{\htau} $. 
    \end{enumerate}
\end{lemma}
In the fixed $Q$ setting, if we assume that the factorial effects that are not included in $\bbM$ are all zero, Lemma \ref{lem:wls-property} implies the EHW variance estimator \eqref{eqn:hV-2} or \eqref{eqn:hV-22} has the same asymptotic statistical property as the direct variance estimator \eqref{eqn:hV-1}, which agrees with the conclusion of \cite{zhao2021regression}.

\subsection{Extension of post-selection inference to vector parameters}\label{sec:extend-to-vector}

In this subsection, we present an extension of Theorem \ref{thm:be-perfect-ms} to a vector of causal parameters:
\begin{align*}
   \Gamma =  (\gamma_1,\dots,\gamma_L)^\top, \quad \text{ where } \gamma_l =  f_l^\top \overline{Y}.
\end{align*}
For convenience we can stack $ f_1,\dots, f_L$ into a weighting matrix $F = ( f_1,\dots, f_L)$ and write
\begin{align}
    \Gamma = F^\top \overline{Y}. 
\end{align}
We will focus on linear projections of $\Gamma$, defined as $\gamma_b =  b^\top \Gamma $ for a given $b\in \bbR^L$. 
Naturally, we can apply forward selection and construct RLS-based estimators for $\Gamma$:
\begin{align}
    \hGamma_\textsc{r} = (\hgamma_{1,\textsc{r}},\dots,\hgamma_{L,\textsc{r}})^\top,\quad \hV_{\hGamma,\textsc{r}} = F[\hat{\bbM}]^\top 
 \hV_\hY F[\hat{\bbM}],\label{eqn:Gamma-r}
\end{align}
where  
\begin{align}
F[\hat{\bbM}]  = 
    Q^{-1}G(\cdot,  {\hat{\bbM}})G(\cdot,  {\hat{\bbM}})^\top F.
\end{align}
For $\gamma_b$, an estimator based on \eqref{eqn:Gamma-r} is
\begin{align}
    \hgamma_{b,\textsc{r}} = b^\top \hGamma_\textsc{r}, \quad \hv^2_{b,\textsc{r}} = b^\top \hV_{\hGamma,\textsc{r}} b.  
\end{align}
For standard factorial effects, we can use WLS to obtain the robust covariance matrix (see Section \ref{sec:wls}).
For one single $b$, we can actually apply Theorem \ref{thm:be-perfect-ms} with 
\begin{align}
     f_b  = Fb = \sum_{l=1}^L b_l  f_l.
\end{align}
Define $  f_b^\star = F[\bbM^\star] b $. We then have the following theorem:
\begin{theorem}[Statistical properties linear projections of $\Gamma$]\label{thm:be-perfect-ms-vector}
 Assume Conditions \ref{cond:uniform-design}-\ref{cond:heredity}. Let $N\to\infty$. Then  
\begin{align*}
    \frac{\hgamma_{b,\textsc{r}} - \gamma}{v_{b,\textsc{r}}} \rightsquigarrow \cN(0,1)
\end{align*}
where $v^{2}_{b,\textsc{r}} =  f_b^{\star\top} V_\hY  f_b^\star$. Further  assume $\| f_b^\star\|_\infty = O(Q^{-1})$. The variance estimator $\hv_{b,\textsc{r}}^2$ is conservative in the sense
    \begin{align*}
        N(\hv_{b,\textsc{r}}^2 - v_{b,\textsc{r},\mathrm{lim}}^2) \xrightarrow{{\mathrm{P}}} 0, \quad v_{b,\textsc{r},\mathrm{lim}}^2 \ge v_{b,\textsc{r}}^2,
    \end{align*}
    where $v_{b,\textsc{r},\mathrm{lim}}^2 =   f_b^{\star\top} D_\hY  f_b^\star$ is the limiting value of $\hv_{b,\textsc{r}}^2$. 
\end{theorem}

The proof of Theorem \ref{thm:be-perfect-ms-vector} is similar to that of Theorem \ref{thm:be-perfect-ms}, which is based on Lemma \ref{lem:BE-perfect} and thus omitted here. 
Moreover, for a fixed integer $L$, Theorem \ref{thm:be-perfect-ms-vector} implies joint normality of $\hGamma_\textsc{r}$, a result due to the Cram\'{e}r--Wold theorem. We summarize the result as the following corollary and omit the proof:
\begin{corollary}\label{cor:joint-clt}
    Assume a fixed $L$. Assume Conditions \ref{cond:uniform-design}-\ref{cond:heredity}. We have
    \begin{align*}
        V_{\hGamma,\textsc{r}}^{-1/2}(\hGamma_{\textsc{r}} - \Gamma) \rightsquigarrow \cN(0, I_L),
    \end{align*}
    where $V_{\hGamma,\textsc{r}} = F[\bbM^\star]^\top V_\hY F[\bbM^\star]$. Further assume $\max_{\|b\|_2 = 1}\| f_b^\star\|_\infty = O(Q^{-1})$. The variance estimator $\hv_{b,\textsc{r}}^2$ is conservative in the sense that 
    \begin{align*}
        N(\hV_{\hGamma,\textsc{r}} - V_{\hGamma,\textsc{r},\mathrm{lim}}) \xrightarrow{{\mathrm{P}}} 0, \quad V_{\hGamma,\textsc{r},\mathrm{lim}} \succcurlyeq V_{\hGamma,\textsc{r}},
    \end{align*}
    where $V_{\hGamma,\textsc{r},\mathrm{lim}} =  F[\bbM^\star]^\top D_\hY F[\bbM^\star]^\top$ is the limiting value of $\hV_{\hGamma,\textsc{r}} $. 
\end{corollary}

\commenting{
\subsection{Discussion of several  selection methods in finite population factorial experiments}\label{sec:more-selection}


In this subsection, we discuss several alternative selection methods that can be adopted with the forward selection algorithm. Many of them have simple forms due to the orthogonality of the factorial design matrix. 

\begin{itemize}
\item Lasso: Lasso solves the following problem: 
\begin{align*}
    \hat{\bbM}_\textsc{l} = \{\cK: \htau_{\textsc{l}, \cK}\neq 0\}, \quad  \htau_{\textsc{l}, \cK} = \min_{\tau' \in \mathbb{R}^H} \frac{1}{2}\sum_{z\in\cT}  w_i(Y_i - g_i^\top \tau')^2 + \lambda_\textsc{l} \|\tau'\|_1.
\end{align*}
Due to the orthogonality of the design, Lasso leads to a direct soft-thresholding solution:
\begin{align}
    \htau_{\textsc{l}, \cK} = \left\{
    \begin{array}{cc}
        0, & |\htau_\cK| \le \lambda_\textsc{l}; \\
        \htau_\cK - \lambda_\textsc{l}, & \htau_\cK > \lambda_\textsc{l};\\
        \htau_\cK + \lambda_\textsc{l}, & \htau_\cK < - \lambda_\textsc{l}.\\
    \end{array}
    \right.
\end{align}
Hence, we have 
\begin{align}
    \hat{\bbM}_\textsc{l} = \{\cK: |\htau_{\cK}|\ge \lambda_\textsc{l}\}.
\end{align}
\item AIC: the AIC criterion directly penalizes the size of the working model:
\begin{align}
    (\htau_{\textsc{a}}, \hat{\bbM}_{\textsc{a}}) = &\underset{\tau' \in \mathbb{R}^H,\bbM\subset [K]}{\arg\min}~ \sum_{z\in\cT}  w_i(Y_i - g_{i,\bbM}^\top \tau'(\bbM))^2 + \lambda_{\textsc{a}} |\bbM|\\
     = &\underset{\tau' \in \mathbb{R}^H,\bbM\subset [K]}{\arg\min}~ \sum_{\cK\in\bbM^c}   \htau_\cK^2 + \lambda_{\textsc{a}} |\bbM|.\label{eqn:AIC}
\end{align}
From \eqref{eqn:AIC} we can see that
\begin{align*}
    \hat{\bbM}_\textsc{a} = \{\cK:|\htau_\cK| \ge \sqrt{\lambda_{\textsc{a}}}\}.
\end{align*}
\item BIC: the BIC criterion is similar to AIC, only with the difference on the penalization level:
\begin{align*}
    (\htau_{\textsc{b}}, \hat{\bbM}_{\textsc{b}}) = \underset{\tau \in \mathbb{R}^H,\bbM\subset [K]}{\arg\min}~ \sum_{z\in\cT}  w_i(Y_i - g_{i,\bbM}^\top \tau(\bbM))^2 + \lambda_\textsc{b} |\bbM| \log N,
\end{align*}
which gives
\begin{align*}
    \hat{\bbM}_\textsc{b} = \{\cK:|\htau_\cK| \ge \sqrt{\lambda_{\textsc{b}}\log N}\}.
\end{align*}
\item SIS: SIS picks the largest $m$ marginal scores:  
\begin{align*}
    \hat{\bbM}_{\textsc{s}} = \{\cK:|\htau_\cK| \text{ is among the largest $m$ ones}\}.
\end{align*}
\end{itemize}
With more delicate assumptions and tuning parameter choices, these methods can also be justified theoretically for selection consistency and post-selection inference. We omit the details.

\subsection{Discussion: more general centering values}

\subsubsection{Unsaturated weighted least square: a closed form expression}

In this section we first derive the closed-form expression for unsaturated WLS estimation, then verify the nice targeting property we mentioned in the previous section. 

First we need to introduce a transformation matrix $\bP_{\Delta\delta_{[K]}}$, with columns and rows indexed by subsets $\{\cK\subset [K]\}$ of the $K$ factors. Generally, it is used to reveal the relationship between designs with different configurations of centering factors $\delta_{[K]}$ and $\delta'_{[K]} = \delta_{[K]}+\Delta\delta_{[K]}$. The transformation is actually linear:
\begin{align}
    \left(f_{\delta'_{[K]}}(z^*_\cK)\right)_{\cK\subset [K]} = \left(f_{\delta_{[K]}}(z^*_\cK)\right)_{\cK\subset [K]} \bP_{\Delta\delta_{[K]}}.
\end{align}

The closed form of $\bP_{\Delta\delta_{[K]}}$ is easy to derive.  Note that for all $\cK'\subset [K]$, we have 
\begin{align}
    f_{\delta'_{[K]}}(z^*_{\cK'}) = \sum_{\cK\subset \cK'}f_{\delta_{[K]}}(z^*_{\cK}) \prod_{k\in \cK'\backslash \cK}(\Delta\delta)_k, \notag
\end{align}
which implies the element of $\bP_{\Delta\delta_{[K]}}$ indexed by $(\cK, \cK')$ is given by
\begin{align}\label{eqn:Pmatrix}
\bP_{\Delta\delta_{[K]}}(\cK,\cK')=\left\{
\begin{array}{ccc}
    \prod_{k\in \cK'\backslash \cK}(\Delta\delta)_k &, & \cK\subset\cK', \\
     0 &, & \cK\subsetneq \cK'. 
\end{array}
\right.
\end{align}

Define $\bQ_{\Delta\delta_{[K]}} = \bP_{\Delta\delta_{[K]}}^{-1}$ to be the inverse. Note that $\bQ_{\Delta\delta_{[K]}}$ is simply taking out a $\Delta\delta_{[K]}$ vector from a group of centering factors, so by symmetry we have 
\begin{align}\label{eqn:Qmatrix}
    \bQ_{\Delta\delta_{[K]}}(\cK,\cK') = \left\{
\begin{array}{ccc}
    (-1)^{|\cK'|-|\cK|}\prod_{k\in \cK'\backslash \cK}(\Delta\delta)_k &, & \cK\subset\cK', \\
     0 &, & \cK\subsetneq \cK'. 
\end{array}
\right.
\end{align}

We shall give an example of the above matrix in the three-factor case, which appears (incompletely) in the appendix of \cite{zhao2021regression}. Let $A'=A-\delta_A$, $B'=B-\delta_B$, $C'=C-\delta_C$. 
\begin{align*}
    \begin{pmatrix}
    1\\
    A\\
    B\\
    C\\
    AB\\
    AC\\
    BC\\
    ABC
    \end{pmatrix} = \bP_{\Delta\delta_{[K]}}^\top 
    \begin{pmatrix}
    1\\
    A'\\
    B'\\
    C'\\
    A'B'\\
    A'C'\\
    B'C'\\
    A'B'C'
    \end{pmatrix}=
    \begin{pmatrix}
    1        & 0 & 0 & 0 & 0 & 0 & 0 & 0 \\
    \delta_A & 1 & 0 & 0 & 0 & 0 & 0 & 0 \\
    \delta_B & 0 & 1 & 0 & 0 & 0 & 0 & 0 \\
    \delta_C & 0 & 0 & 1 & 0 & 0 & 0 & 0 \\
    \delta_A\delta_B & \delta_B & \delta_A & 0 & 1 & 0 & 0 & 0\\ 
    \delta_A\delta_C & \delta_C & 0 & \delta_A & 0 & 1 & 0 & 0\\ 
    \delta_B\delta_C & 0 & \delta_C & \delta_B & 0 & 0 & 1 & 0\\ 
    \delta_A\delta_B\delta_C & \delta_B\delta_C & \delta_A\delta_C & \delta_A\delta_B & \delta_C & \delta_B & \delta_A & 1\\
    \end{pmatrix}
    \begin{pmatrix}
    1\\
    A'\\
    B'\\
    C'\\
    A'B'\\
    A'C'\\
    B'C'\\
    A'B'C'
    \end{pmatrix}.
\end{align*}

The following theorem shows that $\bP_{\Delta\delta_{[K]}}$ and $\bQ_{\Delta\delta_{[K]}}$ totally determines the structure of $\bD_h$.

\begin{theorem}\label{thm:closeD}
Consider weighted least squares with centering factors $\delta_{[K]}$ and weights proportional to the size of each stratum. Let $\Delta\delta_{[K]} = \delta_{[K]}-(1/2)_{k=1}^K$. The unsaturated regression  on up to all $m$-level main/interactions terms has a coefficient vector:
\begin{align}\label{eqn:wls-res}
    (\tilde{\tau}_{\cK})_{\{|\cK|\le m\}} = ({\tau}_{\cK})_{\{|\cK|\le m\}} + \bD_h\cdot({\tau}_{\cK})_{\{|\cK|> m\}},
\end{align}
where $\bD_h$ is given by 
\begin{align*}
    \bD_h = \bP_{\Delta\delta_{[K]}}\left(\{\cK\subset [m]\}, \{\cK\subset [m]\}\right) \cdot \bQ_{\Delta\delta_{[K]}}\left(\{\cK\subset [m]\}, \{\cK\subset [K] \backslash[m]\}\right).
\end{align*}
\end{theorem}

\begin{corollary}\label{cor:closeD}
The matrix $\bD$ has a closed-form expression:
\begin{enumerate}
    \item For $\cK\subsetneq \cK'$,
    \begin{align}
    \bD_h(\cK, \cK') = 0.
    \end{align}
     
    \item For $\cK\subset \cK'$, let $|\cK|=k$, $|\cK'|=k'$, with $k\le m <k'$,
    \begin{align}
    \bD_h(\cK, \cK') = \sum_{l=0}^{m-k} (-1)^{k'-k+1-l} \begin{pmatrix}
    k'-k+1 \\
    l
    \end{pmatrix}
    \prod_{t\in \cK'\backslash \cK}\left(\delta_t-\frac{1}{2}\right).
    \end{align}
\end{enumerate}

\end{corollary}

\begin{proof}
This result can be derived through a careful calculation based on the definition of $\bP$ and $\bQ$ from \eqref{eqn:Pmatrix} and \eqref{eqn:Qmatrix} along with Theorem \ref{thm:closeD} thus omitted here. 
\end{proof}

\subsubsection{A sufficient condition for sign consistency in population WLS regression}

\begin{definition}[Active interaction number]\label{def:sparsity}
For every $z_k$ of the $K$ factors, there are $s_k$ factors that have nonzero interaction with $z_k$, where $s_k\in [K-1]$ is a nonnegative integer associated with $K$. We call $s_k$ the active interaction number of factor $z_k$. The maximal active interaction number is subsequently defined as $s_K = \max_{k\in [K]} s_k$.
\end{definition}

This definition is mainly devoted to finer technical purposes in Theorem \ref{thm:suffcond}. 

\begin{theorem}\label{thm:suffcond}
Assume we run weighted least square under the setting depicted in  Theorem \ref{thm:closeD}. Define the maximal decaying rate $c_K=\max_{l\in[K]} c_l$. Recall the predefined maximal active interaction number $s_K$ from Definition \ref{def:sparsity}. If we have 
\begin{align}
s_Kc_K\max_{k=1,\dots,K} |\delta_k-1/2|<\ln 2, \label{cond:sufficient}
\end{align}
then the unsaturated regression coefficients
$(\tilde{\tau}_{\cK})_{\{|\cK|\le m\}}$ and the corresponding saturated regression coefficients  $({\tau}_{\cK})_{\{|\cK|\le m\}}$ from \eqref{eqn:wls-res} have same signs on every term.

\end{theorem}

Condition \eqref{cond:sufficient} unifies the property of factorial effects and the information of the design pattern(the centering factors $\delta_{[K]}$). The product of $s_K$ and $c_K$ demonstrates a trade-off between the active interaction number and the hierarchy structure. Sparser interactions require a slower decaying rate and vice versa. Besides, the product of $s_Kc_K$ and $\max_{k=1, \dots, K} |\delta_k-1/2|$ shows that if $\delta_k$ lies closer to $1/2$, fewer restrictions are needed on the effect structure. This aligns with the result in \cite{zhao2021regression}: when $\delta_k=1/2$ holds for all $k=1,\dots, K$,  $\bD_h=\boldsymbol{0}$, so that forward selection always works.


}

\subsection{General results on consistency of forward selection}\label{sec:general-consistency}


In this section, we provide some theoretical insights into the forward factor selection algorithm (Algorithm \ref{alg:forward-ms}). This section starts from a more broad discussion where we allow the S-step to be general procedures that satisfy certain conditions. We will show Bonferroni corrected marginal $t$-test is a special case of these procedures. 
\commenting{
Recall that the validity of a test requires that given the null is true, the probability of rejection should be less than the pre-specified significance level. The consistency of a test, on the other hand, requires that when the alternative is true, we have probability tending to 1 to reject the null.  Following the tradition in the realm of  selection \citep{wasserman2009high}, we introduce the following definitions:
\begin{definition}[Terminologies for  selection]
The following terminologies are wrapped up for the discussion of  selection:
\begin{itemize}
    \item Type I error:  commit a false positive. Because the selected positives are contained in $\hat{\bbM}$, this translates into $\hat{\bbM}\cap\bbM^{\star c}\neq \varnothing$. 
    \item Type II error:  commit a false negative. Because the selected negatives are contained in $\hat{\bbM}$, this translates into $\hat{\bbM}^c\cap\bbM^{\star}\neq \varnothing$. 
\end{itemize}
\end{definition}

Clearly size measures the ability of under-selection; because zero $q(\hat{\bbM})$ means $\hat{\bbM}\subset \bbM^\star$. Power measures the ability of over-selection because a power of 1 means $\bbM^\star\subset\hat{\bbM}$. We have the following graph to visualize the idea:

\begin{figure}[th]
\centering

\begin{tikzpicture}[fill]
\scope
\clip (-2,-2) rectangle (2,2)
      (1,0) circle (1);
\fill[gray] (0,0) circle (1);
\endscope
\scope
\clip (-2,-2) rectangle (2,2)
      (0,0) circle (1);
\fill[lightgray] (1,0) circle (1);
\endscope
\draw (0,0) circle (1) (0,1)  node [text=black,above] {$\displaystyle \hat{{\bbM}}$}
      (1,0) circle (1) (1,1)  node [text=black,above] {$\displaystyle {\bbM}^{\star}$}
      (-2,-2) rectangle (3,2);
      
\draw (1.25,0.25) node [anchor=north west][inner sep=0.75pt]   [align=left] {II};
\draw (-0.5,0.25) node [anchor=north west][inner sep=0.75pt]   [align=left] {I};

\end{tikzpicture}
\label{fig:error-visual}
\caption{Visualization of the two types of errors} selection

\end{figure}
}

We start with some regularization conditions to characterize a ``good" layer-wise S-step and ensure the H-step is compatible with the structure of the true factorial effects.  We use ${\bbM}^\star_{d,+}$ denotes the pruned set of effects on the $d$-th layer based on the true model $\bbM_{d-1}^\star$ on the previous layer; that is,
\begin{align*}
    {\bbM}^\star_{d,+} = \texttt{H}({\bbM}^\star_{d-1}).
\end{align*}
These discussions motivate the following assumption on the layer-wise selection procedure $\hat{\texttt{S}}(\cdot)$:
\begin{assumption}[Validity and consistency of the selection operator]\label{asp:valid-consistent}
We denote 
\begin{align*}
    \tilde{\bbM}_{d} = \hat{\text{\em \texttt{S}}}(\bbM^{\star}_{d,+};\{Y_i,Z_i\}^N_{i=1}),
\end{align*}
where $\bbM^{\star}_{d,+} = \text{\em \texttt{H}}(\bbM^{\star}_{d-1})$ is defined as above. Let $\{\alpha_d\}_{d=1}^D$ be a sequence of significance levels in $(0,1)$. We assume that the following \emph{validity} and \emph{consistency} property hold for $\hat{\text{\em \texttt{S}}}(\cdot)$:  
\begin{gather}
    \text{Validity: } \limsup_{N\to\infty}\Prob{\tilde{\bbM}_{d}\cap\bbM^{\star c}_d\neq \varnothing }\le \alpha_d,  \label{eqn:validity}\\
    \text{Consistency: } \limsup_{N\to\infty} D\sum_{d=1}^{D} \Prob{\tilde{\bbM}^c_{d}\cap\bbM^{\star }_{d}\neq\varnothing} = 0.  \label{eqn:consistency}
\end{gather}
\end{assumption}

Assumption \ref{asp:valid-consistent} can be verified for many selection procedures. In Theorem \ref{thm:marginal-t} we will show it holds for the layer-wise Bonferroni corrected marginal testing procedure in Algorithm \ref{alg:forward-ms}. Moreover, in the high dimensional super population study, a combination of data splitting, adaptation of $\ell_1$ regularization, and marginal t tests can also fulfill such a requirement \citep{wasserman2009high}.

Besides, we assume the $\texttt{H}(\cdot)$ operator respects the structure of the nonzero factorial effects:
\begin{assumption}[H-heredity]\label{asp:H-heredity}
For $d=1,\cdots,D-1$, we have
\begin{align*}
\bbM^\star_{d+1}\subset\texttt{\em H}(\bbM^\star_{d}).
\end{align*}
\end{assumption}

One special case of $\texttt{H}(\cdot)$ operator satisfying Assumption \ref{asp:H-heredity} is naively adding all the higher-order interactions regardless of the lower-order selection results. Besides, if we have evidence that the effects have a particular hierarchical structure, applying the heredity principles can improve selection accuracy as well as interpretability of the selection results.

\begin{theorem}[selection consistency]\label{thm:ms-consistency}
Under Assumption \ref{asp:valid-consistent} and \ref{asp:H-heredity}, the forward selection procedure \eqref{eqn:track-model} has the following properties:
\begin{enumerate}[label = (\roman*)]
    \item {\em Type I error control.} Forward selection controls the Type I error rate, in the sense that 
    \begin{align}
        \underset{N\to\infty}{\lim\sup}~\Prob{\hat{\bbM}_d\cap{\bbM_d^\star}^c\neq\varnothing \text{ for some } d\in [D]}\le \alpha = \sum_{d=1}^D \alpha_d.
    \end{align}
    \item {\em  selection consistency.} Further assume $\alpha=\alpha_N\to0$. The forward procedure consistently selects all the nonzero effects up to $D$ levels with probability tending to 1: 
    \begin{align}
        \underset{N \to \infty}{\lim\sup}~\Prob{\hat{\bbM}_d = \bbM_d^\star \text{ for all }d\in[D]}=1.
    \end{align}
\end{enumerate}
\end{theorem}

Theorem \ref{thm:ms-consistency} consists of two parts. First, one can control the type I error rate, which is defined as the probability of over-selecting at least one zero effect. The definition is introduced and elaborated in more detail in \cite{wasserman2009high} for model selection in linear regression. Second, if the tuning parameter $\alpha = \sum_{d=1}^D\alpha_d$ vanishes asymptotically, one can achieve selection consistency up to $D$ levels of effects.  To apply Theorem \ref{thm:ms-consistency} to specific procedures, the key step is to justify Assumption \ref{asp:valid-consistent} and  Assumption \ref{asp:H-heredity}, which we will do for Bonferroni corrected marginal $t$-tests as an example in the proof of Theorem \ref{thm:marginal-t} (see Section \ref{sec:pf-marginal-t}).

Moreover, the scaling of $\alpha$ plays an important role in theoretical discussion. To achieve selection consistency, we hope $\alpha$ decays as fast as possible; ideally, if $\alpha$ equals zero then we do not commit any type I error (or equivalently, we will never select redundant effects). However, for many data-dependent selection procedures, $\alpha$ can only decay at certain rates because a fast decaying $\alpha$ means a higher possibility of rejection, thus leading to strict under-selection. Therefore, in the tuning process, $\alpha_d$ should be scaled properly if one wants to achieve selection consistency. Nevertheless, even if the tuning is hard and selection consistency can not be achieved, we still have many strategies to exploit the advantage of the forward selection procedure; see Section \ref{sec:imperfect-ms} for more discussion. 

Lastly, as we have commented in Section \ref{sec::forward-ms-alg}, in practice people have many alternative methods for the S-step. They are attractive in factorial experiments because many lead to simple form solutions due to the orthogonality of factorial designs. For example, Lasso is a commonly adopted strategy for variable selection in linear models \citep{zhao2006model}. It solves the following penalized WLS problem in factorial settings: 
\begin{align*}
    \hat{\bbM}_\textsc{l} = \{\cK: \htau_{\textsc{l}, \cK}\neq 0\}, \quad  \htau_{\textsc{l}, \cK} = \min_{\tau' \in \mathbb{R}^H} \frac{1}{2}\sum_{z\in\cT}  w_i(Y_i - g_i^\top \tau')^2 + \lambda_\textsc{l} \|\tau'\|_1.
\end{align*}
Due to the orthogonality of $G$, the resulting $\hat{\bbM}$ has a closed-form solution \citep{hastie2009elements}:
\begin{align}
    \hat{\bbM}_\textsc{l} = \{\cK: |\htau_{\cK}|\ge \lambda_\textsc{l}\}.
\end{align}
Other methods, such as information criteria (AIC, BIC, etc.) \citep{bai2022asymptotics}, sure independence selection \citep{fan2008sure}, etc., are also applicable. With more delicate assumptions and tuning parameter choices, these methods can be justified theoretically for selection consistency and post-selection inference. We omit the details.

\subsection{Application to inference on the best arm in factorial experiments} \label{sec:select-best}

In Section \ref{sec:inference-perfect}, we consider the problem of making inference on a single factorial causal effect $\gamma = f^\top \overline{Y}$. As an application of the proposed framework, we study the problem of inference on the ``best'' effect under a constraint on the number of active factors. In our context, we define the best effect as the effect with the highest level.  In what follows, Section \ref{sec::best-arm-procedure} introduces our setup and an inferential procedure, and Section \ref{sec::best-arm-theory} presents theoretical guarantees.

\subsubsection{Inference on the ordered effects in factorial experiments}\label{sec::best-arm-procedure}


In many real-world problems, we ask many questions about inferring the ordered values of a set of causal effects. For example, in agricultural studies, if a researcher aims to identify the best combination of fertilizer type, irrigation level, and pesticide usage to maximize the yield of a particular crop, she can use a factorial design with $K = 3$ factors and choose the weighting vectors introduced in Section \ref{Sec:inference-perfect-screeenig} to be the canonical bases to identify the maximal mean of the potential yields across all possible factor combinations. 

Mathematically, we have a set of causal effects $ \Gamma $ defined by pre-specified weighting vectors $  f_1, \ldots,   f_L$, where $L$ can be  potentially large:
\begin{align}
    \Gamma = \{\gamma_1,\dots, \gamma_L\} \text{ where }  \gamma_l =  f_l^\top \overline{Y}. 
\end{align}
We aim to perform statistical inference on their ordered values 
\begin{align}\label{eqn:ordered-gamma}
    \gamma_{(1)}\geq \ldots \geq \gamma_{(l_0)}
\end{align}
with $l_0<L$ being a fixed positive integer. 
In particular, if we choose $l_0 = 1$ and $\{ f_l\}_{l\in[L]} = \{\bse(z)\}_{z\in\cT}$ to be the set of the canonical bases 
\begin{align}
    \{\bse(z): \bse(z) = (0,\dots,0,\underbrace{1}_{\text{index $z$}},0,\dots,0)^\top\}_{z\in\cT},
\end{align}
then our inferential targets include the maximal potential outcome means:
\begin{align}\label{eqn:best-Yz}
\overline{Y}_{(1)} = \max_{z\in\cT} \overline{Y}(z).
\end{align}

A more practical consideration in factorial experiments is to incorporate structural constraints into the choices of $\{ f_l\}_{l\in[L]}$, as it might be unnecessary or infeasible to consider all treatment combinations $\cT$ due to the question of interest or resource constraints, especially when $K$ is large. For example, in the conjoint survey experiments regarding preferences for presidential candidates (\cite{hainmueller2014causal}, also in Section \ref{sec:real-data}), we are more interested in a particular subset of combinations of candidate traits instead of all possibilities. This suggests that we should take a subset $\cT'$ from $\cT$ for comparison. By focusing on $\{ f_l \}_{l\in[L]}$ that is most relevant, the inferential target $\max_{z\in\cT'} \overline{Y}(z)$ allows us to use the available data to decide if the best causal parameter among those practically interesting ones has a non-zero causal effect.

Two challenges exist in delivering valid statistical inference on $\gamma_{(1)}, \ldots, \gamma_{(l_0)}$ in factorial experiments. On the one hand, sample analogs of the ordered parameters, $(\hat{\gamma}_{(1)}, \ldots,\hat{\gamma}_{(l_0)} )$, are often biased estimates of $({\gamma}_{(1)}, \ldots,{\gamma}_{(l_0)} )$ due to the well-known winner's curse phenomenon \citep{andrews2019inference,guo2021sharp, wei2023inference}. On the other hand, although one might argue that existing approaches can be applied to remove the winner's curse bias in $\hat{\gamma}_{(l)}$, these approaches do not account for the special structural constraint in factorial experiments. Rigorous statistical guarantees have been lacking in our context due to the unique presence of both large $L$ and large $Q$ in factorial designs.

To simultaneously address the above challenges, we propose a procedure that tailors the tie-set identification approach proposed in \cite{claggett2014meta} and  \cite{wei2023inference} to our current problem setup. We focus on making inferences on the first ordered value $\gamma_{(1)}$ to simplify discussion, and our approach extends naturally to other ordered values. The proposed procedure is provided in Algorithm \ref{alg:select-tie}.

\begin{algorithm}[!ht]
\DontPrintSemicolon
\SetKwInput{KwInput}{Input}                
\SetKwInput{KwOutput}{Output}  
\SetKwFunction{MADES}{MADE-S}
\SetKwProg{Fn}{Function}{:}{\KwRet}

  \KwInput{Factorial data $(Y_i, {Z}_i)$; prespecified integer $D$; initial model for factorial effects $\hat{\bbM} = \{\varnothing\}$; prespecified significance level $\{\alpha_d\}_{d=1}^D$; set of weighting vectors $\{ f_l\}_{l\in[L]}$; thresholds $ \eta $.}
  \KwOutput{Selected working model $\hat{\bbM}$. }
    
    Perform forward selection with Algorithm \ref{alg:forward-ms} and obtain a working model $\hat{\bbM}$.
    
    Obtain RLS-based estimates: use Equation \eqref{eqn:fM} and definition of $\hY_\textsc{r}$ \eqref{eqn:RLS-1} to compute 
      \begin{align*}
            f_l[\hat{\bbM}] = Q^{-1} G(\cdot, \hat{\bbM}) G(\cdot, \hat{\bbM})^\top f_l, \quad \hgamma_l =  f_l^\top \hat{Y}_{\textsc{r}}  =  f_l[\hat{\bbM}]^\top\hY, \quad l\in[L]. 
      \end{align*}
      
    Record the set of effects close to $\hat{\gamma}_{(1)}$:
    \begin{align*}
          \hat{\cL}_{1} = \lt\{l\in[L]\mid |\hgamma_l- \hat{\gamma}_{(1)} | \le \eta\rt\}
      \end{align*} 
    where $\eta$ is a tuning parameter that can be selected using the algorithm provided in \citet[][Appendix C.1]{wei2023inference}. 
    
    Define
    \begin{align*}
     f_{\hat{\cL}_1}[\hat{\bbM}] =(Q|\hat{\cL}_{1}|)^{-1}\sum_{l\in \hat{\cL}_1}G(\cdot,{\hat{\bbM}}) G({\cdot, \hat{\bbM}})^\top f_l.
    \end{align*}
   Generate point estimate and variance estimator for $\gamma_{(1)}$:
      \begin{gather*}
         \hY_{(1)} = \frac{1}{|\hat{\cL}_{1}|} \sum_{l\in\hat{\cL}_1} \hgamma_{l} =  f_{\hat{\cL}_1}[\hat{\bbM}]^\top\hY , \quad \hat{v}^2_{(1)} =  f_{\hat{\cL}_1}[\hat{\bbM}]^\top \hV_Y  f_{\hat{\cL}_1}[\hat{\bbM}].  
      \end{gather*}

    \KwRet{$\hat{\cL}_1, \hY_{(1)}, \hv^2_{(1)}$.}
    
\caption{Inference on best causal effect(s)}
\label{alg:select-tie}
\end{algorithm}

Algorithm \ref{alg:select-tie} consists of three major components. First, we need to construct $\hgamma_l =  f_l^\top\hY_{\textsc{R}}$ with feature selection (see Step 1-2). These RLS-based estimators  enjoy great benefits for large $Q$ and small $N_0$ regimes based on our previous discussion. Second, we construct $\hat{\cL}_1$ to include the estimates that are close to $\hgamma_{(1)}$ (see Step 3). Intuitively, these collected estimates are different due to random error.  We will show that with proper tuning, this procedure will include all the $l$ for which $\gamma_{l}$ are statistically indistinguishable from $\gamma_{(1)}$ with high probability. Third, we construct estimators by averaging over $\hat{\cL}_1$ (see Step 4). By averaging the estimates over the selected $\hat{\cL}_1$ we alleviate the impact of randomness and obtain accurate estimates for the maximal effect.

\subsubsection{Theoretical guarantees}\label{sec::best-arm-theory}

In the following, we present the theoretical guarantees for Algorithm \ref{alg:select-tie}. We introduce the following notation $\cL_{1}$ to include all effects that stay in a local neighborhood of $\gamma_{(1)}$:
\begin{align}\label{eqn:near-tie}
    \cL_{1} = \lt\{l\in[L] \mid |\gamma_l - \gamma_{(1)}| =  O(N^{-\delta_3})\rt\},  \text{ for some } \delta_3 > 0.
\end{align}
A well-known fact is that the naive estimator $\max_{z\in[Q]}\hY(z)$ is an overly optimistic estimate for $\gamma_{(1)}$ when  $\cL_1$ contains more than one element \citep{andrews2019inference, wei2023inference}. 
Define
\begin{align*}
    d_h = \max_{z\in\cL_{1} } |\gamma_l - \gamma_{(1)}|, \quad d^\star_h = \min_{z\notin\cL_{1}} |\gamma_l - \gamma_{(1)}|.
\end{align*}
 as within- and between-group distances, respectively. We work under the following condition: 
\begin{condition}[Order of $d_h$, $d_h^\star$ and $\eta$]\label{cond:distance}
Assume the within and between group distances satisfy: 
$$
d_h^\star= \Theta(N^{\delta_1}), \quad \eta = \Theta(N^{\delta_2}), \quad d_h = \Theta(N^{\delta_3}).
$$ 
with $\delta_3 \le -1/2 < \delta_2 < \delta_1 \le 0$.
\end{condition}
Define the population counterpart of $ f_{\hat{\cL}_1}[\hat{\bbM}]$ as 
\begin{align*}
     f^\star_{(1)}=(Q|{\cL}_{1}|)^{-1}\sum_{l\in {\cL}_{1}}G(\cdot,{{\bbM}^\star}) G({\cdot, {\bbM}^\star})^\top f_l.
\end{align*}

We establish the following result for the procedure provided in Algorithm \ref{alg:select-tie}. 
\begin{theorem}[Asymptotic results on the estimated effects using Algorithm \ref{alg:select-tie}]\label{thm:infer-order} 
Recall $ \delta_2 $ from Condition \ref{cond:distance} and $ \delta'' $ from Condition \ref{cond:order}(iii). Assume Condition \ref{cond:uniform-design}--\ref{cond:heredity} and \ref{cond:distance}.
Let $N \rightarrow \infty$. 
 If
\begin{gather}
{N^{-(1+2\delta_2-\delta'')}}\to 0, \label{eqn:asp-condition-1}\\ 
L\cdot|\cL_1|\cdot{N}^{-\frac{1-\delta''}{2}} \to 0, \label{eqn:asp-condition-2}
\end{gather}
with $ \delta_2 $ from Condition \ref{cond:distance} and $ \delta'' $ from Condition \ref{cond:order}(iii). Then
\begin{align*}
      \frac{ \hgamma_{(1)}  -   \gamma_{(1)}}{v_{(1)}} \rightsquigarrow \cN(0, 1), 
\end{align*}
where $v_{(1)}^2 =  f_{{\cL}_1}[\bbM^\star]^{\top} V_Y  f_{{\cL}_1}[\bbM^\star]^{\top} $.
Moreover,  $\hat{v}_{(1)}^2$ is conservative in the sense that 
\begin{align*}
    {N} (\hat{v}_{(1)}^2 - v^2_{(1),\mathrm{lim}})\xrightarrow{{\mathrm{P}}} 0,  ~ v^2_{(1),\mathrm{lim}} \ge v_{(1)}^2,
\end{align*}
where $v^2_{(1),\mathrm{lim}} =  f_{\cL_1}[\bbM^\star]^{\top} D_\hY  f_{\cL_1}[\bbM^\star]^{\top}$ is the limiting value of $v_{(1)}^2 $.
\end{theorem}

The conditions \eqref{eqn:asp-condition-1} and \eqref{eqn:asp-condition-2} in Theorem \ref{thm:infer-order} are mild and reveal a trade-off between some mathematical quantities. For the first asymptotic condition in
\eqref{eqn:asp-condition-1},
when the size of the targeted working model is small compared to $N$, say $\delta'' = 0$ (meaning $|\bbM^\star|$ does not grow with $N$), \eqref{eqn:asp-condition-1} always holds. More generally, \eqref{eqn:asp-condition-1} is easier to satisfy with a larger between-group distance (larger $\delta_2$) and smaller true working model size (smaller $\delta''$). The second condition \eqref{eqn:asp-condition-2} reflects the trade-off among the total number of interested parameters (given by $L$, which is also $|\cT'|$), the size of the neighborhood of $\gamma_{(1)}$ (given by $|\cL_1|$), and the size of the true working model (captured by $\delta''$). The smaller these quantities are, the easier inference will be. Moreover, \eqref{eqn:asp-condition-2} requires that the number of parameters of interest and the size of the local neighborhood  $\cL_1$ should be asymptotically vanishing compared to the total sample size $N$ for the purpose of inference.

Theorem \ref{thm:infer-order} also suggests the benefits of factor selection compared to procedures where no selection is involved following similar reasoning provided in Remark \ref{rmk:clt-conditions}. More precisely, without selection, 
one requires $Q$ to be small compared to $N$ or $\{ f_l\}_{l\in[L]}$ are dense, which is violated in large $Q$ setups and many practical scenarios such as \eqref{eqn:best-Yz}. 

\commenting{
Comparing Theorem \ref{thm:infer-order} with Corollary \ref{cor:infer-order}, we can conclude some benefits of selection:
\begin{itemize}
    \item Sufficient conditions for CLT. With selection, when the size of the targeted working model is small (say $\delta_4 = 0$) and selection is consistent, one only needs
\begin{align*}
|\cT'||\cT_1|\lt({|\bbM^\star|}/{N}\rt)^{1/2} \to 0. \see{because ${N^{-(1+2\delta_2)}}$ always converge to $0$ as $N\to\infty$}
\end{align*}
The condition relies on the scaling of $N$ instead of a particular set of $N(z)$'s. In other words, we can incorporate information from other treatment arms to facilitate inference. On the contrary, without selection, one requires $Q$ to be small compared to $N$, which might not be true if each arm contains a limited number of units.

\item Length of confidence intervals. The length of confidence intervals with selection has an upper bound
\begin{align*}
    v^2_{(1),\mathrm{lim}}  
     = \sum_{z\in\cT} [ f_{(1)}(z)]^2 N(z)^{-1}S(z,z)
     \le  \frac{|\bbM^\star||\cT_1|}{Q} \max_{z\in\cT}\lt\{\frac{S(z,z)}{N({z})}\rt\}.
\end{align*}
Meanwhile, the length of confidence intervals with selection is given by 
$$
\frac{1}{|\cT_1|}\sum_{z\in\cT_1}N(z)^{-1} {S}(z,z).
$$
When $Q$ is large and the working model is sparse, selection leads to a much shorter confidence interval and improves statistical power for inferring the maximum.

\end{itemize}
}

As a final comment, Theorem \ref{thm:infer-order} relies on the selection consistency property of Theorem \ref{thm:marginal-t}, which are ensured by Conditions \ref{cond:uniform-design}-\ref{cond:heredity}. Without selection consistency, there might be additional sources of bias due to the uncertainty induced by the selection step and possible under-selection results. Nevertheless, one can consider applying the over-selection strategy (Strategy \ref{str:select-by-heredity} in Section \ref{sec:imperfect-selection}) to facilitate inference on the best factorial effects.

\section{Proofs} \label{sec:technical-pfs}

In this section, we present the technical proofs for the results across the whole paper. Section \ref{sec:preliminary-pf} presents some preliminary probabilistic results that are useful in randomized experiments which are mainly attributed to \cite{shi2022berry}.  The main proof starts from Section \ref{sec:pf-ms-consistency}.

\subsection{Preliminaries: some probabilistic results in randomized experiments}\label{sec:preliminary-pf}

In this subsection, we present some preliminary probability results that are crucial for our theoretical discussion. Consider an estimator of the form
\begin{align*}
    \hgamma = Q^{-1}\sum_{z\in\cT} w(z)\hat{Y}(z),
\end{align*}
with the variance estimator
\begin{align*}
    \hat{v}^2 = Q^{-2}\sum_{z\in\cT} w(z)^2 \hat{S}(z,z).
\end{align*}

\citet{li2017general} showed that 
\begin{align}\label{eqn:exp-var-hY}
    \bbE\{\hY\} = \overline{Y},~ V_\hY = \Var{\hY} = D_\hY - N^{-1}S.
\end{align}
Then \eqref{eqn:exp-var-hY} further leads to the following facts:
\begin{align}
    \bbE\{\hgamma\} &= \sum_{z\in\cT}  f(z)\overline{Y}(z) = \gamma, \label{eqn:mean-hgamma}\\
    \Var{\hgamma} &= \sum_{z\in\cT} f(z)^2N({z})^{-1}S(z,z) - N^{-1} f^\top S  f, \label{eqn:var-hgamma}\\
    \bbE\{\hat{v}^2\} & =  \sum_{z\in\cT}  f(z)^2N({z})^{-1}S(z,z). \label{eqn:mean-varR}
\end{align}

We have the following variance estimation results and Berry--Esseen bounds:

\begin{lemma}[Variance concentration and Berry--Esseen bounds]\label{lem:BE-finite-pop}
Define $\gamma = \bbE\{\hgamma\}$, $v^2 = \operatorname{Var}(\hgamma)$ and $v^2_{\lim} = \bbE\{\hat{v}^2\}$. 
Suppose the following conditions hold: 
\begin{itemize}
    \item  Nondegenerate variance. There exists a $\sigma_w > 0$, such that
    \begin{align}\label{eqn:nondegenerate-var}
        Q^{-2}\sum_{z=1}^Q w(z)^2 N_{z}^{-1} S(z,z) \le \sigma_w^2 v^2.
    \end{align}
    
    \item Bounded fourth moments. There exists a $ \Delta > 0$ such that
    \begin{align}\label{eqn:bounded-moments}
    \max_{z\in[Q]}\frac{1}{N}\sum_{i=1}^N \{Y_i(z) - \overline{Y}(z)\}^4 \le \Delta^4.
    \end{align}
\end{itemize}

Then we have the following conclusions:
\begin{enumerate}
    \item The variance estimator is conservative for the true variance:
    $v^2_{\lim} \ge v^2$. Moreover, the following tail bound holds:
    \begin{align}\label{eqn:tail-vhatR}
        \Prob{N |\hat{v}^2-v^2_{\lim} | > t} 
        \le \frac{C\overline{c}^3\underline{c}^{-4} \|w\|_\infty^2 \Delta^4 }{QN_0}\cdot \frac{1}{t^2}.
    \end{align}

    \item We have a Berry--Esseen bound with the true variance:
    \begin{align}\label{eqn:uniform-design-be}
       \sup_{t\in\bbR}\lt|\Prob{\frac{\hgamma-\gamma}{v} \le t} - \Phi(t)\rt|  \le  2C\sigma_w   \frac{\underline{c}^{-1}  \|w\|_{\infty}   \max_{i\in[N],z\in[Q]}|Y_i(z)-\overline{Y}(z)|}{\|w\|_2\sqrt{\overline{c}^{-1} \min_{z\in[Q]} S(z,z)}\cdot \sqrt{ N_0}} .
    \end{align}
    
    \item We have a Berry--Esseen bound with the estimated variance: for any $\epsilon_N \in (0, 1/2]$,
    \begin{align*}
        \sup_{t\in\bbR}\lt|{\mathrm{P}}\lt\{\frac{\hgamma - \gamma}{\hat{v}} \le t\rt\} - \Phi\lt(\frac{v_{\lim}}{v}t\rt)\rt| &\le
        \epsilon_N +  \frac{C\overline{c}^3\underline{c}^{-4} \|w\|_\infty^2 \Delta^4 }{QN_0}\cdot \frac{1}{(Nv^2\epsilon_N)^2} \\
        &+ 2C\sigma_w   \frac{\underline{c}^{-1}  \|w\|_{\infty}   \max_{i\in[N],z\in[Q]}|Y_i(z)-\overline{Y}(z)|}{\|w\|_2\sqrt{\overline{c}^{-1} \min_{z\in[Q]} S(z,z)}\cdot \sqrt{N_0}}.
    \end{align*}
\end{enumerate}

\end{lemma}

\begin{proof}[Proof of Lemma \ref{lem:BE-finite-pop}]
\begin{enumerate}
    \item See Lemma S13 of \cite{shi2022berry}.

    \item See Theorem 1 of \cite{shi2022berry}.
    
    \item 
    First we show a useful result: for $|a|\le 1/2$ and any $b\in\bbR$, 
    \begin{align}\label{form:Phi-bD}
        \sup_{t\in\bbR}|\Phi\{(1+a)t + b\} - \Phi\{t\}| \le |a| + |b|.
    \end{align}
    \eqref{form:Phi-bD} is particularly useful for small choices of $a$ and $b$. Intuitively, it evaluates the change of $\Phi$ under a small affine perturbation of $t$. 
    
    The proof of \eqref{form:Phi-bD} is based on a simple step of the mean value theorem: for any $t\in\bbR$, there exists a value $\xi \in [t, (1+a)t + b]$ such that
    \begin{align*}
        &|\Phi\{(1+a)t + b\} - \Phi\{t\}| \\
        =& |\phi(\xi)\cdot (at + b)|\\
        =& |\phi(\xi)\cdot at| + |\phi(\xi)\cdot b|\\
        =& |a| \cdot |\phi(\xi)\cdot t|\cdot \ind{|t|\le 1} + |a| \cdot |\phi(\xi)\cdot t|\cdot \ind{|t| > 1} + |\phi(\xi)\cdot b|\\
        \le& \frac{1}{\sqrt{2\pi}}|a|\cdot \ind{|t|\le 1} + \frac{1}{\sqrt{2\pi}} |a| |t|\cdot\exp(-t^2/8)\cdot\ind{|t| > 1} +  {\frac{1}{\sqrt{2\pi}}}|b|\\
        \le& |a| + |b|.
    \end{align*}
    We consider $t\ge 0$ because $t<0$ can be handled similarly. For any $\epsilon_N > 0$, We have
    \begin{align*}
        \Prob{\frac{\hgamma - \gamma}{\hat{v}} \le t} &= \Prob{\frac{\hgamma - \gamma}{{v}} \le \frac{\hat{v}}{{v}}t}\\
        &= \Prob{\frac{\hgamma - \gamma}{{v}} \le \frac{\hat{v}}{{v}}t, \lt|\frac{\hat{v} - v_{\lim}}{v}\rt| \le \epsilon_N} + \Prob{\frac{\hgamma - \gamma}{{v}} \le \frac{\hat{v}}{{v}}t, \lt|\frac{\hat{v} - v_{\lim}}{v}\rt| > \epsilon_N}.
    \end{align*}
    Then we can show that
    \begin{align}
        \Prob{\frac{\hgamma - \gamma}{\hat{v}} \le t} &\le 
        \Prob{\frac{\hgamma - \gamma}{{v}} \le \frac{\hat{v}}{{v}}t, \lt|\frac{\hat{v} - v_{\lim}}{v}\rt| \le \epsilon_N} + \Prob{ \lt|\frac{\hat{v} - v_{\lim}}{v}\rt| > \epsilon_N}\\
        &\le \Prob{\frac{\hgamma - \gamma}{{v}} \le \lt(\frac{{v}_{\lim}}{v} + \epsilon_N\rt)t} + \Prob{ \lt|\frac{\hat{v} - v_{\lim}}{v}\rt| > \epsilon_N}.\label{eqn:decompse-tail}
    \end{align}
    
    For the first term in \eqref{eqn:decompse-tail}, we have
    \begin{align*}
        &\sup_{t\ge 0}\lt|\Prob{\frac{\hgamma - \gamma}{{v}}   \le \lt(\frac{{v}_{\lim}}{v} + \epsilon_N\rt)t} - \Phi\lt\{\lt(\frac{{v}_{\lim}}{v} + \epsilon_N\rt)t \rt\}\rt| \\
        &\le 2C\sigma_w   \frac{\underline{c}^{-1}  \|w\|_{\infty}   \max_{i\in[N],z\in[Q]}|Y_i(z)-\overline{Y}(z)|}{\|w\|_2\sqrt{\overline{c}^{-1} \min_{z\in[Q]} S(z,z)}\cdot \sqrt{ N_0}}.
    \end{align*}
    For the second term in \eqref{eqn:decompse-tail}, using the variance estimation results in Part 1, we have
    \begin{align*}
        \Prob{ \lt|\frac{\hat{v} - v_{\lim}}{v}\rt| \ge \epsilon_N} &\le  \Prob{ \lt|\frac{\hat{v} - v_{\lim}}{v}\rt|\cdot\lt|\frac{\hat{v} + v_{\lim}}{v}\rt| \ge \epsilon_N} \see{because $v_{\lim}$ is conservative}\\
        & =  \Prob{ \lt|\frac{N\hat{v}^2 - Nv^2_{\lim}}{Nv^2}\rt| \ge \epsilon_N}   \\
        & \le \frac{C\overline{c}^3\underline{c}^{-4} \|w\|_\infty^2 \Delta^4 }{QN_0}\cdot \frac{1}{(Nv^2\epsilon_N)^2}.
    \end{align*}
    Besides, by \eqref{form:Phi-bD}, when $\epsilon_N\le 1/2$, we also have
    \begin{align*}
        \sup_{t\in\bbR}\lt|\Phi\lt\{\lt(\frac{{v}_{\lim}}{v} + \epsilon_N \rt)t \rt\} - \Phi\lt(\frac{v_{\lim}}{v}t\rt)\rt| \le \frac{v\epsilon_N}{v_{\lim}} \le \epsilon_N. 
    \end{align*}
    Using all the parts above, we can show that for any $t\ge 0$,
    \begin{align}\label{eqn:upper-tail}
        \Prob{\frac{\hgamma - \gamma}{\hat{v}} \le t} &\le \Phi\lt(\frac{v_{\lim}}{v}t\rt) + \epsilon_N +  \frac{C\overline{c}^3\underline{c}^{-4} \|w\|_\infty^2 \Delta^4 }{QN_0}\cdot \frac{1}{(Nv^2\epsilon_N)^2} \\
        &+ 2C\sigma_w   \frac{\underline{c}^{-1}  \|w\|_{\infty}   \max_{i\in[N],z\in[Q]}|Y_i(z)-\overline{Y}(z)|}{\|w\|_2\sqrt{\overline{c}^{-1} \min_{z\in[Q]} S(z,z)}\cdot \sqrt{N_0}}.
    \end{align}
    
    On the other hand, we can show that
    \begin{align}
        \Prob{\frac{\hgamma - \gamma}{\hat{v}} \le t} &\ge 
        \Prob{\frac{\hgamma - \gamma}{{v}} \le \frac{\hat{v}}{{v}}t, \lt|\frac{\hat{v} - v_{\lim}}{v}\rt| \le \epsilon_N} \notag\\
        &\ge \Prob{\frac{\hgamma - \gamma}{{v}} \le \lt(\frac{{v}_{\lim}}{v} - \epsilon_N\rt)t} - \Prob{ \lt|\frac{\hat{v} - v_{\lim}}{v}\rt| \ge \epsilon_N}.\label{eqn:upper-bd}
    \end{align}
    By \eqref{form:Phi-bD}, when $\epsilon_N\le 1/2$, we also have
    \begin{align*}
        \sup_{t\in\bbR}\lt|\Phi\lt\{\lt(\frac{{v}_{\lim}}{v} - \epsilon_N\rt)t \rt\} - \Phi\lt(\frac{v_{\lim}}{v}t\rt)\rt| \le \epsilon_N. 
    \end{align*}
    So we can derive a lower bound analogous to \eqref{eqn:upper-tail}. Note that the results can be analogously generalized to $t\le 0$. Using the upper bound \eqref{eqn:upper-tail} and its lower bound counterpart, we can show that for any $t\ge 0$, $\epsilon_N \le 1/2$, 
    \begin{align*}
        \sup_{t\in\bbR}\lt|\Prob{\frac{\hgamma - \gamma}{\hat{v}} \le t} - \Phi\lt(\frac{v_{\lim}}{v}t\rt)\rt| &\le
        \epsilon_N +  \frac{C\overline{c}^3\underline{c}^{-4} \|w\|_\infty^2 \Delta^4 }{QN_0}\cdot \frac{1}{(Nv^2\epsilon_N)^2} \\
        &+ 2C\sigma_w   \frac{\underline{c}^{-1}  \|w\|_{\infty}   \max_{i\in[N],z\in[Q]}|Y_i(z)-\overline{Y}(z)|}{\|w\|_2\sqrt{\overline{c}^{-1} \min_{z\in[Q]} S(z,z)}\cdot \sqrt{ N_0}}.
    \end{align*}
\end{enumerate}

\end{proof}

The following corollary shows a Berry--Esseen bound for the studentized statistic in the special case where $w = (w(z))_{z\in[Q]}$ is a contrast vector for factorial effects. That is, $ w = g_\cK$ for some $\cK \in \bbK $. 

\begin{corollary}\label{cor:factorial-student-be}
Assume Condition \eqref{eqn:nondegenerate-var} and \eqref{eqn:bounded-moments} hold. Let $ w = g_\cK$ for some $\cK \in \bbK $. Then we have a Berry--Esseen bound with the estimated variance:
\begin{align*}
    \sup_{t\in\bbR}\lt|\Prob{\frac{\htau_\cK - \tau_\cK}{\hat{v}} \le t} - \Phi\lt(\frac{v_{\lim}}{v}t\rt)\rt| &\le
          2\lt(\frac{C\sigma_w^4\overline{c}^5\underline{c}^{-6}  \Delta^4 }{\{\min_{z\in\cT}S(z,z)\}^2}\rt)^{1/3}\cdot\frac{1}{(QN_0)^{1/3}} \\
        &+ 2C\sigma_w   \frac{\underline{c}^{-1}     \max_{i\in[N],z\in[Q]}|Y_i(z)-\overline{Y}(z)|}{ \sqrt{\overline{c}^{-1} \min_{z\in[Q]} S(z,z)}}\cdot\frac{1}{(QN_0)^{1/2}}.
\end{align*}
\end{corollary}

\begin{proof}[Proof of Corollary \ref{cor:factorial-student-be}]
\textbf{Lower bound for $Nv^2$.} Note that
$\|w\|_2^2 = Q$ and $\|w\|_\infty = 1$. 
Using Condition \eqref{eqn:nondegenerate-var}, we have
\begin{align*}
    Nv^2 &\ge N\sigma_w^{-2} Q^{-2}\sum_{z=1}^Q w(z)^2 N_{z}^{-1} S(z,z) \\
    &\ge (\underline{c}QN_0)\cdot\sigma_w^{-2} \overline{c}^{-1}Q^{-1}N_{0}^{-1} \min_{z\in\cT} S(z,z) \cdot (Q^{-1}\|w\|_2^2)\\
    &= \sigma_w^{-2}\underline{c}\overline{c}^{-1}\min_{z\in\cT}S(z,z).
\end{align*}

Therefore, the Berry--Esseen bound becomes
\begin{align*}
    \sup_{t\in\bbR}\lt|\Prob{\frac{\htau_\cK - \tau_\cK}{\hat{v}} \le t} - \Phi\lt(\frac{v_{\lim}}{v}t\rt)\rt| &\le
        \epsilon_N +  \frac{C\sigma_w^4\overline{c}^5\underline{c}^{-6}  \Delta^4 }{(QN_0)\{\min_{z\in\cT}S(z,z)\}^2}\cdot \frac{1}{\epsilon_N^2} \\
        &+ 2C\sigma_w   \frac{\underline{c}^{-1}     \max_{i\in[N],z\in[Q]}|Y_i(z)-\overline{Y}(z)|}{\sqrt{\overline{c}^{-1} \min_{z\in[Q]} S(z,z)}\cdot \sqrt{QN_0}}.
\end{align*}

\textbf{Optimize the summation of the first and second term.}
By taking derivative over $\epsilon_N $ on the upper bound and solving for the zero point, we know that when 
\begin{align*}
    \epsilon_N = \lt(\frac{2C\sigma_w^4\overline{c}^5\underline{c}^{-6}  \Delta^4 }{(QN_0)\{\min_{z\in\cT}S(z,z)\}^2}\rt)^{1/3},
\end{align*}
the upper bound is minimized and 
\begin{align*}
    \sup_{t\in\bbR}\lt|\Prob{\frac{\htau_\cK - \tau_\cK}{\hat{v}} \le t} - \Phi\lt(\frac{v_{\lim}}{v}t\rt)\rt| &\le
          2\lt(\frac{C\sigma_w^4\overline{c}^5\underline{c}^{-6}  \Delta^4 }{\{\min_{z\in\cT}S(z,z)\}^2}\rt)^{1/3}\cdot\frac{1}{(QN_0)^{1/3}} \\
        &+ 2C\sigma_w   \frac{\underline{c}^{-1}     \max_{i\in[N],z\in[Q]}|Y_i(z)-\overline{Y}(z)|}{ \sqrt{\overline{c}^{-1} \min_{z\in[Q]} S(z,z)}}\cdot\frac{1}{(QN_0)^{1/2}}.
\end{align*}
\end{proof}

Additionally, we have a Berry--Esseen bounds after selection on the effects:

\begin{lemma}[Berry Esseen bound with  selection]\label{lem:tail-perfect-ms} 
Assume there exists $\sigma_w>0$ such that
\begin{align}\label{eqn:nondegenerate-var-tw}
    \sum_{z=1}^Q \lt({ f}[\bbM](z)\rt)^2 N_{z}^{-1} S(z,z) \le \sigma_w^2 v^2(\bbM).
\end{align}
Then  
\begin{align}\label{eqn:tail-perfect-ms}
    &\sup_{t\in\bbR}\lt|\Prob{\frac{\hgamma[\hat{\bbM}]-\gamma[\bbM]}{v(\bbM)} \le t} - \Phi(t)\rt| \notag\\
    &\le 2\Prob{\hat{\bbM}\neq \bbM} +  2C\sigma_w   \frac{\underline{c}^{-1} \max_{i\in[N],z\in[Q]}|Y_i(z)-\overline{Y}(z)|}{\sqrt{\overline{c}^{-1} \min_{z\in[Q]} S(z,z)}\cdot \sqrt{N_0}}\cdot  \frac{\|{ f}[\bbM]\|_\infty}{\|{ f}[\bbM]\|_2} .
\end{align}
\end{lemma}

\begin{proof}[Proof of Lemma \ref{lem:tail-perfect-ms}]
With the selected working model we have
\begin{align*}
    &\sup_{t\in\bbR}\lt|\Prob{\frac{\hgamma[\hat{\bbM}]-\gamma[\bbM]}{v(\bbM)} \le t} - \Phi(t)\rt| \\
    = &\sup_{t\in\bbR}\lt|\Prob{\frac{\hgamma[\hat{\bbM}]-\gamma[\bbM]}{v(\bbM)} \le t, \hat{\bbM} = \bbM} - \Phi(t) + \Prob{\frac{\hgamma[\hat{\bbM}]-\gamma[\bbM]}{v(\bbM)} \le t, \hat{\bbM} \neq \bbM} \rt|\\
    \le &\sup_{t\in\bbR}\lt|\Prob{\frac{\hgamma[\hat{\bbM}]-\gamma[\bbM]}{v(\bbM)} \le t, \hat{\bbM} = \bbM} - \Phi(t)\rt| +  \Prob{\frac{\hgamma[\hat{\bbM}]-\gamma[\bbM]}{v(\bbM)} \le t, \hat{\bbM} \neq \bbM}  \\
    = &\sup_{t\in\bbR}\lt|\Prob{\frac{\hgamma[{\bbM}]-\gamma[\bbM]}{v(\bbM)} \le t, \hat{\bbM} = \bbM} - \Phi(t)\rt| +  \Prob{\frac{\hgamma[\hat{\bbM}]-\gamma[\bbM]}{v(\bbM)} \le t, \hat{\bbM} \neq \bbM}  \\
    \le & \sup_{t\in\bbR}\lt|\Prob{\frac{\hgamma[{\bbM}]-\gamma[\bbM]}{v(\bbM)} \le t} - \Phi(t)\rt| +  2\Prob{ \hat{\bbM} \neq \bbM}.
\end{align*}

Now we have 
\begin{align*}
    \hgamma(\bbM) &=  f^\top G(\cdot, \bbM) \htau(\bbM) \\
    & =   f^\top G(\cdot, \bbM) G(\cdot, \bbM)^\top \hY \\
    & = { f}[\bbM]^\top \hY.
\end{align*}

By Theorem 1 of \cite{shi2022berry},  we have a Berry--Esseen bound with the true variance:
\begin{align*}
    \sup_{t\in\bbR}\lt|\Prob{\frac{\hgamma( {\bbM})-\gamma[\bbM]}{v(\bbM)} \le t} - \Phi(t)\rt| 
    \le  2C\sigma_w   \frac{\|{ f}[\bbM]\|_\infty\underline{c}^{-1} \max_{i\in[N],z\in[Q]}|Y_i(z)-\overline{Y}(z)|}{\|{ f}[\bbM] \|_2\sqrt{\overline{c}^{-1} \min_{z\in[Q]} S(z,z)}\cdot \sqrt{ N_0}}.  
\end{align*}
    
\end{proof}

A crucial quantity that appeared in Lemma \ref{lem:tail-perfect-ms} is the ratio of norms:
\begin{align}\label{eqn:ratio-bsf}
    \frac{\|{ f}[\bbM]\|_\infty}{\|{ f}[\bbM]\|_2}. 
\end{align}
The following Lemma \ref{lem:ratio-bsf} provides an explicit bound on  \eqref{eqn:ratio-bsf} which reveals how the ratio is controlled with respect to the size of the working model.
\begin{lemma}\label{lem:ratio-bsf}
For  $ f[\bbM] \neq 0$, we have
\begin{align}\label{eqn:infty-over-2}
    \frac{\| f[\bbM]\|_\infty}{\| f[\bbM]\|_2}  \le \lt(\frac{|\bbM|}{Q}\rt)^{1/2}.
\end{align}
\end{lemma}
\begin{proof}[Proof of Lemma \ref{lem:ratio-bsf}]
Because the LHS of \eqref{eqn:infty-over-2} is a ratio, based on the definition of $ f^\star$ \eqref{eqn:fM} we can assume $\| f\|_2 = 1$ without loss of generality. Due to the orthogonality of $G$, we can use the columns of $G$ as bases and express $ f$ as 
    \begin{align*}
         f = \frac{1}{\sqrt{Q}}G(\cdot,\bbM)b_1 + \frac{1}{\sqrt{Q}}G(\cdot,\bbM^c) b_2,
    \end{align*}
    where $b_1 \in \bbR^{|\bbM|}$ and $b_2\in\bbR^{|\bbM^{c}|}$ and $\|(b_1^\top,b_2^\top)^\top\|_2 = 1 $.
    Then  
    \begin{align*}
         f[\bbM] = Q^{-1}G(\cdot,\bbM) G(\cdot,\bbM)^\top  f = \frac{1}{\sqrt{Q}}G(\cdot,\bbM)b_1.  
    \end{align*}
    Hence
    \begin{align*}
        \| f[\bbM]\|_\infty \le \frac{1}{\sqrt{Q}}\|b_1\|_1,\quad  \| f[\bbM]\|_2 = \|b_1\|_2, \quad \frac{\| f[\bbM]\|_\infty}{\| f[\bbM]\|_2} \le \frac{1}{\sqrt{Q}}\cdot \frac{\|b_1\|_1}{\|b_1\|_2} \le \lt(\frac{|\bbM|}{Q}\rt)^{1/2}.
    \end{align*}
\end{proof}

\subsection{Proof of Theorem \ref{thm:ms-consistency}}\label{sec:pf-ms-consistency}

\begin{proof}[Proof of Theorem \ref{thm:ms-consistency}]
We introduce several key events that will play a crucial role in the proof: for $D_0\in [D]$, define
\begin{gather}
    \textit{Under-selection: }\cE_\textsc{u}(D_0) = \{\hat{\bbM}_d\subset\bbM^\star_d, d\in[D_0]\},\\
    \textit{Strict under-selection: }\cE_\textsc{su}(D_0) = \{\hat{\bbM}_d\subset\bbM^\star_d, d\in[D_0]; \text{ there exists } d\in[D_0], \hat{\bbM}_d\subsetneq\bbM^\star_d\}.
\end{gather}
\noindent\textbf{High-level idea of the proof.} To prove selection consistency,  we will prove two facts: 
\begin{align}
    \Prob{\cE_\textsc{u}(D) \text{ holds}} \to 1,\quad \Prob{\cE_\textsc{su}(D) \text{ holds}} \to 0.
\end{align}
Combining these two results, we can conclude asymptotic selection consistency. 

We start from the strict under-selection probability.  

\noindent\textbf{Step 1: Prove that asymptotically, there is no strict under-selection.}

By definition,  
\begin{align*}
    \Prob{\cE_\textsc{su}(1) \text{ holds}} = \Prob{\tilde{\bbM}_{1}\subsetneq \bbM^{\star }_{1}} \le \Prob{\tilde{\bbM}_1^c\cap{\bbM_1^\star} \neq \varnothing}.
\end{align*}
We now derive a recursive bound for $\Prob{\cE_\textsc{su}(D_0+1) \text{ holds}}$ where $1\le D_0 \le D-1$. We have decomposition
\begin{align*}
    \cE_\textsc{su}(D_0+1) &= \left\{\hat{\bbM}_d\subset\bbM^\star_d, d\le D_0+1\right\} - \left\{\hat{\bbM}_d=\bbM^\star_d, d\le D_0+1\right\} \\
    &=  \cE_\textsc{su,1}(D_0+1) \cup \cE_\textsc{su,2}(D_0+1),
\end{align*}
where
\begin{gather}
    \cE_\textsc{su,1}(D_0+1) = \left\{\hat{\bbM}_d\subset\bbM^\star_d, d\le D_0+1\right\} - \left\{\hat{\bbM}_d=\bbM^\star_d, d\le D_0; \hat{\bbM}_{D_0+1}\subset\bbM^\star_{D_0+1}\right\},\\
    \cE_\textsc{su,2}(D_0+1) =  \lt\{\hat{\bbM}_d=\bbM^\star_d, d\le D_0; \hat{\bbM}_{D_0+1}\subset\bbM^\star_{D_0+1}\rt\}  -  \lt\{\hat{\bbM}_d=\bbM^\star_d, d\le D_0+1\rt\}.  
\end{gather}

For $\cE_\textsc{su,1}(D_0+1)$, we have
    \begin{align} 
        \Prob{\cE_\textsc{su,1}(D_0+1) \text{ holds}}&= \Prob{\left\{\hat{\bbM}_d\subset\bbM^\star_d, d\le D_0+1\right\} - \left\{\hat{\bbM}_d=\bbM^\star_d, d\le D_0; \hat{\bbM}_{D_0+1}\subset\bbM^\star_{D_0+1}\right\}}\notag\\
        &\le \Prob{\forall d\in[D_0+1], \hat{\bbM}_{d}\subset\bbM^\star_{d}; \exists d\in[D_0], \hat{\bbM}_d\subsetneq\bbM^\star_d}\notag\\
        &\le \Prob{\forall d\in[D_0], \hat{\bbM}_{d}\subset\bbM^\star_{d};\exists d\in[D_0], \hat{\bbM}_d\subsetneq\bbM^\star_d}\\
        &= \Prob{\cE_\textsc{su}(D_0) \text{ holds}}.\label{eqn:step-1}
    \end{align}

    For $\cE_\textsc{su,2}(D_0+1)$, we notice that $\hat{\bbM}_{D_0+1}$ is generated based on $\hat{\bbM}_{D_0}$ and the set of estimates over the preselected effect set  $\hat{\bbM}_{D_0+1,+}$. Under Assumption \ref{asp:H-heredity}, on the event $\hat{\bbM}_d = \bbM_d^\star$ we have
    \begin{align*}
        \hat{\bbM}_{d+1} = \tilde{\bbM}_{d+1}.
    \end{align*}
    Hence we can compute
    \begin{align}
        \Prob{\cE_\textsc{su,2}(D_0+1) \text{ holds}}
        = & \Prob{\hat{\bbM}_d=\bbM^\star_d, d\le D_0; \hat{\bbM}_{D_0+1}\subsetneq \bbM^{\star }_{D_0+1}} \notag\\
        = & \Prob{\hat{\bbM}_d=\bbM^\star_d, d\le D_0; \tilde{\bbM}_{D_0+1}\subsetneq \bbM^{\star }_{D_0+1}} \notag\\
        \le&  \Prob{\tilde{\bbM}^c_{D_0+1}\cap\bbM^{\star }_{D_0+1}\neq\varnothing}. \label{eqn:step-2}
    \end{align}
   Now \eqref{eqn:step-1} and \eqref{eqn:step-2} together suggest that
   \begin{align}
    &\Prob{\cE_\textsc{su}(D_0 + 1) \text{ holds}}\notag\\
        \le &\Prob{\cE_\textsc{su}(D_0) \text{ holds}} + \Prob{\tilde{\bbM}^c_{D_0+1}\cap\bbM^{\star }_{D_0+1}\neq\varnothing}\notag\\
        \le & \cdots \le \sum_{d=1}^{D_0+1} \Prob{\tilde{\bbM}^c_{D_0+1}\cap\bbM^{\star }_{D_0+1}\neq\varnothing }. \label{eqn:su-bound}
   \end{align}
    Taking $D_0 = D-1$ in \eqref{eqn:su-bound} and applying Assumption \ref{asp:valid-consistent}, we conclude
    \begin{align*}
        \Prob{\cE_\textsc{su}(D)\text{ holds}} \to 0.
    \end{align*}
    
    \noindent\textbf{Step 2: Prove the first part of Theorem \ref{thm:ms-consistency} and give a probability bound for under-selection.}
    We compute the probability of under-selection:
    \begin{align*}
        &\Prob{\cE_\textsc{u}(D) \text{ fails} }\\
        = & \Prob{\cE_\textsc{u}(1) \text{ fails}} + \sum_{D_0=2}^D \Prob{\cE_\textsc{u}(D_0-1) \text{ holds}; \cE_\textsc{u}(D_0) \text{ fails}} \\
        = & \underbrace{\vphantom{\sum_{D_0=2}^D}\Prob{\cE_\textsc{u}(1) \text{ fails}}}_{\ostar_1} + \underbrace{\sum_{D_0=2}^D \Prob{\cE_\textsc{u}(D_0-1) \text{ holds}; \cE_\textsc{u}(D_0) \text{ fails}}}_{\ostar_2} \\
        + & \underbrace{\sum_{D_0=2}^D \Prob{\cE_\textsc{su}(D_0-1) \text{ holds}; \cE_\textsc{u}(D_0) \text{ fails}}}_{\ostar_3}.
    \end{align*}
    For $\ostar_1$, by definition of $\cE_\textsc{u}(1) $ we have
    \begin{align}\label{eqn:ostar-1}
        \ostar_1 = \Prob{\cE_\textsc{u}(1) \text{ fails}} = \Prob{\hat{\bbM}_1\cap{\bbM_1^\star}^c\neq\varnothing} = \Prob{\tilde{\bbM}_1\cap{\bbM_1^\star}^c\neq\varnothing}.
    \end{align}
    For $\ostar_2$, we have
    \begin{align}\label{eqn:ostar-2}
        \ostar_2 
    &\le  \sum_{D_0=2}^D \Prob{ \hat{\bbM}_d=\bbM^\star_d,d\in[D_0-1]; \tilde{\bbM}_{D_0}\cap\bbM_{D_0}^{\star c}\neq\varnothing} \le \sum_{D_0=2}^D\Prob{\tilde{\bbM}_{D_0}\cap\bbM_{D_0}^{\star c}\neq\varnothing}.
    \end{align}
    The inequalities in \eqref{eqn:ostar-2} are because on the given event, $\hat{\bbM}_{D_0,+} = \mathrm{H}(\hat{\bbM}_{D_0-1}) = \mathrm{H}({\bbM}^\star_{D_0-1}) = {\bbM}^\star_{D_0,+}$ and $\hat{\bbM}_{D_0} = \hat{\mathrm{S}}(\hat{\bbM}_{D_0,+}) = \tilde{\bbM}_{D_0}$.
   From \eqref{eqn:ostar-1} and \eqref{eqn:ostar-2}, by Assumption \ref{asp:valid-consistent},
    \begin{align}\label{eqn:ostar-1-2}
        \underset{{N\to\infty}}{\lim\sup}~(\ostar_1 + \ostar_2) = \sum_{D_0=1}^D\Prob{\tilde{\bbM}_{D_0}\cap\bbM_{D_0}^{\star c}\neq\varnothing} \le \sum_{D_0=1}^D\alpha_{D_0} = \alpha.  
    \end{align}
   For $\ostar_3$, we have
    \begin{align}
       \ostar_3
       \le &
       \sum_{D_0=2}^D \Prob{\cE_\textsc{su}(D_0-1) \text{ holds}} \\
       \le & \sum_{D_0=2}^D \sum_{d=1}^{D_0-1}\Prob{\tilde{\bbM}^c_{d}\cap\bbM^{\star }_{d}\neq\varnothing} \see{using \eqref{eqn:su-bound}} \\
       = & \sum_{d=1}^{D-1}(D-d)\Prob{\tilde{\bbM}^c_{d}\cap\bbM^{\star }_{d}\neq\varnothing} \to 0. \see{using Assumption \ref{asp:valid-consistent}}\label{eqn:ostar-3}
    \end{align}
    
Therefore, by \eqref{eqn:ostar-1-2} and \eqref{eqn:ostar-3}, the probability of failure of under-selection gets controlled under $\alpha$ asymptotically.

As a side product, we have obtained the finite sample bounds:
\begin{align*}
    \Prob{\cE_\textsc{u}(D) \text{ fails} } \le \sum_{D_0=1}^D\Prob{\tilde{\bbM}_{D_0}\cap\bbM_{D_0}^{\star c}\neq\varnothing} + \sum_{d=1}^{D-1}(D-d)\Prob{\tilde{\bbM}^c_{d}\cap\bbM^{\star }_{d}\neq\varnothing}. 
\end{align*}

\noindent\textbf{Step 3. Proof of the second part of Theorem \ref{thm:ms-consistency} and conclude selection consistency.} Under $\alpha=\alpha(N)\to0$, the first part of the result implies that with probability tending to 1, we have under-selection:
\begin{align*}
     \Prob{\cE_\textsc{u}(D) \text{ holds}} \to 1.
\end{align*}
By \eqref{eqn:su-bound} and Assumption \ref{asp:valid-consistent}, strict under-selection will not happen asymptotically:
\begin{align*}
    \Prob{\cE_\textsc{su}(D) \text{ holds}} \to 0.
\end{align*}
Therefore, we conclude the consistency of the selection procedure.

\end{proof}

\subsection{Proof of Theorem \ref{thm:marginal-t}}\label{sec:pf-marginal-t}

We state and prove a more general version of Theorem \ref{thm:marginal-t}:
\begin{theorem}[Bonferroni corrected marginal t test] \label{thm:marginal-t-general}
Let $\tilde{\bbM}_{d} = \hat{\texttt{\em S}}(\bbM^{\star}_{d,+})$ where $\bbM^{\star}_{d,+} = \texttt{\em P}(\bbM^{\star}_{d-1})$. Assume Conditions \ref{cond:uniform-design},  \ref{cond:order}, \ref{cond:regularity} and \ref{cond:heredity}. Then we have
the following results for the  selection procedure based on Bonferroni corrected marginal $t$-tests:
\begin{enumerate}[label= (\roman*)]
    \item (Validity) $\limsup_{N\to\infty}\sum_{d=1}^D \Prob{\tilde{\bbM}_{d}\cap\bbM^{\star c}_d\neq \varnothing}
        \le \sum_{d=1}^D \alpha_d = \alpha$.
    \item (Consistency) $\lim\sup_{N\to\infty}D\sum_{d=1}^D \Prob{\tilde{\bbM}^c_{d}\cap\bbM^{\star }_d\neq \varnothing} = 0$.
    \item (Type I error control) Overall the procedure achieves type I error rate control:
    \begin{align*}
    \underset{N\to\infty}{\lim\sup}~\Prob{\hat{\bbM}\cap(\cup_{d=1}^D{\bbM_d^\star})^c\neq \varnothing}
    \le \alpha.
    \end{align*}
    \item (Selection consistency) When $\delta' $ is strictly positive, we have $\max_{d\in[D]}\alpha_d\to0$ and
    \begin{align*}
    \underset{N\to\infty}{\lim}~\Prob{\hat{\bbM} = \bigcup_{d=1}^D{\bbM_d^\star}  }
     = 1.
    \end{align*}
\end{enumerate}

\end{theorem}

Theorem \ref{thm:marginal-t-general}(i) and (ii) justified that forward selection based on Bonferroni corrected marginal t tests satisfy Assumption \ref{asp:valid-consistent} and \ref{asp:H-heredity} respectively, which build up the basis for applying Theorem \ref{thm:ms-consistency}. Theorem \ref{thm:marginal-t-general}(iii) guarantees type I error control under the significance level $\alpha$. When we let $\alpha$ decay to zero, Theorem \ref{thm:marginal-t-general}(iii) implies that we will not include redundant terms into the selected working model. Theorem \ref{thm:marginal-t-general}(iv) further states a stronger result with vanishing $\alpha$ - selection consistency can be achieved asymptotically.

\begin{proof}[Proof of Theorem \ref{thm:marginal-t}]
\begin{enumerate}[label = (\roman*)]
    \item First, we show the validity of the algorithm:
    \begin{align*}
        \Prob{\tilde{\bbM}_{d}\cap\bbM^{\star c}_d\neq \varnothing} 
        &= \Prob{\exists \cK\in\bbM^{\star }_{d,+}\backslash\bbM^{\star }_{d}, \lt|\frac{\htau_{\cK}}{\hat{v}_{\cK,\textsc{r}}}\rt|\ge\Phi^{-1}\lt(1-\frac{\alpha_d}{2|\bbM^{\star }_{d,+}|} \rt)}\\
        &\le \sum_{\cK\in\bbM^{\star }_{d,+}\backslash\bbM^{\star }_{d}} \Prob{\lt|\frac{\htau_{\cK}}{\hat{v}_{\cK,\textsc{r}}}\rt|\ge\Phi^{-1}\lt(1-\frac{\alpha_d}{2|\bbM^{\star }_{d,+}|} \rt)}\\
        &\le \sum_{\cK\in\bbM^{\star }_{d,+}\backslash\bbM^{\star }_{d}} ~\lt(\frac{\alpha_d}{|\bbM^{\star }_{d,+}|} + \frac{\tilde{C}}{(QN_0)^{1/3}}\rt)\see{by Corollary \ref{cor:factorial-student-be}}\\
        & \le \lt(\alpha_d + \frac{\tilde{C}|\bbM^\star_{d,+}|}{N^{1/3}}\rt).
    \end{align*}
    Hence,
    \begin{align*}
        \sum_{d=1}^D \Prob{\tilde{\bbM}_{d}\cap\bbM^{\star c}_d\neq \varnothing}
        \le 
        \sum_{d=1}^D \lt(\alpha_d + \frac{\tilde{C}|\bbM^\star_{d,+}|}{N^{1/3}}\rt).
    \end{align*}
   Due to the effect heredity condition \ref{cond:heredity}, we have
   \begin{align*}
      |\bbM^\star_{1,+}| = |\bbM^\star_{1}| ,
      \quad
      |\bbM^\star_{d,+}| \le K|\bbM^\star_{d-1}|. 
   \end{align*}
   Hence
   \begin{align}
       \limsup_{N\to\infty}\sum_{d=1}^D \Prob{\tilde{\bbM}_{d}\cap\bbM^{\star c}_d\neq \varnothing}
        \le 
       \alpha + \limsup_{N\to\infty}\frac{K\tilde{C}|\bbM^\star|}{N^{1/3}} = \alpha. \see{using Condition \ref{cond:order}(iii)} 
   \end{align}
    
    \item Second, we show the consistency of the algorithm. Assume the nonzero $\tau_{\cK}$'s are positive. If some are negative, one can simply modify the direction of some of the inequalities and still validate the proof. We have
    \begin{align*}
        &\Prob{\tilde{\bbM}^c_{d}\cap\bbM^{\star }_d\neq \varnothing} \\
        &= \Prob{\exists \cK\in \bbM^{\star }_{d}, \lt|\frac{\htau_{\cK}}{\hat{v}_{\cK,\textsc{r}}}\rt|\le\Phi^{-1}\lt(1-\frac{\alpha_d}{2|\bbM^{\star }_{d,+}|} \rt)}\\
        &\le \sum_{\cK\in\bbM^{\star }_{d}} \Prob{\lt|\frac{\htau_{\cK}}{\hat{v}_{\cK,\textsc{r}}}\rt|\le\Phi^{-1}\lt(1-\frac{\alpha_d}{2|\bbM^{\star }_{d,+}|} \rt)}\\
        &\le \sum_{\cK\in\bbM^{\star }_{d}} \Prob{\lt|\frac{\htau_{\cK}}{ v_{\cK.\textsc{r}}}\rt|\le\frac{\hat{v}_{\cK,\textsc{r}}}{v_{\cK.\textsc{r}}}\Phi^{-1}\lt(1-\frac{\alpha_d}{2|\bbM^{\star }_{d,+}|} \rt)}\\
        &\le \sum_{\cK\in\bbM^{\star }_{d}} \Prob{\lt|\frac{\htau_{\cK}}{ v_{\cK.\textsc{r}}}\rt|\le\lt\{1+\frac{\tilde{C}}{(QN_0)^{1/3}}\rt\}\Phi^{-1}\lt(1-\frac{\alpha_d}{2|\bbM^{\star }_{d,+}|} \rt)} + \Prob{\frac{\hat{v}_{\cK,\textsc{r}}}{v_{\cK.\textsc{r}}}> 1+\frac{\tilde{C}}{(QN_0)^{1/3}}}.
    \end{align*}
    For simplicity, let
    \begin{align*}
        Z_d^\star = \Phi^{-1}\lt(1-\frac{\alpha_d}{2|\bbM^{\star }_{d,+}|} \rt). 
    \end{align*}
    Then
    \begin{align}
        &\Prob{\tilde{\bbM}^c_{d}\cap\bbM^{\star }_d\neq \varnothing} \notag\\
        &\le  \sum_{\cK\in\bbM^{\star }_{d}} \lt(\Prob{-Z_d^\star - \frac{\tau_\cK}{v_{\cK.\textsc{r}}}\le\frac{\htau_{\cK}}{ v_{\cK.\textsc{r}}} - \frac{\tau_\cK}{v_{\cK.\textsc{r}}}\le Z_d^\star - \frac{\tau_\cK}{v_{\cK.\textsc{r}}}} + \frac{\tilde{C}}{(QN_0)^{1/3}}\rt) \label{eqn:typeII-bound}\\
        & =  \sum_{\cK\in\bbM_d^\star} \Phi\lt\{r_\cK^{-1}\lt(Z_d^\star - \frac{\tau_\cK}{v_{\cK.\textsc{r}}}\rt)\rt\} - \Phi\lt\{r_\cK^{-1}\lt(-Z_d^\star - \frac{\tau_\cK}{v_{\cK.\textsc{r}}}\rt)\rt\}(\triangleq \ostar)\\
        & + \frac{\tilde{C}|\bbM^\star_d|}{(QN_0)^{1/3}}.\label{eqn:typeII-limit}
    \end{align}
    The inequality from \eqref{eqn:typeII-bound} is derived as follows: first, with marginal $t$-tests in the selection step, the event
    \begin{align}
        \{\tilde{\bbM}^c_{d}\cap\bbM^{\star }_d\neq \varnothing\}
    \end{align}
    is equivalent to 
    \begin{align}\label{eqn:target-event}
        \lt\{-Z_d^\star - \frac{\tau_\cK}{v_{\cK.\textsc{r}}}\le\frac{\htau_{\cK}}{ v_{\cK.\textsc{r}}} - \frac{\tau_\cK}{v_{\cK.\textsc{r}}}\le Z_d^\star - \frac{\tau_\cK}{v_{\cK.\textsc{r}}}\rt\}.
    \end{align}
    Second, we apply the Bonferroni bound and the Berry-Esseen bounds given by Lemma \ref{lem:BE-finite-pop} to \eqref{eqn:target-event}, then the inequality is obtained. 
     With Condition \ref{cond:order}, we have
    \begin{gather}
        Z_d^\star =  \Theta\lt(\sqrt{2\ln\frac{2|\bbM_{d,+}^\star|}{\alpha_d}}\rt)  = \Theta(\sqrt{(\delta'+\delta''/3)\ln N}),\\ \lt|\frac{\tau_\cK}{v_{\cK.\textsc{r}}}\rt| =  \Theta(N^{1/2+\delta}) = \Theta(N^{\delta_0})\see{by defining $\delta_0 = 1/2+\delta > 0 $}.
    \end{gather}
    Because $\delta>-1/2$ and $\delta'\ge 0$, we have $|\frac{\tau_\cK}{v_{\cK.\textsc{r}}}| \to \infty$ and $Z_d^\star/(|\frac{\tau_\cK}{v_{\cK.\textsc{r}}}|) \to 0$.
    Therefore,
    \begin{align*}
        \Phi\lt\{r_\cK^{-1}\lt(Z_d^\star - \frac{\tau_\cK}{v_{\cK.\textsc{r}}}\rt)\rt\} - \Phi\lt\{r_\cK^{-1}\lt(-Z_d^\star - \frac{\tau_\cK}{v_{\cK.\textsc{r}}}\rt)\rt\} = \Theta(N^{-\delta_0}\exp\{-N^{2\delta_0}/2\}).
    \end{align*}
    Now applying Condition \ref{cond:order} again, we have
    \begin{align*}
        D\sum_{d=1}^D \Prob{\tilde{\bbM}^c_{d}\cap\bbM^{\star }_d\neq \varnothing} = \Theta\lt(D|\bbM^\star|N^{-\delta_0}\exp\{-N^{2\delta_0}/2\} +  {D|\bbM^\star|}/{N^{1/3}}\rt) = o(1).
    \end{align*}
    
    \item The Type I error rate control comes from Theorem \ref{thm:ms-consistency}. 
    
    \item The selection consistency result follows from Theorem \ref{thm:ms-consistency}.

\end{enumerate}
 
\end{proof}

\subsection{Statement and the proof of Lemma \ref{lem:fstar-f}}

The following lemma is the key to our inferential results, which gives an alternative identification of the causal parameter. 

\begin{lemma}\label{lem:fstar-f}
    Given $\bbM^\star$ is the true  working model,
we have $  ( f^{\star})^{\top}\overline{Y} =  f^\top\overline{Y}, ~\text{for all }  f\in\bbR^Q$.
\end{lemma}
\begin{proof}[Proof of Lemma \ref{lem:fstar-f}]
This identity holds for the true working model, not a general model, suggested by the following algebraic facts:
\begin{align}
     f^\top \overline{Y} &=   f^\top \{Q^{-1}G(\cdot, \bbM^\star)G(\cdot, \bbM^\star)^\top + Q^{-1}G(\cdot, \bbM^{\star c})G(\cdot, \bbM^{\star c})^\top\} \overline{Y} \see{orthogonality of $G$}\\
    & = ( f^\star)^\top \overline{Y} + G(\cdot, \bbM^{\star c}) \tau(\bbM^{\star c} ) \see{definition of $ f^\star$ based on \eqref{eqn:fM}}\\
    & = ( f^\star)^\top \overline{Y}. \see{using $\tau(\bbM^{\star c}) = 0$}
\end{align}
\end{proof}

\subsection{Proof of Theorem \ref{thm:be-perfect-ms}}
Theorem \ref{thm:be-perfect-ms} is a direct result of Theorem \ref{thm:marginal-t}, Lemma \ref{lem:BE-finite-pop} and the following Berry--Esseen bound:
\begin{lemma}[Berry--Esseen bound under selection consistency]\label{lem:BE-perfect}
Assume \eqref{eqn:nondegenerate-var-tw}. 
Then  
\begin{align}\label{eqn:be-perfect-ms}
    &\sup_{t\in\bbR}\lt|\Prob{\frac{\hgamma(\hat{\bbM})-\gamma}{v(\hat{\bbM})} \le t} - \Phi(t)\rt| \notag\\
    &\le 2\Prob{\hat{\bbM}\neq \bbM^\star} +  2C\sigma_w   \frac{\underline{c}^{-1} \max_{i\in[N],z\in[Q]}|Y_i(z)-\overline{Y}(z)|}{\sqrt{\overline{c}^{-1} \min_{z\in[Q]} S(z,z)}\cdot \sqrt{N_0}}\cdot  \frac{\|{ f}[\bbM^\star]\|_\infty}{\|{ f}[\bbM^\star]\|_2} .
\end{align}
\end{lemma}
\begin{proof}[Proof of Lemma \ref{lem:BE-perfect}]
This lemma is a direct application of Lemma \ref{lem:tail-perfect-ms}.
First, we check that 
\begin{align*}
    \gamma(\bbM^\star) = \gamma.
\end{align*}
From the definition of $\gamma$ \eqref{eqn:mean-hgamma}, we have
\begin{align*}
    \gamma &=  f^\top \overline{Y}\\
    &=  f^\top G\tau = f^\top G(\cdot, \bbM^\star)\tau(\bbM^\star)  \\
    &=Q^{-1} f^\top G(\cdot, \bbM^\star)G(\cdot,\bbM^\star)^\top \overline{Y} = \gamma(\bbM^\star).
\end{align*}

\end{proof}

Now apply Lemma \ref{lem:tail-perfect-ms} with $\bbM = \bbM^\star$ to get the result of Theorem \ref{thm:be-perfect-ms}.

\subsection{Statement and the proof of Lemma \ref{lem:RLS}}
The following lemma gives the closed-form solution of the RLS estimator \eqref{eqn:RLS-1}. 
\begin{lemma}\label{lem:RLS}
$\hY_{\textsc{r}}$ from \eqref{eqn:RLS-1} can be expressed as:
\begin{align}\label{eqn:RLS-2}
    \hY_{\textsc{r}} = Q^{-1}G(\cdot,  {\hat{\bbM}})G(\cdot,  {\hat{\bbM}})^\top \hY.
\end{align}
If $\hat{\bbM} = \bbM^\star$, then $\E{\hY_{\textsc{r}}}= \overline{Y}$.
\end{lemma}

\begin{proof}[Proof of Lemma \ref{lem:RLS}]

Due to the orthogonality of $G$, we have the following decomposition:
\begin{align*}
    \hY = Q^{-1}G(\cdot,  {\hat{\bbM}})G(\cdot,  {\hat{\bbM}})^\top \hY + Q^{-1}G(\cdot,  {\hat{\bbM}^c})G(\cdot,  {\hat{\bbM}^c})^\top \hY.
\end{align*}
By the constraint in \eqref{eqn:RLS-1}, we have 
\begin{align*}
    \|\hY - \mu\|^2 = \|Q^{-1}G(\cdot,  {\hat{\bbM}^c})G(\cdot,  {\hat{\bbM}^c})^\top \hY\|^2 + \|Q^{-1}G(\cdot,  {\hat{\bbM}})G(\cdot,  {\hat{\bbM}})^\top \hY - \mu\|^2,
\end{align*}
which is minimized at 
\begin{align*}
    \hmu = \hY_{\textsc{r}} =  Q^{-1}G(\cdot,  {\hat{\bbM}})G(\cdot,  {\hat{\bbM}})^\top \hY.
\end{align*}
Besides, $\hmu$ satisfies the constraint in \eqref{eqn:RLS-1}. 

Next we verify $\E{\hY_{\textsc{r}}} = \overline{Y}$ if $\hat{\bbM} = \bbM^\star$. Utilizing the orthogonality of $G$ again, we have
\begin{align*}
    \overline{Y} = Q^{-1}G(\cdot,  {{\bbM^\star}})G(\cdot,  {{\bbM^\star}})^\top \overline{Y} + Q^{-1}G(\cdot,  {{\bbM^\star}^c})G(\cdot,  {{\bbM^\star}^c})^\top \overline{Y}
\end{align*}
\end{proof}

\subsection{Proof of Proposition \ref{prop:ARE-comparison}}

\begin{proof}[Proof of Proposition \ref{prop:ARE-comparison}]
    (i) Based on the definition of $v_\textsc{r}^2$ and $v^2$, we have
    \begin{align*}
        \frac{v_\textsc{r}^2}{v^2} = \frac{ f^{\star \top} V_\hY  f^\star}{ f^\top V_\hY  f} = \frac{\| f^\star\|_2^2}{\| f\|_2^2} 
    \end{align*}
    because  $\kappa(V_\hY) = 1$.
    We further compute
    \begin{align*}
        \frac{\| f^\star\|_2^2}{\| f\|_2^2} = \frac{ f^\top \{Q^{-1}G(\cdot,\bbM^\star)G(\cdot,\bbM^\star)^\top\}  f}{ f^\top   f} \le 1
    \end{align*}
    where the inequality holds because of the following dominance relationship:
    \begin{align}
Q^{-1}G(\cdot,\bbM^\star)G(\cdot,\bbM^\star)^\top \preccurlyeq I_Q.
    \end{align}

    \noindent (ii) Because the order of the nonzero elements in $ f$ is not crucial here, we assume the first $s^\star$ coordinates of $ f$ are nonzero while the rest are zero without loss of generality. We can compute 
    \begin{align}\label{eqn:vr-over-v}
        \frac{v_\textsc{r}^2}{v^2} = \frac{ f^{\star \top} V_\hY  f^\star}{ f^\top V_\hY  f} \le \kappa(V_\hY)\cdot \frac{\| f^\star\|_2^2}{\| f\|_2^2}. 
    \end{align}
    For $ f^\star$, we have
    \begin{align*}
        \| f^\star\|_2 &= \|Q^{-1}G(\cdot,\bbM^\star) G(\cdot,\bbM^\star)^\top  f\|_2\\
        &=\lt\|Q^{-1}G(\cdot,\bbM^\star) G(\cdot,\bbM^\star)^\top \lt\{\sum_{s=1}^{s^\star}  f(s)\bse_s\rt\}\rt\|_2\\
        &\le \sum_{s=1}^{s^\star} | f(s)|\|Q^{-1}G(\cdot,\bbM^\star) G(\cdot,\bbM^\star)^\top \bse_s\|_2\\
        &=\lt(\frac{|\bbM^\star|}{Q}\rt)^{1/2}\sum_{s=1}^{s^\star} | f(s)| = \lt(\frac{|\bbM^\star|}{Q}\rt)^{1/2} \| f\|_1. 
    \end{align*}
    Then we have
    \begin{align}\label{eqn:fstar-over-f}
      \frac{\| f^\star\|_2^2}{\| f\|_2^2} \le \frac{|\bbM^\star|}{Q}\frac{\| f\|_1^2}{\| f\|_2^2} \le \frac{s^\star|\bbM^\star|}{Q}.
    \end{align}
    Combining \eqref{eqn:vr-over-v} and \eqref{eqn:fstar-over-f}, we conclude the result.
\end{proof}

As an extension of Proposition \ref{prop:ARE-comparison}, we compare the asymptotic lengths of confidence intervals in the following Proposition \ref{prop:ASY-CI-LEN-comparison}.
\begin{proposition}[Asymptotic length of confidence interval comparison]\label{prop:ASY-CI-LEN-comparison} 
Assume that both $\hat{\gamma}$ and $\hat{\gamma}_{\textsc{r}}$ converge to normal distributions with variances $v^2$ and $v^2_{\textsc{r}}$ as the sample size tends to infinity. Assume the variance estimators are consistent: $N(\hv^2 - v^2_{\mathrm{lim}}) = o_{\mathrm{P}}(1)$, $N(\hv_{\textsc{r}}^2 - v^2_{\textsc{r}, \mathrm{lim}}) = o_{\mathrm{P}}(1)$.
\begin{enumerate}[label = (\roman*)]
    \item If the condition number of $D_\hY$ satisfies $\kappa(D_\hY)=1$, we have
 \begin{align*}
    \frac{v_{\textsc{r}, \mathrm{lim}}^2}{v^2_{\mathrm{lim}}} \le 1. 
\end{align*}
    \item Let $s^\star$ denote the number of nonzero elements in $ f$, then we have
\begin{align*}
    \frac{v_{\textsc{r}, \mathrm{lim}}^2}{v^2_{\mathrm{lim}}} \le  \kappa(D_\hY) \cdot\frac{s^\star |\bbM^\star|}{Q}.
\end{align*}
\end{enumerate}
\end{proposition}

\subsection{Proof of Theorem \ref{thm:strategy-I}}
\begin{proof}[Proof of Theorem \ref{thm:strategy-I}]
According to Condition \ref{cond:under-selection} and Theorem \ref{thm:marginal-t}, with Strategy \ref{str:exclude-all}, 
\begin{align*}
    \Prob{\hat{\bbM} = \cup_{d=1}^{d^\star}\bbM^\star_d} \to 1. 
\end{align*}
We will apply Lemma \ref{lem:BE-perfect} with 
\begin{align*}
\bbM = \underline{\bbM}^\star = \cup_{d=1}^{d^\star}\bbM_d^\star.
\end{align*}
We only need to verify $\gamma = \gamma[\bbM]$ under the orthogonality condition \eqref{eqn:orthogonality}.
\begin{align*}
    \gamma &=  f^\top \overline{Y}\\
    &=  f^\top G\tau = f^\top G(\cdot, \underline{\bbM}^\star)\tau(\underline{\bbM}^\star) +  f^\top G(\cdot, \underline{\bbM}^{\star c})\tau(\underline{\bbM}^{\star c}).
\end{align*}
Now by \eqref{eqn:orthogonality}, $ f^\top G(\cdot, \bbM^c) = 0$. Hence
\begin{align*}
    \gamma = Q^{-1} f^\top G(\cdot, \cup_{d=1}^{d^\star}\bbM_d^\star)G(\cdot,\cup_{d=1}^{d^\star}\bbM_d^\star)^\top \overline{Y} = \gamma.
\end{align*}
\end{proof}

\subsection{Proof of Theorem \ref{thm:strategy-II}}

\begin{proof}[Proof of Theorem \ref{thm:strategy-II}]
This proof can be finished by applying Lemmas \ref{lem:tail-perfect-ms} and \ref{lem:ratio-bsf} with $\bbM=\overline{\bbM}^{\star}$ and checking $\gamma[\overline{\bbM}^{\star}] = \gamma$, which is omitted here. 
\end{proof}

\subsection{Proof of Proposition \ref{prop:ARE-comparison}}

\begin{proof}[Proof of Proposition \ref{prop:ARE-comparison}]
(i) Assume $ V_\hY = Q^{-2}G\Lambda G^\top$ where $\Lambda$ is a diagonal matrix in $\bbR^{Q\times Q}$.  We directly compute
\begin{align*}
    \frac{v_{\textsc{r}}^2}{v^2} = \frac{ f^{\star\top} V_\hY  f^\star}{ f^\top V_\hY  f} &= \frac{ f^\top 
    \{Q^{-1}G(\cdot,\bbM^\star)G(\cdot,\bbM^\star)^\top\}  
    \{Q^{-2}G\Lambda G^\top\} 
    \{Q^{-1}G(\cdot,\bbM^\star)G(\cdot,\bbM^\star)^\top\}  f}{ f^\top\{Q^{-2}G\Lambda G^\top\}  f}\\
    &= \frac{ f^\top 
    \{Q^{-2}G(\cdot,\bbM^\star) \Lambda(\bbM^\star,\bbM^\star) G(\cdot,\bbM^\star)^\top\}  f}{ f^\top\{Q^{-2}G\Lambda G^\top\}  f}\le 1.
\end{align*}

(ii) Because the order of the nonzero elements in $ f^\star$ is not crucial, we assume only the first $s^\star$ elements of $ f$ are nonzero. That is,
\begin{align}\label{eqn:expand-bsf}
     f = f_1 \bse_1 + \cdots + f_{s^\star} \bse_{s^\star}. 
\end{align}
We can verify that 
\begin{align}\label{eqn:bse}
    \|Q^{-1}G(\cdot,\bbM^\star)G(\cdot,\bbM^\star)^\top\bse_k\|_2 = \frac{|\bbM^\star|}{Q}, \quad \forall~ k\in[Q]. 
\end{align}
Therefore, 
\begin{align*}
    \frac{v_{\textsc{r}}^2}{v^2} = \frac{ f^{\star\top} V_\hY  f^\star}{ f^\top V_\hY  f} \le \frac{\varrho_{\max}(V_\hY)\| f^\star\|_2^2}{\varrho_{\min}(V_\hY)\| f\|_2^2} = \kappa(V_{\hY})\cdot\frac{\| f^\star\|_2^2}{\| f\|_2^2}.
\end{align*}
On the one hand, using $Q^{-1}G(\cdot,\bbM^\star)G(\cdot,\bbM^\star)^\top \preccurlyeq I_{Q} $, we have
\begin{align}\label{eqn:ratio-bsf-bound-1}
    \frac{\| f^\star\|_2^2}{\| f\|_2^2} \le 1.
\end{align}
On the other hand, using \eqref{eqn:expand-bsf} and \eqref{eqn:bse}, we have
\begin{align}\label{eqn:ratio-bsf-bound-2}
    \frac{\| f^\star\|_2^2}{\| f\|_2^2} \le \frac{\| f\|_1^2}{\| f\|_2^2} \cdot \frac{|\bbM^\star|}{Q} \le \frac{s^\star|\bbM^\star|}{Q}.
\end{align}
Combining \eqref{eqn:ratio-bsf-bound-1} and \eqref{eqn:ratio-bsf-bound-2} concludes the proof. 
\end{proof}

\subsection{Proof of Theorem \ref{thm:infer-order}}\label{sec:infer-order}

For simplicity, we focus on the case given by \eqref{eqn:best-Yz}. The general proof can be completed similarly.
We begin with the following lemma:
\begin{lemma}[Consistency of the selected tie sets]\label{lem:wt-converge}
Assume Conditions \ref{cond:uniform-design}, \ref{cond:regularity} and \ref{cond:distance}.  There exists universal constants $C,C'>0$, such that
when $N>n(\delta_1,\delta_2,\delta_3)$, we have 
\begin{align*}
   \Prob{\hat{\cT}_{1} = {\cT}_{1} } 
   \ge&  1 - \Prob{\hat{\bbM}\neq \bbM^\star}\\
    -& C|\cT'||\cT_{1}|\Bigg\{\sqrt{\frac{\bar{c}\Delta|\bbM^\star|}{N^{1+2\delta_2}}}\exp\lt(-\frac{C'N^{1+2\delta_2}}{\bar{c}\Delta|\bbM^\star|}\rt)\\
    +& \sigma   \frac{ \underline{c}^{-1/2} \max_{i\in[N],z\in[Q]}|Y_i(z)-\overline{Y}(z)|}{ {\overline{c}^{-1/2} \{\min_{z\in[Q]} S(z,z)}\}^{1/2}}\cdot \sqrt{\frac{|\bbM^\star|}{N}}\Bigg\}.
\end{align*}

\end{lemma}

Lemma \ref{lem:wt-converge} establishes a finite sample bound to quantify the performance of the tie set selection step in Algorithm \ref{alg:select-tie}. The bound in Lemma \ref{lem:wt-converge} implies that the performance of tie selection depends on several elements:
\begin{itemize}
\item Quality of effect selection. Ideally, we hope selection consistency can be achieved. In other words, the misselection probability $\Prob{\hat{\bbM}\neq \bbM^\star}$ is small in an asymptotic sense. 

\item Size of the tie $|\cT_1|$ and the number of factor combinations considered $|\cT'|$. These two quantities play a natural role because one can expect the difficulty of selection will increase if there are too many combinations present in the first tie or involved in comparison.

\item Size of between-group distance $d^\star_h$. If the gap between $\overline{Y}_{(1)}$ and the remaining order values are large, $\eta = \Theta(N^{\delta_2})$ is allowed to take larger values and the term 
\begin{align*}
\sqrt{\frac{\bar{c}\Delta|\bbM^\star|}{N^{1+2\delta_2}}}\exp\lt(-\frac{C'N^{1+2\delta_2}}{\bar{c}\Delta|\bbM^\star|}\rt)
\end{align*}
can become smaller in magnitude.

\item Population level property of potential outcomes. The scale of the centered potential outcomes $|Y_i(z) - \overline{Y}(z)|$ should be controlled, and the population variance $S(z,z)$ should be non-degenerate. 

\item The relative scale between number of nonzero effects $|\bbM^\star|$ and the total number of units $N$. The larger $N$ is compared to $|\bbM^\star|$, the easier for us to draw valid asymptotic conclusions. 
\end{itemize}

\begin{proof}[Proof of Lemma \ref{lem:wt-converge}]
The high-level idea of the proof is: we first prove the non-asymptotic bounds over the random event $\hat{\bbM} = \bbM^\star$, then make up for the cost of $\hat{\bbM} \neq \bbM^\star$. Over $\hat{\bbM} = \bbM^\star$, we have 
\begin{align*}
    \hat{Y}_{\textsc{r}} = \hat{Y}^\star_{\textsc{r}} = G(\cdot,{\bbM^\star}) \htau(\bbM^\star) = Q^{-1}G(\cdot,{\bbM^\star})G(\cdot,{\bbM^\star})^\top \hY.
\end{align*}

We apply Lemma \ref{lem:tail-perfect-ms} to establish a Berry--Esseen bound for each $\hY^\star_{\textsc{r}}(z)$. Note that
\begin{align}\label{eqn:hY-r-star}
    \hY^\star_{\textsc{r}}(z) =  f_{z}^{\star\top} \hY, ~ f_{z}^{\star\top} = Q^{-1}G(z, \bbM^\star) G(\cdot,\bbM^\star)^\top.
\end{align}
By calculation we have
\begin{align*}
    \| f_{z}^\star\|_\infty = Q^{-1}|\bbM^\star|, ~\| f_{z}^\star\|_2 = \sqrt{Q^{-1} |\bbM^\star|}.
\end{align*}
Also, we can show that
\begin{align*}
    \sum_{z'=1}^Q  { f}_{z}(z')^2 N_{z'}^{-1} S(z',z') \le \sigma^2 v^2(\bbM).
\end{align*}
and obtain
\begin{align}\label{eqn:BE-marginal-t}
  \sup_{t\in\bbR} \lt|\Prob{\frac{\hY^\star_{\textsc{r}}(z) - \overline{Y}(z)}{v_N}\le t} - \Phi(t)\rt| \le 2C\sigma   \frac{ \underline{c}^{-1} \max_{i\in[N],z\in[Q]}|Y_i(z)-\overline{Y}(z)|}{ {\overline{c}^{-1/2} \{\min_{z\in[Q]} S(z,z)}\}^{1/2}} \sqrt{\frac{|\bbM^\star|}{QN_0}}.
\end{align}

   \textbf{A probabilistic bound on the order statistics.} We show a bound on
\begin{align*}
    \Prob{\max_{z \in  \cT'\backslash\cT_{1}}\hY^\star_{\textsc{r}}(z) <\min_{z\in\cT_{1}}\hY^\star_{\textsc{r}}(z) \le \max_{z\in\cT_{1}}\hY^\star_{\textsc{r}}(z)}.
\end{align*}

It is known that \citep[Exercise 2.2]{wainwright2019high}
\begin{align*}
    1-\Phi(x) = \int_x^\infty \phi(t) dt \le \frac{1}{x}\int_x^\infty t\phi(t) dt \le \frac{1}{\sqrt{2\pi}x}\lt\{\exp\lt(-\frac{x^2}{2}\rt)\rt\}.
\end{align*}

Hence  
\begin{align}\label{eqn:dev-bound}
       &\Prob{\sqrt{N}\lt|\hY^\star_{\textsc{r}}(z) - \overline{Y}(z)\rt|\ge {\sqrt{N}d_h^\star} }\notag\\
   \le    &\frac{v_N}{\sqrt{2\pi }d_h^\star}\cdot \exp\lt(-\frac{ d_h^{\star 2}}{2v_N^2}\rt) + 2C\sigma   \frac{ \underline{c}^{-1} \max_{i\in[N],z\in[Q]}|Y_i(z)-\overline{Y}(z)|}{ {\overline{c}^{-1/2} \{\min_{z\in[Q]} S(z,z)}\}^{1/2}}\cdot \sqrt{\frac{|\bbM^\star|}{N_0Q}}.
\end{align}

Therefore, for all $z\in\cT'\backslash\cT_1$ and $z'\in\cT_{1}$,
\begin{align*}
    &\Prob{\hY^\star_{\textsc{r}}(z') - \hY^\star_{\textsc{r}}(z)<0} \\
    &= \Prob{\sqrt{N}(\hY^\star_{\textsc{r}}(z') - \overline{Y}(z')) - \sqrt{N}(\hY^\star_{\textsc{r}}(z) - \overline{Y}(z))<\sqrt{N}(\overline{Y}(z) - \overline{Y}(z'))} \\
    & \le \Prob{\sqrt{N}(\hY^\star_{\textsc{r}}(z') - \overline{Y}(z')) - \sqrt{N}(\hY^\star_{\textsc{r}}(z) - \overline{Y}(z))<-2\sqrt{N}d_h^\star} \\
    & = {\mathrm{P}}\Big\{\sqrt{N}(\hY^\star_{\textsc{r}}(z') - \overline{Y}(z')) - \sqrt{N}(\hY^\star_{\textsc{r}}(z) - \overline{Y}(z))<-2\sqrt{N}d_h^\star, \\
    &\phantom{hello}\sqrt{N}(\hY^\star_{\textsc{r}}(z) - \overline{Y}(z))< {\sqrt{N}d_h^\star}\Big\} \\
    & + {\mathrm{P}}\Big\{\sqrt{N}(\hY^\star_{\textsc{r}}(z') - \overline{Y}(z')) - \sqrt{N}(\hY^\star_{\textsc{r}}(z) - \overline{Y}(z))<-2\sqrt{N}d_h^\star, \\
    &\phantom{hello}\sqrt{N}(\hY^\star_{\textsc{r}}(z) - \overline{Y}(z))\ge {\sqrt{N}d_h^\star} \Big\} \\
    & \le \Prob{\sqrt{N}(\hY^\star_{\textsc{r}}(z') - \overline{Y}(z')) < - {\sqrt{N}d_h^\star} }
     + \Prob{\sqrt{N}(\hY^\star_{\textsc{r}}(z) - \overline{Y}(z))\ge{\sqrt{N}d_h^\star}}.
\end{align*}

Using \eqref{eqn:dev-bound} we have
\begin{align*}
    &\Prob{\hY^\star_{\textsc{r}}(z') - \hY^\star_{\textsc{r}}(z) < 0} \\
    \le& \frac{\sqrt{\bar{c}\Delta|\bbM^\star|}}{\sqrt{2\pi N_0Q }d_h^\star}\cdot \exp\lt(-\frac{ N_0Q d_h^{\star 2}}{2\bar{c}\bar{s}|\bbM^\star|}\rt) + 2C\sigma   \frac{ \underline{c}^{-1} \max_{i\in[N],z\in[Q]}|Y_i(z)-\overline{Y}(z)|}{ {\overline{c}^{-1/2} \{\min_{z\in[Q]} S(z,z)}\}^{1/2}}\cdot \sqrt{\frac{|\bbM^\star|}{N_0Q}}.
\end{align*}

Now a union bound gives 
\begin{align*}
    &\Prob{\min_{z'\in\cT_1}\hY^\star_{\textsc{r}}(z') - \max_{z\in\cT'\backslash\cT_1}\hY^\star_{\textsc{r}}(z) < 0}\\
    &\ge 1 - |\cT_{1}||\cT'|\lt\{\frac{\sqrt{\bar{c}\bar{s}|\bbM^\star|}}{\sqrt{2\pi N_0Q }d_h^\star}\cdot \exp\lt(-\frac{ N_0Q d_h^{\star 2}}{2\bar{c}\bar{s}|\bbM^\star|}\rt) + 2C\sigma   \frac{ \underline{c}^{-1} \max_{i\in[N],z\in[Q]}|Y_i(z)-\overline{Y}(z)|}{ {\overline{c}^{-1/2} \{\min_{z\in[Q]} S(z,z)}\}^{1/2}}\cdot \sqrt{\frac{|\bbM^\star|}{N_0Q}}\rt\}.
\end{align*}
Now using that $d_h^\star =  \Theta(N^{\delta_1})$, $Nd_h^{\star2} =  \Theta(N^{1+2\delta_1})$ with $1+2\delta_1>0$. The first term in the bracket has the following order
\begin{align*}
\frac{\sqrt{\bar{c}\bar{s}|\bbM^\star|}}{\sqrt{2\pi N_0Q }d_h^\star}\cdot \exp\lt(-\frac{ N_0Q d_h^{\star 2}}{2\bar{c}\bar{s}|\bbM^\star|}\rt) = \Theta\lt(\sqrt{\frac{\bar{c}\bar{s}|\bbM^\star|}{N^{1+2\delta_1}}}\exp\lt\{-\frac{C'N^{1+2\delta_1}}{\bar{c}\bar{s}|\bbM^\star|}\rt\}\rt)
\end{align*}
where $C'>0$ is a universal constant due to Condition \ref{cond:order}.Note that $\delta_2 > \delta_1$. Thus when $N$ is large enough, we have
\begin{align}
    &\Prob{\min_{z'\in\cT_1}\hY^\star_{\textsc{r}}(z') - \max_{z\in\cT'\backslash\cT_1}\hY^\star_{\textsc{r}}(z) < 0}
    \notag\\
    \ge &1 -  C|\cT_{1}||\cT'|\lt\{\sqrt{\frac{\bar{c}\bar{s}|\bbM^\star|}{N^{1+2\delta_1}}}\exp\lt\{-\frac{C'N^{1+2\delta_1}}{\bar{c}\bar{s}|\bbM^\star|}\rt\} + \sigma   \frac{ \underline{c}^{-1} \max_{i\in[N],z\in[Q]}|Y_i(z)-\overline{Y}(z)|}{ {\overline{c}^{-1/2} \{\min_{z\in[Q]} S(z,z)}\}^{1/2}}\cdot \sqrt{\frac{|\bbM^\star|}{N_0Q}}\rt\}.\label{eqn:order}
\end{align}

\textbf{Nice separation.} Consider the following random index:
    \begin{align*}
        \tilde{z} \in \underset{z\in\cT'}{\arg\max}~\hY^\star_{\textsc{r}}(z).
    \end{align*}
For any $\bar{\epsilon}>0$, 
\begin{align}
    &\Prob{\min_{z\notin \cT_{1}}|\hY^\star_{\textsc{r}}(z) - \hY^\star_{\textsc{r}}(\tilde{z})|/\eta \ge 2\bar{\epsilon}} \notag\\
    & \ge \Prob{\min_{z\notin \cT_{1}, z'\in \cT_{1}}|\hY^\star_{\textsc{r}}(z) - \hY^\star_{\textsc{r}}({z}')|/\eta \ge 2\bar{\epsilon}, \tilde{z}\in \cT_{1}}\notag\\
    & \ge \Prob{\min_{z\notin \cT_{1}, z'\in \cT_{1}}|\hY^\star_{\textsc{r}}(z) - \hY^\star_{\textsc{r}}({z}')|/\eta \ge 2\bar{\epsilon}} + \Prob{\tilde{z}\in \cT_{1}} - 1\notag\\
    & \ge \Prob{\tilde{z}\in \cT_{1}} - \sum_{z\notin \cT_{1}, z'\in\cT_{1}}\Prob{|\hY^\star_{\textsc{r}}(z) - \hY^\star_{\textsc{r}}(\tilde{z}')|/\eta \le 2\bar{\epsilon}}. \label{eqn:separation-1}
\end{align}
To proceed we have the following bound:
\begin{align*}
    & \Prob{|\hY^\star_{\textsc{r}}(z) - \hY^\star_{\textsc{r}}(z')|/\eta \le 2\bar{\epsilon}} \\
    = & \Prob{|\{\hY^\star_{\textsc{r}}(z)-\overline{Y}(z)} - \{\hY^\star_{\textsc{r}}(z')-\overline{Y}(z')\} - \{\overline{Y}(z)-\overline{Y}(z')\}| \le 2\bar{\epsilon}\eta\}\\
    \le &  \Prob{|\overline{Y}(z)-\overline{Y}(z')|-|\hY^\star_{\textsc{r}}(z)-\overline{Y}(z)| - |\hY^\star_{\textsc{r}}(z')-\overline{Y}(z')| \le 2\bar{\epsilon}\eta}\\
    \le & \Prob{|\hY^\star_{\textsc{r}}(z)-\overline{Y}(z)| + |\hY^\star_{\textsc{r}}(z')-\overline{Y}(z')|\ge 2d_h^\star - 2\bar{\epsilon}\eta }\\
    \le & \Prob{|\hY^\star_{\textsc{r}}(z)-\overline{Y}(z)|\ge  {d_h^\star - \bar{\epsilon}\eta} } + \Prob{|\hY^\star_{\textsc{r}}(z')-\overline{Y}(z')|\ge {d_h^\star - \bar{\epsilon}\eta} } \\
    & \see{because $z\notin\cT_{1}$ and $z'\in\cT_{1}$}\\
    \le & 4\Bigg\{ \frac{\sqrt{\bar{c}\Delta|\bbM^\star|}}{\sqrt{2\pi N_0Q }(d_h^\star - \overline{\epsilon}\eta)}\cdot \exp\lt(-\frac{ N_0Q (d_h^\star - \overline{\epsilon}\eta)^2}{2\bar{c}\bar{s}|\bbM^\star|}\rt) \\
    + & 2C\sigma   \frac{ \underline{c}^{-1} \max_{i\in[N],z\in[Q]}|Y_i(z)-\overline{Y}(z)|}{ {\overline{c}^{-1/2} \{\min_{z\in[Q]} S(z,z)}\}^{1/2}}\cdot \sqrt{\frac{|\bbM^\star|}{N_0Q}}\Bigg\}. \\
    & \see{This is deduced analogously to the proof in the previous part}
\end{align*}
By Condition \ref{cond:distance}, we know that when $N$ is large enough,
\begin{align*}
    d_h^\star - \bar{\epsilon}\eta > d_h^\star/2.
\end{align*}
Hence, for $N>N(\delta_1,\delta_2)$, we have
\begin{align*}
    &\sum_{z\notin \cT_{1}, z'\in\cT_{1}}\Prob{|\hY^\star_{\textsc{r}}(z) - \hY^\star_{\textsc{r}}(z')|/\eta \le 2\bar{\epsilon}} \\
    \le & 4|\cT_{1}| |\cT'|\lt\{ \frac{\sqrt{2\bar{c}\bar{s}|\bbM^\star|}}{\sqrt{\pi N_0Q }{d_h^\star}}\cdot \exp\lt(-\frac{ N_0Q d_h^{\star 2}}{8\bar{c}\bar{s}|\bbM^\star|}\rt)
    +  2C\sigma   \frac{ \underline{c}^{-1} \max_{i\in[N],z\in[Q]}|Y_i(z)-\overline{Y}(z)|}{ {\overline{c}^{-1/2} \{\min_{z\in[Q]} S(z,z)}\}^{1/2}}\cdot \sqrt{\frac{|\bbM^\star|}{N_0Q}}\rt\}.
\end{align*}

Combined with \eqref{eqn:separation-1},  we have:
\begin{align}
    &\Prob{\min_{z\notin \cT_{1}}|\hY^\star_{\textsc{r}}(z) - \hY^\star_{\textsc{r}}(\tilde{z})|/\eta \ge 2\bar{\epsilon}} \notag\\
    \ge & \Prob{\tilde{m}\in \cT_{1}} -   \underbrace{4|\cT_{1}| |\cT'|  \frac{\sqrt{2\bar{c}\bar{s}|\bbM^\star|}}{\sqrt{\pi N_0Q }{d_h^\star}}\cdot \exp\lt(-\frac{ N_0Q d_h^{\star 2}}{8\bar{c}\bar{s}|\bbM^\star|}\rt)}_{\text{Term I}}\notag\\
    - &  \underbrace{4|\cT_{1}| |\cT'| 2C\sigma   \frac{ \underline{c}^{-1} \max_{i\in[N],z\in[Q]}|Y_i(z)-\overline{Y}(z)|}{ {\overline{c}^{-1/2} \{\min_{z\in[Q]} S(z,z)}\}^{1/2}}\cdot \sqrt{\frac{|\bbM^\star|}{N_0Q}}}_{\text{Term II}}. \label{eqn:large}
\end{align}

Analogous to the discussion in the previous part, when $N$ is sufficiently large, we can show
\begin{align*}
    &\Prob{\min_{z\notin \cT_{1}}|\hY^\star_{\textsc{r}}(z) - \hY^\star_{\textsc{r}}(\tilde{z})|/\eta \ge 2\bar{\epsilon}} \\
    \ge & \Prob{\tilde{m}\in \cT_{1}} - C|\cT_{1}| |\cT'| \lt\{\sqrt{\frac{\bar{c}\bar{s}|\bbM^\star|}{N^{1+2\delta_2}}}\exp\lt\{-\frac{C'N^{1+2\delta_2}}{\bar{c}\bar{s}|\bbM^\star|}\rt\} + \sigma   \frac{ \underline{c}^{-1} \max_{i\in[N],z\in[Q]}|Y_i(z)-\overline{Y}(z)|}{ {\overline{c}^{-1/2} \{\min_{z\in[Q]} S(z,z)}\}^{1/2}}\cdot \sqrt{\frac{|\bbM^\star|}{N_0Q}}\rt\}.
\end{align*}

Similarly we can show for any $z\in\cT_{1}$ and $\underline{\epsilon}>0$,
\begin{align*} 
    &\Prob{\max_{z\in\cT_{1}}|\hY^\star_{\textsc{r}}(z) - \hY^\star_{\textsc{r}}(\tilde{z})|/\eta \le 2\underline{\epsilon}} \\
    & \ge \Prob{\tilde{z}\in\cT_{1}}  
     - \sum_{z\neq z'\in\cT_{1}}\Prob{|\hY^\star_{\textsc{r}}(z) - \hY^\star_{\textsc{r}}({z'})|/\eta > 2\underline{\epsilon}}.\notag
\end{align*}
Then we have for $z\neq z'\in\cT_{1}$,
\begin{align*}
    &\Prob{|\hY^\star_{\textsc{r}}(z) - \hY^\star_{\textsc{r}}({z'})|/\eta > 2\underline{\epsilon}} \\
    & \le \Prob{|\hY^\star_{\textsc{r}}(z)-\overline{Y}(z)|\ge  {\underline{\epsilon}\eta - d_h}   } + \Prob{|\hY^\star_{\textsc{r}}(z') - \overline{Y}({z'})|\ge  {\underline{\epsilon}\eta - d_h}   }\\
    &\le 4\Bigg\{ \frac{\sqrt{\bar{c}\bar{s}|\bbM^\star|}}{\sqrt{2\pi N_0Q }({\underline{\epsilon}\eta - d_h})}\cdot \exp\lt(-\frac{ N_0Q ({\underline{\epsilon}\eta - d_h})^{  2}}{2\bar{c}\bar{s}|\bbM^\star|}\rt)\\
    &+ 2C\sigma   \frac{ \underline{c}^{-1} \max_{i\in[N],z\in[Q]}|Y_i(z)-\overline{Y}(z)|}{ {\overline{c}^{-1/2} \{\min_{z\in[Q]} S(z,z)}\}^{1/2}}\cdot \sqrt{\frac{|\bbM^\star|}{N_0Q}}\Bigg\}.
\end{align*}
By the scaling of the parameters, when $N_0$ is large enough $N>N(\delta_2,\delta_3)$, $\underline{\epsilon}\eta-d_h >\underline{\epsilon}\eta/2$. Therefore,
\begin{align*}
    &\Prob{|\hY^\star_{\textsc{r}}(z) - \hY^\star_{\textsc{r}}({z'})|/\eta > 2\underline{\epsilon}} \\
    \le & 4\lt\{ \frac{\sqrt{2\bar{c}\bar{s} |\bbM^\star|}}{\sqrt{\pi N_0Q }({\underline{\epsilon}\eta})}\cdot \exp\lt(-\frac{ N_0Q ({\underline{\epsilon}\eta})^{  2}}{8\bar{c}\bar{s} |\bbM^\star|}\rt)
    +  2C\sigma   \frac{ \underline{c}^{-1} \max_{i\in[N],z\in[Q]}|Y_i(z)-\overline{Y}(z)|}{ {\overline{c}^{-1/2} \{\min_{z\in[Q]} S(z,z)}\}^{1/2}}\cdot \sqrt{\frac{|\bbM^\star|}{N_0Q}}\rt\}.
\end{align*}

Hence we  have:
\begin{align*}
    &\Prob{\max_{z\in\cT_{1}}|\hY^\star_{\textsc{r}}(z) - \hY^\star_{\textsc{r}}(\tilde{z})|/\eta \le 2\underline{\epsilon}} \\
    \ge & \Prob{\tilde{z}\in \cT_{1}} -   \underbrace{4|\cT_{1}| |\cT'|  \frac{\sqrt{2\bar{c}\bar{s}|\bbM^\star|}}{\sqrt{\pi N_0Q }({\underline{\epsilon}\eta})}\cdot \exp\lt(-\frac{ N_0Q ({\underline{\epsilon}\eta})^{\star 2}}{8\bar{c}\bar{s}|\bbM^\star|}\rt)}_{\text{Term I}}\notag\\
    - &  \underbrace{4|\cT_{1}| |\cT'| 2C\sigma   \frac{ \underline{c}^{-1} \max_{i\in[N],z\in[Q]}|Y_i(z)-\overline{Y}(z)|}{ {\overline{c}^{-1/2} \{\min_{z\in[Q]} S(z,z)}\}^{1/2}}\cdot \sqrt{\frac{|\bbM^\star|}{N_0Q}}}_{\text{Term II}}.
\end{align*}
Again, by the conditions, we can show
\begin{align*}
    &\Prob{\max_{z\in\cT_{1}}|\hY^\star_{\textsc{r}}(z) - \hY^\star_{\textsc{r}}(\tilde{z})|/\eta \le 2\underline{\epsilon}} \\
    \ge & \Prob{\tilde{z}\in \cT_{1}} 
    - C|\cT_{1}| |\cT'| \lt\{\sqrt{\frac{\bar{c}\bar{s}|\bbM^\star|}{N^{1+2\delta_2}}}\exp\lt\{-\frac{C'N^{1+2\delta_2}}{\bar{c}\bar{s}|\bbM^\star|}\rt\} + \sigma   \frac{ \underline{c}^{-1} \max_{i\in[N],z\in[Q]}|Y_i(z)-\overline{Y}(z)|}{ {\overline{c}^{-1/2} \{\min_{z\in[Q]} S(z,z)}\}^{1/2}}\cdot \sqrt{\frac{|\bbM^\star|}{N_0Q}}\rt\}.
\end{align*}

Applying \eqref{eqn:order} we know that
\begin{align*}
    &{\mathrm{P}}\{\tilde{z}\in\cT_{1}\} \\
    \ge& 1 - C|\cT'||\cT_{1}|\lt\{\sqrt{\frac{\bar{c}\bar{s}|\bbM^\star|}{N^{1+2\delta_2}}}\exp\lt(-\frac{C'N^{1+2\delta_2}}{\bar{c}\bar{s}|\bbM^\star|}\rt)+\sigma   \frac{ \underline{c}^{-1} \max_{i\in[N],z\in[Q]}|Y_i(z)-\overline{Y}(z)|}{ {\overline{c}^{-1/2} \{\min_{z\in[Q]} S(z,z)}\}^{1/2}}\cdot \sqrt{\frac{|\bbM^\star|}{N_0Q}}\rt\}.
\end{align*}

\textbf{Aggregating parts.} Using all the results above, we can show that, when $N>n(\delta_1,\delta_2,\delta_3)$, we have
\begin{align}
     &\Prob{\max_{z\in\cT_{1}}|\hY^\star_{\textsc{r}}(z) - \hY^\star_{\textsc{r}}(\tilde{z})|\le \underline{\epsilon}\eta, \min_{z\notin\cT_{1}}|\hY^\star_{\textsc{r}}(z) - \hY^\star_{\textsc{r}}(\tilde{z})|\ge \bar{\epsilon}\eta } \notag\\
     \ge & 1 -  C|\cT'||\cT_{1}|\lt\{\sqrt{\frac{\bar{c}\bar{s}|\bbM^\star|}{N^{1+2\delta_2}}}\exp\lt(-\frac{C'N^{1+2\delta_2}}{\bar{c}\bar{s}|\bbM^\star|}\rt)+\sigma   \frac{ \underline{c}^{-1} \max_{i\in[N],z\in[Q]}|Y_i(z)-\overline{Y}(z)|}{ {\overline{c}^{-1/2} \{\min_{z\in[Q]} S(z,z)}\}^{1/2}}\cdot \sqrt{\frac{|\bbM^\star|}{N_0Q}}\rt\}. \label{eqn:key-bound}
\end{align}

\commenting{
\textbf{Making adjustment for the failure of selection consistency.}
Now to translate the above results into a non-asymptotic bound for the algorithm output $\hY_{\textsc{r}}$, we only need to adjust for failure of  selection:
\begin{align*}
     &\Prob{\max_{z\in\cT_{K_0;h}}|\hY_{\textsc{r}}(z) - \hY_{\textsc{r}}(\tilde{z})|\le \underline{\epsilon}\eta, \min_{z\notin\cT_{K_0;h}}|\hY_{\textsc{r}}(z) - \hY_{\textsc{r}}(\tilde{z})|\ge \bar{\epsilon}\eta } \\
     \ge & 1 - 2(1 - \Prob{\tilde{z}\in \cT_{K_0;h}}) 
     -  16|\cT_{K_0;h}| |\Gamma_{K_0}| \frac{C \max_{i\in[N],z\in\cT}|\breve{Y}_i(z)| }{\sqrt{\underline{\lambda}}}\cdot \sqrt{\frac{|\bbM^\star|}{Q N_0}} - {\mathrm{P}}\{\hat{\bbM} \neq {\bbM}^\star\}.
\end{align*}
}

    \textbf{Bounding the factor level combination selection probability.} For the selected set $\hat{\cT}_1$, we have
    \begingroup
    \allowdisplaybreaks
    \begin{align*}
        &\Prob{\hat{\cT}_{1} = {\cT}_{1} } \\
        = & {\mathrm{P}}\Bigg\{|\hY_{\textsc{r}}(z)-\max_{z\in\cT'}\hY_{\textsc{r}}(z)| \le \underline{\epsilon}\eta ,  \text{ for }z\in\cT_{1} ; \\
        &\phantom{ {\mathrm{P}} } |\hY_{\textsc{r}}(z)-\max_{z\in\cT'}\hY_{\textsc{r}}(z)| > \underline{\epsilon}\eta ,  \text{ for }z\notin\cT_{1}\Bigg\} \\
        \ge & {\mathrm{P}}\Bigg\{  |\hY^\star_{\textsc{r}}(z)-\max_{z\in\cT'}\hY^\star_{\textsc{r}}(z)| \le \underline{\epsilon}\eta ,  \text{ for }z\in\cT_{1} ; \\
        &\phantom{ {\mathrm{P}}\Bigg\{  } |\hY^\star_{\textsc{r}}(z)-\max_{z\in\cT'}\hY^\star_{\textsc{r}}(z)| > \underline{\epsilon}\eta ,  \text{ for }z\notin\cT_{1}\Bigg\} - {\mathrm{P}}\{\hat{\bbM}\neq \bbM^\star\}\\
        = & {\mathrm{P}}\Bigg\{    |\hY^\star_{\textsc{r}}(z)-\hY^\star_{\textsc{r}}(\tilde{z})| \le \underline{\epsilon}\eta ,  \text{ for }z\in\cT_{1} ; \\
        &\phantom{ {\mathrm{P}}\Bigg\{  } |\hY^\star_{\textsc{r}}(z)-\hY^\star_{\textsc{r}}(\tilde{z})| > \underline{\epsilon}\eta ,  \text{ for }z\notin\cT_{1}\Bigg\}  - {\mathrm{P}}\{\hat{\bbM}\neq \bbM^\star\}\\
        &\see{where we introduce random index $\tilde{z}$ to record the position that achieves maximum} \\
       \ge & {\mathrm{P}}\Bigg\{    |\hY^\star_{\textsc{r}}(z)-\hY^\star_{\textsc{r}}(\tilde{z})| \le \underline{\epsilon}\eta ,  \text{ for }z\in\cT_{1} ; \\
        &\phantom{ {\mathrm{P}}\Bigg\{  } |\hY^\star_{\textsc{r}}(z)-\hY^\star_{\textsc{r}}(\tilde{z})| > \overline{\epsilon}\eta ,  \text{ for }z\notin\cT_{1}\Bigg\}  - {\mathrm{P}}\{\hat{\bbM}\neq \bbM^\star\}\\
        &\see{simply using the fact that $\overline{\epsilon} > \underline{\epsilon}$} \\
        = & {\mathrm{P}}\Bigg\{    \max_{z\in\cT_{1}}|\hY^\star_{\textsc{r}}(z)-\hY^\star_{\textsc{r}}(\tilde{z})| \le \underline{\epsilon}\eta;  \min_{z\notin\cT_{1}}|\hY^\star_{\textsc{r}}(z)-\hY^\star_{\textsc{r}}(\tilde{z})| > \overline{\epsilon}\eta\Bigg\} \\
        &- {\mathrm{P}}\{\hat{\bbM}\neq \bbM^\star\}\\
        \ge & 1 - \sum_{h=1}^{H_0}\lt(1 - \Prob{\max_{z\in\cT_{1}}|\hY^\star_{\textsc{r}}(z)-\hY^\star_{\textsc{r}}(\tilde{z})| \le \underline{\epsilon}\eta;  \min_{z\notin\cT_{1}}|\hY^\star_{\textsc{r}}(z)-\hY^\star_{\textsc{r}}(\tilde{z})| > \overline{\epsilon}\eta}\rt) \\
        & - {\mathrm{P}}\{\hat{\bbM}\neq \bbM^\star\}\\
        \ge & 1 - {\mathrm{P}}\{\hat{\bbM}\neq \bbM^\star\} \\
        -&C|\cT'||\cT_{1}|\lt\{\sqrt{\frac{\bar{c}\bar{s}|\bbM^\star|}{N^{1+2\delta_2}}}\exp\lt(-\frac{C'N^{1+2\delta_2}}{\bar{c}\bar{s}|\bbM^\star|}\rt)+\sigma   \frac{ \underline{c}^{-1} \max_{i\in[N],z\in[Q]}|Y_i(z)-\overline{Y}(z)|}{ {\overline{c}^{-1/2} \{\min_{z\in[Q]} S(z,z)}\}^{1/2}}\cdot \sqrt{\frac{|\bbM^\star|}{N_0Q}}\rt\}.
    \end{align*}
    \endgroup

\end{proof}

Lemma \ref{lem:wt-converge} suggests that under the conditions assumed in Theorem \ref{thm:infer-order}, we select the first tie set consistently as $N\to\infty$. Now Theorem \ref{thm:infer-order} is a direct result of Lemma \ref{lem:BE-perfect} and Lemma \ref{lem:wt-converge}.

\end{document}